# Gonogo: An R Implementation of Test Methods to Perform, Analyze and Simulate Sensitivity Experiments

Paul A. Roediger (proediger@comcast.net)

**Abstract**

This work provides documentation for a suite of R functions contained in gonogo.R. The functions provide sensitivity testing practitioners and researchers with an ability to conduct, analyze and simulate various sensitivity experiments involving binary responses and a single stimulus level (e.g., drug dosage, drop height, velocity, etc.). Included are the modern Neyer and 3pod adaptive procedures, as well as the Bruceton and Langlie. The latter two benchmark procedures are capable of being performed according to generalized up-down transformed-response rules. Each procedure is designated phase-one of a three-phase experiment. The goal of phase-one is to achieve overlapping data. The two additional (and optional) refinement phases utilize the D-optimal criteria and the Robbins-Monro-Joseph procedure. The goals of the two refinement phases are to situate testing in the vicinity of the median and tails of the latent response distribution, respectively.

**Comments**

This documentation is 58 pages in length and contains 31 figures, 40 tables and 2 flow diagrams. The subject of much of the paper, the gonogo.R file, contains 118 functions plus 2 constants and is available online.

## Table of Contents

## 1. Background

In 1994 Barry Neyer published what has become known as the Neyer D-Optimal procedure (Neyer 1994). Unlike some existing methods, one part of it (the D-optimal part 3) required a computer program to calculate the sequence of stress/response pairs. Some years later the procedure became available to sensitivity testing practitioners in proprietary software called SenTest™ (Neyer Software LLC, SenTest™, Version 1.0, http://neyersoftware.com/SensitivityTest/SensitivityTestFlyer.htm). His contribution can hardly be overestimated. The present author had a technical role in the 2009 issuance of Change 1 to MIL-STD-331C, which added an Appendix G entitled "Statistical Methods to Determine the Initiation Probability of One-shot Devices". Since the standard, widely used as a benchmark for choosing a sensitivity testing procedure, was slated to include a description of the Neyer test procedure, it was prudent to dig into the methodology and develop an alternative, non-commercial implementation. Barry graciously answered most of the many questions asked of him during this endeavor. The current version of the standard is MIL-STD-331D (United States Department of Defense 2017).

In 2012, a draft copy of a new three-phase procedure, dubbed 3pod was circulated. Since 3pod's phase II was Neyer's part 3, it was a logical step to add phase III to the already existing generic Neyer parts 1 and 2 (phase I) and part 3 (phase II) and add another phase I option to include 3pod. This new R implementation of the two procedures was called gonogo. This upgrade couldn't have taken place without having many dialogues with Jeff Wu and Danny Wang. Since then, 3pod has been modified twice. The first was a modification to phase 1 (called 3podM), and the second to phase III, ensuring it gives consistent results when mlo and mhi are translated and/or scaled. The latter property of phase III is called "equivariance" and the combined modifications constitute 3pod2.0 (Wang, Tian and Wu 2020). Gonogo's implementation of 3pod is 3pod2.0, but we normally refer to it simply as 3pod.

Lastly, to keep gonogo in synch with a future revision of Appendix G of MIL-STD-331D, two other historically important procedures were added, namely the Bruceton and Langlie methods. Choosing one of these methods permits phase I to be a single "up-down" or "one shot" transformed response test (abbreviated UDTR or OSTR), or two such transformed tests (both Bruceton, or both Langlie) performed in succession. The reasoning to build in the option to do two Bruceton tests (or two Langlie tests), one "homing in" to $L_p$, the other to $L_{1-p}$ ($p \neq .5$), is that such tests are described (and recommended) by Einbinder (1973), and they introduce phase III's "homing in" ability early on in phase I.

All four procedures have been thus cast into a 3pod-like mold consisting of 3 phases, I, II and III. The original Neyer test is technically now a gonogo Neyer with no phase III; and a Bruceton (or Langlie) test is technically a gonogo Bruceton (or Langlie) phase I test.

One limitation of gonogo is that it presently only offers a normal or lognormal analysis capability. There are a couple of reasons for this: first, gonogo already has so many features that it seemed too risky to try adding in another one at this time; the second is that these choices have proven to be sufficient for most DoD and industrial applications.

## 2. Installation

The new gonogo R code (gonogo.R) may be downloaded from Jeff Wu's personal website: https://www2.isye.gatech.edu/~jeffwu/sensitivitytesting/. Gonogo was designed to be run in R. If run in RStudio, you will find that some of gonogo's prompts are misplaced in its console window. Therefore, the user may opt to not to use it there. If you have an existing gonogo folder for an earlier version, you can put and run gonogo.R there (the new functions should work fine with tests created with it).

If you're a new R user, create a conveniently located folder, perhaps on your desktop and place gonogo.R there. This will be your gonogo working directory. Launch R (64 bit), navigate to the new directory with **setwd(),** execute the command **rm(list=objects())** (to start with a clean slate), then execute **source("gonogo.R")**. The command **objects()** should reveal a folder with the following 120 new objects:

**Table 1**. 118 functions plus 2 constants (*) comprising gonogo.R

| | | | | | | | |
|---|---|---|---|---|---|---|---|
| abllik | about4 | add3pod | addBorL | addneyr | al15 * | al49 * | bintodec |
| blrb1 | blrb2 | blrb3 | blrb4 | blrb5 | blrb6 | blrb7 | blrb8 |
| bphaseBI | bphaseI | bpI | calcblim | cbs | chabout | clim | clim0 |
| cpq | d.update | dgs | f38 | f3point8 | fgs | figtab | fixw |
| fixw1 | fm.lims | fofL | gd0 | getBd0 | getBxr | getd0 | getxr |
| Gk | glm.lims | glmmle | gonogo | gonogoSim | graf1 | grafl | gxr |
| ifg | intToBitVect | iofL | jlik | jlrcb | kstar | lims | llik |
| lphaseBI | lphaseI | lpI | lrcb | lrmax | lrq.lims | m.update | mdose.p |
| mixed | mkb0 | n.update | nmel3 | nphaseBI | nphaseI | npI | ntau |
| ok1 | otherpoint | pavdf | pdat1 | pdat2 | pdat3 | pfun | phaseBI1 |
| phaseBI2 | phaseBI3 | phaseBII | phaseI1 | phaseI2 | phaseI3 | phaseII | pI1 |
| pI2 | pI3 | picdat | pII | plotmm | prd0 | prtrans | pSdat1 |
| pSdat2 | pSdat3 | ptest | qrda | reset | simp | Sk | skewL |
| sphaseBIII | sphaseIII | spIIIsim | stopQuietly | tauf | udli | ulik | uliknext |
| unbd | wabout | wabout13 | wxdat | xlead0 | xyllik | yinfomat | zpfun |

When closing out the session, make sure you respond YES to the "Save Workspace?" query. Going forward, double-clicking on the .RData icon will be the way to initiate future R sessions using gonogo.

The full calls to the functions are catalogued in Appendix 1. Two of the functions are cloned: "mdose.p" is dose.p from library(MASS); and "stopQuietly" was lifted from StackOverflow (https://stackoverflow.com/questions/14469522/stop-an-r-program-without-error).

## 3. Graphics and Tables

Gonogo.R was developed in R version 3.1.1 (2014-07-10) on an x86_64-w64-mingw32/x64 (64-bit) platform. The graphs and tabular outputs presented have been re-prepared in more recent versions of R (including 3.5.1, 3.62 and most recently 4.03) to ensure consistency.

The reader can reproduce all Figures and most Tables by running the R command(s) cited in the text. The **figtab** function provides an easier way to recreate all Figures (1 – 31) and 16 Tables (4, 10, 12, 14, 16, 19, 20, 21 and 26 – 33) by running the command **figtab(i)** for i=1:31 and i=32:47, respectively. If you see a spinning throbber, the program is waiting for your console input(s) (for those specifics, consult the documentation).

## 4. Three Usages

Gonogo offers three modes of generating sensitivity test data: simulation, via the gonogoSim function; batch, via the gonogo function; and console, also via the gonogo function. Batch mode and console mode are differentiated by the presence or absence of a user supplied Y argument in the call to gonogo, respectively. In console mode, Y is NULL (the default). In batch mode, Y is a vector of responses. An optional X vector of responses may also be provided in batch mode. The default for X is NULL. In batch mode, stresses beyond the ones specified in X will be determined by the procedure being used and rounded per the reso argument.

The different modes will be illustrated in the sections that follow. New users are encouraged to try out all of the sample R commands cited below. In this way quick exposure to the important features of gonogo will be gained.

The gonogoSim function has 17 arguments in total, as indicated by the general call to it:

**gonogoSim(mlo,mhi,sg,n2=0,n3=0,p=0,lam=0,dm=0,ds=0,**

**reso=0,ln=F,plt=0,iseed=-1,IIgo=T,M=1,test=1,BL=c(4,1,0))**

Brief descriptions of the arguments are presented below:

**Table 2**. Arguments to gonogoSim. Only the first three are mandatory.

| Argument | Description |
|---|---|
| mlo ( $\mu_{lo}$ )<br>mhi ( $\mu_{hi}$ )<br>sg ( $\sigma_g$ ) | These three entries are "guesstimates". $\mu_{true}$ and $\sigma_{true}$, which form the basis of the simulations, are computed from mlo, mhi, and sg, plus other arguments defined below (e.g., dm, ds, test, BL and ln). **See section 7.3 for specifics**. |
| n2 | Size of phase II (could be 0) |
| n3 | Size of phase III (could be 0) |
| p | To approximate the stress level $L_p$ - for phase III only |
| lam | The $\lambda$ parameter needed for phase III only (see note below) |
| dm | Deviation from a target mean (tm) based upon mlo, mhi and sg |
| ds | Deviation from a target standard deviation (ts) based upon mlo, mhi and sg |
| reso | Resolution, e.g., .01 |
| ln | TRUE (for a log transformed test) or FALSE (the default) |
| plt | 1 (for a history plot of a test) or 0 (no plot, the default) |
| iseed | Initialization of the random seed (-1, the default, for full random) |
| IIgo | TRUE – Test proceeds unencumbered<br>FALSE – Test stops at end of stage 2 of phase I (test = 1) or phase I (test > 1) |
| M | Factor to scale stresses (usually 1, the default) |
| test | 1, 2, 3 or 4 for the 4 different Phase I's offered:<br>3pod, Neyer, Bruceton or Langlie, respectively |
| BL | Vector of 3 integers (needed for Bruceton or Langlie Tests only) |

**Note**: For more information about lam ( $\lambda$ ), see Wu and Tian (2014), Wang, Tian and Wu (2015), and Wang, Tian and Wu (2020)

Responses (Y's) are obtained by comparing stresses (X's) to simulated strengths. The strengths are randomly chosen from a latent distribution whose true mean and standard deviation are the sum of targeted settings (tm and ts, based on mlo, mhi, sg, test, etc.) plus deviations (dm and ds), respectively (see section 7.3 for specific details). Random responses are obtained via the formulae

$$Y_{stress \geq strength} = 1, \text{ and } Y_{stress < strength} = 0$$

The new gonogo function now has 11 arguments in total, as indicated by the general call to it:

**gonogo(mlo=0,mhi=0,sg=0,newz=T,reso=0,ln=F,test=1,term1=T,BL=NULL,Y=NULL,X=NULL)**

A brief description of the arguments is presented below:

**Table 3**. Elements of the call to gonogo. Only the first three arguments are mandatory

| Argument | Description |
|---|---|
| mlo, mhi, sg | Starting values for the test. Details about the 3 entries depend on the test argument |
| newz | TRUE (the default, for a new test) or FALSE (**continue an existing test defined by z**) |
| reso | Resolution, e.g., .01 |
| ln | FALSE (the default), TRUE (for a log transformed test) |
| test | 1, 2, 3 or 4 for the 4 different Phase I's offered, 3pod, Neyer, Bruceton or Langlie, resp. |
| term1 | TRUE (the default), or FALSE (ensure 3pod's stage I2 of phase 1 ends with overlap) |
| BL | Vector of 3 integers (needed for Bruceton or Langlie Tests only) |
| Y | NULL (Console mode, Default), or THE ENTIRE Response vector (Batch mode) |
| X | NULL (Console or Batch mode, Default), or, if present, a vector of<br>THE FIRST n Stresses (Batch mode). $length(Y) > n$ is OK. |

The variable "z" has special status in gonogo lingo and should be reserved for the following use: when a test, saved in a list called, e.g., XYZ, is suspended, it can be resumed by running z=XYZ followed by running XYZ=gonogo(newz=F) (see Section 7.1, page 30).

### 4.1 Random

The gonogoSim function can generate complete or partial 3pod, Neyer, Bruceton or Langlie tests. In the first example, a complete 3pod test having: phase I starting values of 0, 22 and 3 for mlo, mhi and sigma guess, respectively; a phase II size (n2) of 6; a Skewed Robbins-Monro-Josephs (RMJ) phase III of size 15 (n3) with parameters p and lambda equal to .9, and 1, respectively; and stresses rounded to the nearest .01 will be generated (with a random seed initialized at 42983), saved and plotted by running the command:

**w1=gonogoSim(0,22,3,6,15,.9,1,plt=1,reso=.01,iseed=42983).**

The output of gonogo is saved in a list called w1. The history plot generated appears below:

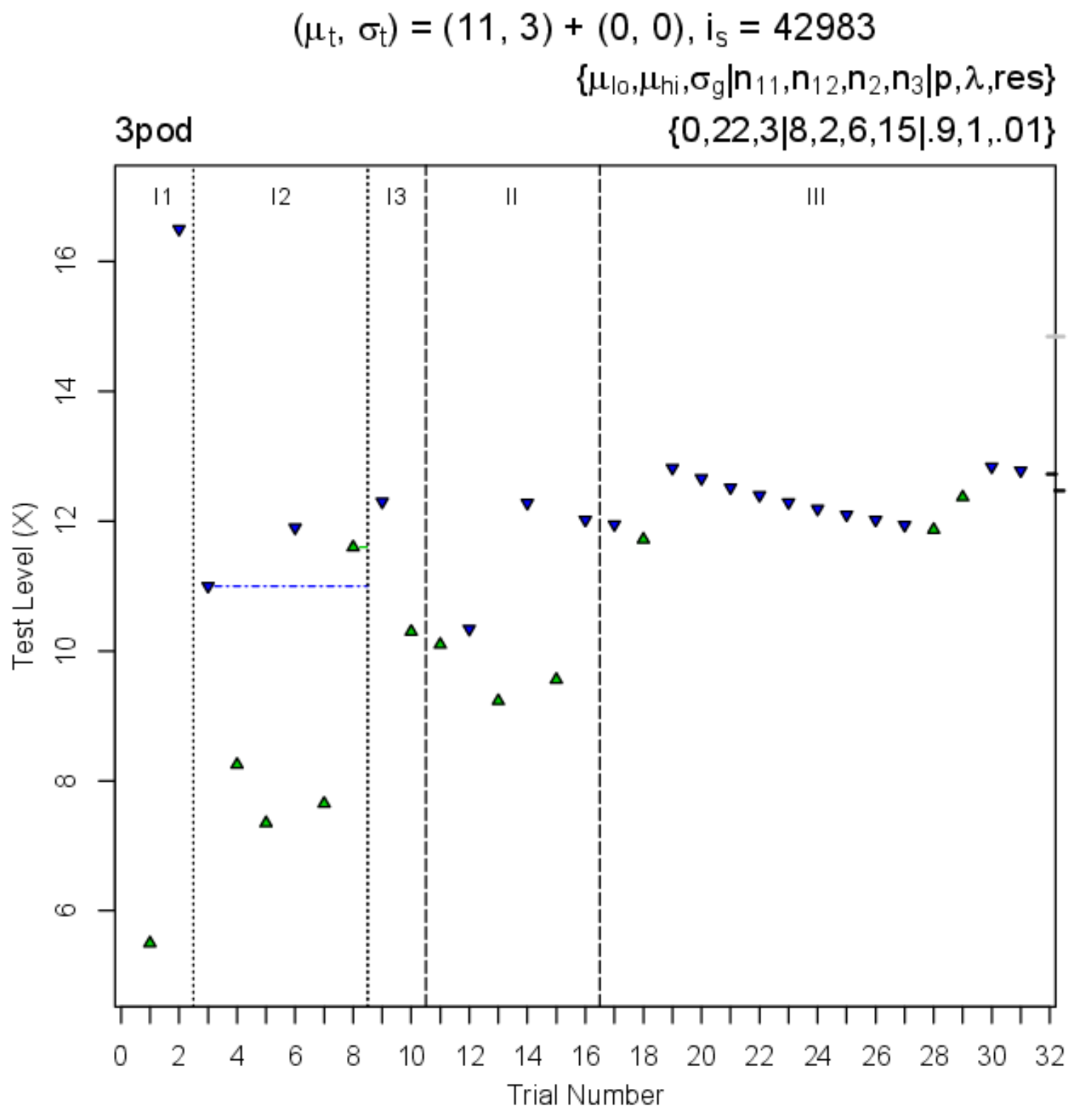

**Figure 1.** History plot of a simulated 3pod.

gonogoSim creates a list that has 25 components. One of them, d0, contains the run specifics. Another, jvec, details how the phase III stresses were computed. The console display of running **w1** is presented below:

**Table 4**. Run details of the simulated 3pod test saved in the list w1

**$d0**

| i | X | Y | COUNT | RX | EX | TX | ID |
|---|---|---|---|---|---|---|---|
| 1 | 5.50 | 0 | 1 | 5.50000 | 5.500000 | 5.50 | I1(iii) |
| 2 | 16.50 | 1 | 1 | 16.50000 | 16.500000 | 16.50 | I1(iii) |
| 3 | 11.00 | 1 | 1 | 11.00000 | 11.000000 | 11.00 | I2(ib) |
| 4 | 8.25 | 0 | 1 | 8.25000 | 8.250000 | 8.25 | I2(ib) |
| 5 | 7.35 | 0 | 1 | 7.35000 | 7.350000 | 7.35 | I2(id) |
| 6 | 11.90 | 1 | 1 | 11.90000 | 11.900000 | 11.90 | I2(id) |
| 7 | 7.65 | 0 | 1 | 7.65000 | 7.650000 | 7.65 | rI2(id) |
| 8 | 11.60 | 0 | 1 | 11.60000 | 11.600000 | 11.60 | rI2(id) |
| 9 | 12.30 | 1 | 1 | 12.30000 | 12.300000 | 12.30 | I3 |
| 10 | 10.30 | 0 | 1 | 10.30000 | 10.300000 | 10.30 | I3 |
| 11 | 10.10 | 0 | 1 | 10.10000 | 10.104985 | 10.10 | II1 |
| 12 | 10.34 | 1 | 1 | 10.34000 | 10.342099 | 10.34 | II2 |
| 13 | 9.23 | 0 | 1 | 9.23000 | 9.229696 | 9.23 | II2 |
| 14 | 12.28 | 1 | 1 | 12.28000 | 12.278542 | 12.28 | II2 |
| 15 | 9.56 | 0 | 1 | 9.56000 | 9.560114 | 9.56 | II2 |
| 16 | 12.02 | 1 | 1 | 12.02000 | 12.020638 | 12.02 | II2 |
| 17 | 11.95 | 1 | 1 | 11.95000 | 11.953786 | 11.95 | III1 |
| 18 | 11.72 | 0 | 1 | 11.72000 | 11.721333 | 11.72 | III2 |
| 19 | 12.82 | 1 | 1 | 12.82000 | 12.822440 | 12.82 | III2 |
| 20 | 12.66 | 1 | 1 | 12.66000 | 12.657951 | 12.66 | III2 |
| 21 | 12.52 | 1 | 1 | 12.52000 | 12.519094 | 12.52 | III2 |
| 22 | 12.40 | 1 | 1 | 12.40000 | 12.395547 | 12.40 | III2 |
| 23 | 12.29 | 1 | 1 | 12.29000 | 12.288679 | 12.29 | III2 |
| 24 | 12.19 | 1 | 1 | 12.19000 | 12.189381 | 12.19 | III2 |
| 25 | 12.10 | 1 | 1 | 12.10000 | 12.098257 | 12.10 | III2 |
| 26 | 12.02 | 1 | 1 | 12.02000 | 12.015730 | 12.02 | III2 |
| 27 | 11.94 | 1 | 1 | 11.94000 | 11.942102 | 11.94 | III2 |
| 28 | 11.87 | 0 | 1 | 11.87000 | 11.867597 | 11.87 | III2 |
| 29 | 12.37 | 0 | 1 | 12.37000 | 12.367699 | 12.37 | III2 |
| 30 | 12.84 | 1 | 1 | 12.84000 | 12.841672 | 12.84 | III2 |
| 31 | 12.78 | 1 | 1 | 12.78000 | 12.780299 | 12.78 | III2 |
| 311 | 0.00 | 0 | 0 | 12.72361 | 0.000000 | 0.00 | III3 |

**$jvec**

| i | j | k | v | u | a | tau2 | nu | b | x | y |
|---|---|---|---|---|---|---|---|---|---|---|
| 1 | 0.000000 | 0.000000 | 0.0000000 | 0.00000000 | 0.0000000 | 1.7916445 | 0 | 0.0000000 | 11.95379 | 1 |
| 2 | 1.281552 | 1.259256 | 0.8455910 | 0.19335891 | 1.4809191 | 1.5052956 | 0 | 0.8455910 | 11.72133 | 0 |
| 3 | 1.281552 | 1.221520 | 0.8529442 | 0.16212021 | 1.2925118 | 1.2957533 | 0 | 0.8529442 | 12.82244 | 1 |
| 4 | 1.281552 | 1.193150 | 0.8586089 | 0.13913674 | 1.1461051 | 1.1362880 | 0 | 0.8586089 | 12.65795 | 1 |
| 5 | 1.281552 | 1.171100 | 0.8630914 | 0.12161481 | 1.0291980 | 1.0111223 | 0 | 0.8630914 | 12.51909 | 1 |
| 6 | 1.281552 | 1.153497 | 0.8667188 | 0.10786537 | 0.9337592 | 0.9104020 | 0 | 0.8667188 | 12.39555 | 1 |
| 7 | 1.281552 | 1.139135 | 0.8697100 | 0.09681697 | 0.8544095 | 0.8276807 | 0 | 0.8697100 | 12.28868 | 1 |
| 8 | 1.281552 | 1.127203 | 0.8722164 | 0.08776164 | 0.7874182 | 0.7585756 | 0 | 0.8722164 | 12.18938 | 1 |
| 9 | 1.281552 | 1.117136 | 0.8743455 | 0.08021496 | 0.7301198 | 0.7000091 | 0 | 0.8743455 | 12.09826 | 1 |
| 10 | 1.281552 | 1.108534 | 0.8761754 | 0.07383544 | 0.6805607 | 0.6497596 | 0 | 0.8761754 | 12.01573 | 1 |
| 11 | 1.281552 | 1.101099 | 0.8777644 | 0.06837606 | 0.6372774 | 0.6061850 | 0 | 0.8777644 | 11.94210 | 1 |
| 12 | 1.281552 | 1.094611 | 0.8791568 | 0.06365402 | 0.5991525 | 0.5680466 | 0 | 0.8791568 | 11.86760 | 0 |
| 13 | 1.281552 | 1.088901 | 0.8803866 | 0.05953140 | 0.5653181 | 0.5343924 | 0 | 0.8803866 | 12.36770 | 0 |
| 14 | 1.281552 | 1.083838 | 0.8814805 | 0.05590230 | 0.5350905 | 0.5044796 | 0 | 0.8814805 | 12.84167 | 1 |
| 15 | 1.281552 | 1.079317 | 0.8824597 | 0.05268411 | 0.5079233 | 0.4777201 | 0 | 0.8824597 | 12.78030 | 1 |
| 16 | 1.281552 | 1.075256 | 0.8833413 | 0.04981154 | 0.4833751 | 0.4536425 | 0 | 0.8833413 | 12.72361 | NA |

| | |
|---|---|
| **$tmu** | 11 |
| **$tsig** | 3 |
| **$mhat** | 11.10749 |
| **$shat** | 1.064955 |
| **$en** | 8 2 6 15 |
| **$about** | "{0,22,3\|8,2,6,15\|.9,1,.01}" |
| **$title** | NULL |
| **$ttl1** | NULL |
| **$ttl2** | "3pod" |
| **$ttl0** | paste("(", mu[t], ", ", sigma[t], ") = (", 11, ", ", 3, ") + (", |

```
                    "0", ", ", "0", "), ", i[s], " = ", 42983, sep = "")
$units  "X"
$p      0.9
$reso   0.01
$ln     FALSE
$lam    1
$test   1
$M      1
$dm     0
$ds     0
$iseed  42983
$BL     NULL
$dud    NULL
$lev    NULL
```

w1$d0 is the first element of the list w1. d0 is an R data frame having 7 columns named X, Y, COUNT, RX, EX, TX and ID. The various columns are retrievable, e.g., w1$d0$X, w1$do$Y, etc. As for the meanings of information stored in each column, the user is referred to the following table:

**Table 5**. The information stored in the data frame w1$d0

| Column | Information |
|---|---|
| X | Actual stress sequence |
| Y | Response sequence (0 or 1) |
| COUNT | Number of occurrences (always 1, except last entry of a completed Phase III is always 0) |
| RX | Recommended Stress (perhaps rounded to some other resolution besides .0001) |
| EX | Exact stress (rounded to .000001) |
| TX | Transformed response (for the log=T option) |
| ID | Phase, stage and step (if 3pod per Wu and Tian (2014), where r indicates reductions in sg), or block and phase (if Neyer, per diagram in Appendix B) or phase (if Bruceton or Langlie) |

## 4.2 Batch

Since all of w1's stresses (X) are the same as its exact stresses (EX), it can be recreated by only knowing its initialization details and the vector of responses (Y). This is the simplest batch mode usage, and it's the quickest way to reproduce w1 from scratch - as it avoids having to manually enter in the stress/response pairs in console mode. If some of w1's stresses (X) deviated from the exact stresses (EX), the X option would have to be invoked in the call to gonogo (see section 5.1.1, pp 12, where a legacy 3pod test is reproduced that required use of the X option).

To reproduce w1, you will also need to make three entries in the console window (*otherwise, the program will be just waiting for these inputs*). They are:

enter an n2 of 6 at the first prompt ("Enter Phase II (D-Optimal) size n2: ");

enter an n3 of 15 at the second prompt ("Enter Phase III (S-RMJ) size n3: "); and enter a p, lambda pair of .9 1 at the third prompt ("Enter p lam: ").

Run **yW=w1$d0$Y**. This defines the response vector for test w1 and assigns it to yW. Then run **w2=gonogo(0,22,3,test=1,reso=.01,Y=yW)**. This reproduces the same plot and run sequence we got before.

The graph produced by creating w2 to recreate w1 is shown below:

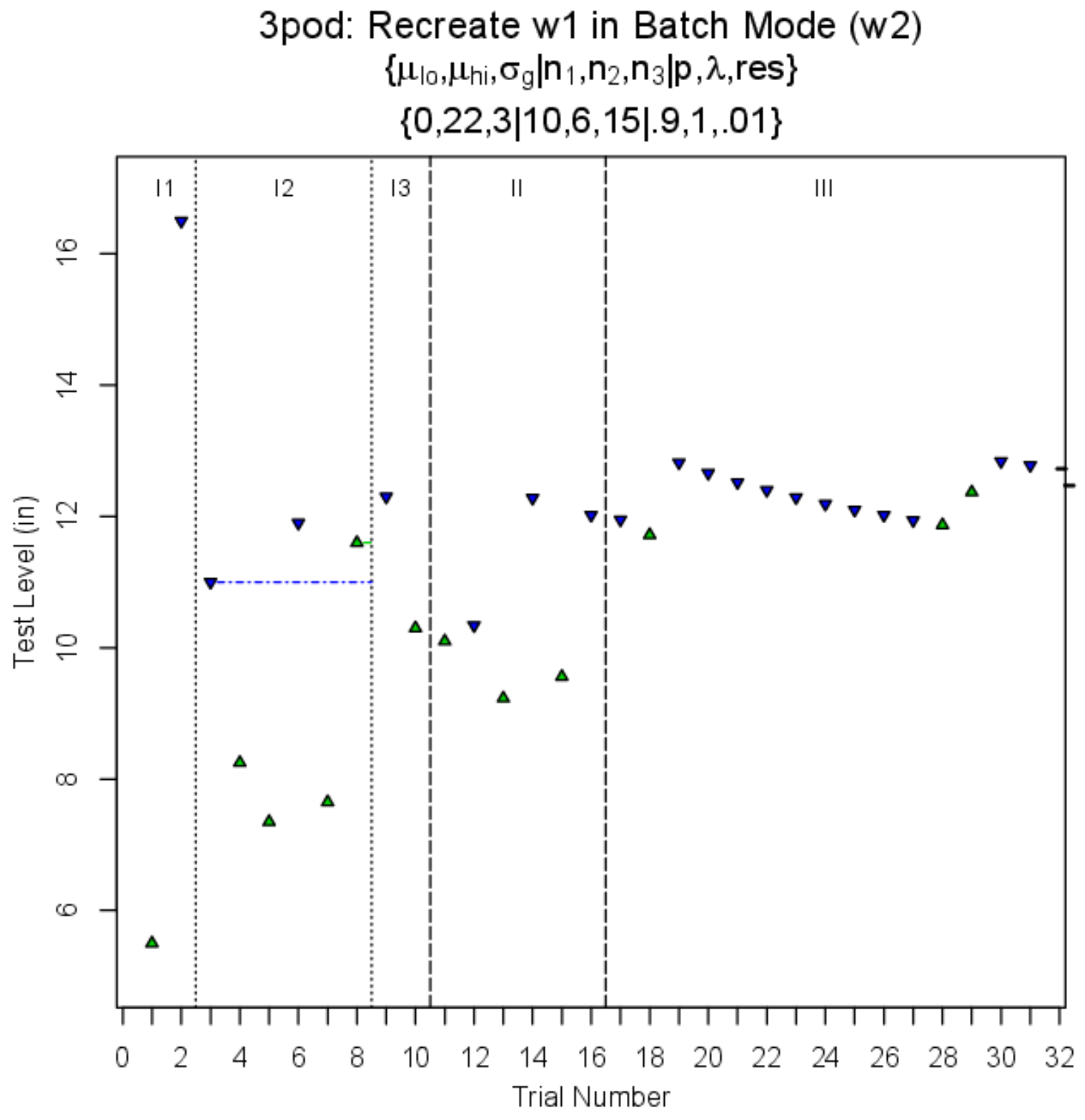

**Figure 2**. Simulated 3pod (w1) reproduced in Batch mode (w2)

Tests produced by executing the gonogo function are saved in a text file called "gonogo.txt". The contents of the text file generated by creating w2 are shown below:

**Table 6**. Contents of the text file gonogo.txt generated by w2

| i, X, Y, COUNT, RX, EX, TX, ID | i, X, Y, COUNT, RX, EX, TX, ID |
|---|---|
| 1, 5.5, 0, 1, 5.5, 5.5, 5.5, I1(iii) | 17, 11.95, 1, 1, 11.95, 11.95379, 11.95, III1 |
| 2, 16.5, 1, 1, 16.5, 16.5, 16.5, I1(iii) | 18, 11.72, 0, 1, 11.72, 11.72133, 11.72, III2 |
| 3, 11, 1, 1, 11, 11, 11, I2(ib) | 19, 12.82, 1, 1, 12.82, 12.82244, 12.82, III2 |
| 4, 8.25, 0, 1, 8.25, 8.25, 8.25, I2(ib) | 20, 12.66, 1, 1, 12.66, 12.65795, 12.66, III2 |
| 5, 7.35, 0, 1, 7.35, 7.35, 7.35, I2(id) | 21, 12.52, 1, 1, 12.52, 12.51909, 12.52, III2 |
| 6, 11.9, 1, 1, 11.9, 11.9, 11.9, I2(id) | 22, 12.4, 1, 1, 12.4, 12.39555, 12.4, III2 |
| 7, 7.65, 0, 1, 7.65, 7.65, 7.65, rI2(id) | 23, 12.29, 1, 1, 12.29, 12.28868, 12.29, III2 |
| 8, 11.6, 0, 1, 11.6, 11.6, 11.6, rI2(id) | 24, 12.19, 1, 1, 12.19, 12.18938, 12.19, III2 |
| 9, 12.3, 1, 1, 12.3, 12.3, 12.3, I3 | 25, 12.1, 1, 1, 12.1, 12.09826, 12.1, III2 |
| 10, 10.3, 0, 1, 10.3, 10.3, 10.3, I3 | 26, 12.02, 1, 1, 12.02, 12.01573, 12.02, III2 |
| 11, 10.1, 0, 1, 10.1, 10.10498, 10.1, II1 | 27, 11.94, 1, 1, 11.94, 11.9421, 11.94, III2 |
| 12, 10.34, 1, 1, 10.34, 10.3421, 10.34, II2 | 28, 11.87, 0, 1, 11.87, 11.8676, 11.87, III2 |
| 13, 9.23, 0, 1, 9.23, 9.2297, 9.23, II2 | 29, 12.37, 0, 1, 12.37, 12.3677, 12.37, III2 |
| 14, 12.28, 1, 1, 12.28, 12.27854, 12.28, II2 | 30, 12.84, 1, 1, 12.84, 12.84167, 12.84, III2 |
| 15, 9.56, 0, 1, 9.56, 9.56011, 9.56, II2 | 31, 12.78, 1, 1, 12.78, 12.7803, 12.78, III2 |
| 16, 12.02, 1, 1, 12.02, 12.02064, 12.02, II2 | 311, 0, 0, 0, 12.72361, 0, 0, III3 |

This file was made to be edited, copied and pasted into an MS Word document, highlighted, and made into a MS Word table (by using Insert, Table, Convert Text to Table). No such file is generated by the gonogoSim function, e.g., when w1 was created.

## 4.3 Console

We could also have used the console mode to reproduce our simulated 3pod test, w1. Then, you would enter each stress/response pair at the keyboard, one pair at a time. This is the mode gonogo is normally used to conduct sensitivity experiments. The command to start the console mode for this example would be the same as for batch mode, except there would be no Y argument. The following command gets such a test started: **w = gonogo(0,22,3,test=1,reso=.01)**. The recreation of w1 in console mode will be left as an exercise for the reader.

## 5. Three Phases

A three-part test design was first described by Neyer (1994). The first part consisted of a modified binary search to "home in" on the mean, and the second part was designed to achieve overlap, thereby allowing unique maximum likelihood estimates (MLE's) of mu and sigma to be obtained. Neyer's third part consisted of a D-optimal procedure designed to refine these estimates.

Wu and Tian (2014) describe a three-phase design, dubbed 3pod, and Wang, Tian and Wu (2020) made improvements, calling it 3pod2.0. It features a modified phase I, called 3podM; a phase II that is the same as Neyer's D-optimal procedure (which Neyer called part 3); and a skewed version of the Robbins-Monro-Joseph procedure for phase III that's now "equivariant" with respect to measurement units. Phase III is designed to "home in" on $L_p$, the stress having a probability of response equal to $p$. In gonogo's 3 phase framework, phase I is a necessary component, whereas Phases II and III may be considered optional add-ons. In this framework, Neyer's first two parts become gonogo's Neyer phase I and his D-optimal third part is gonogo's phase II.

gonogo adopts the three-phase model as its template and implements the improved phase III whenever specified. Four distinct phase I's are offered in gonogo.

## 5.1 Phase I

The gonogo and gonogoSim functions feature four test procedures for phase I: 3podM (Wang, Tian and Wu 2020); Neyer, parts 1 and 2 (Neyer 1994); Bruceton [Dixon and Mood 1948]; or Langlie (1962). The four procedures are specified by the value of the test argument passed into the function, 1, 2, 3 or 4, respectively. The value of test determines how three other arguments, namely mlo, mhi and sg, are interpreted by the function. Details are summarized in the following table:

**Table 7**. gonogo's convention to specify the Phase I procedure you want to run

| test | Procedure | mlo | mhi | sg |
|---|---|---|---|---|
| 1 (default) | 3podM | $\mu_{\min}$ | $\mu_{\max}$ | $\sigma_g$ |
| 2 | Neyer (parts 1 & 2) | $\mu_{\min}$ | $\mu_{\max}$ | $\sigma_g$ |
| 3 | Bruceton | $L_p$ | $L_p$ | $\sigma_g$ |
| 4 | Langlie | $L = L_0$ | $U = L_1$ | 0 |

Notice, for the Bruceton and Langlie cases, mlo and mhi guesses depend upon the quantity $L_p$ which satisfies Pr [Resp =1 |stress= $L_p$ ] = $p$, per Langlie (1962) and Einbinder (1973). $L_0$ and $L_1$ are also known as the "no-fire" point and the "all-fire" point, respectively.

When opting to run a test having a Bruceton or a Langlie phase I, three additional inputs, $nRev$, $i_1$ and $i_2$, are required at a subsequent prompt. The first one, $nRev$, is the user-specified minimum number of reversals in response required to exit phase I, where an $Up$ response followed by a $Down$ response, or a $Down$ response followed by an $Up$ response counts as 1 reversal. Use of $nRev$ as a stopping rule is recommended by Einbinder (1973). An $nRev$ equaling zero is permitted, which means phase I ends when interval overlap is achieved. Otherwise, phase I ends upon achieving overlap AND $nRev$ reversals.

To accommodate Bruceton and Langlie phase I's designed to "home in" to $L_p$'s other than $L_{.50}$, gonogo offers an option to conduct a Bruceton or Langlie phase I that operates on the so-called up-down transformed response (UDTR) according to certain rules described by Einbinder (1973) and Wetherill, Chen and Vasudeva (1966).

Additionally, gonogo permits doing two UDTR tests in succession for phase I. Specifically, Bruceton and Langlie phase I's may consist of a single $L_p$ test, an $L_p$ test followed by an $L_{1-p}$ test ( $p \neq .50$ ), or an $L_{.50}$ test followed by an $L_p$ test ( $p \neq .50$ ). These combinations are in accord with similar one-shot transformed response (OSTR) usages recommended by Einbinder (1973). The various test combinations for 1 or 2 Bruceton (or Langlie) tests for phase I, are determined by selecting two values of $i$, $i_1$ and $i_2$, from the following table:

**Table 8**. Up-Down Transformed Response (UDTR) Sequences for $L_p$ Testing

| $i$ | $Down_{X=1} / Up_{X=0}$ | $Up_{X=1} / Down_{X=0}$ | $p_{X=1} / (1-p)_{X=0}$ |
|---|---|---|---|
| 1 | $X$ | $O$ | .5 |
| 3 | $XX$ | $\{O, XO\}$ | .707107 |
| 5 | $XXX$ | $\{O, XO, XXO\}$ | .793701 |
| 7 | $XXXX$ | $\{O, XO, XXO, XXXO\}$ | .840896 |
| ⋮ | ⋮ | ⋮ | ⋮ |
| 2 | $\{XX, XOX\}$ | $\{O, XOO\}$ | .596968 |
| 4 | $\{XXX, XXOX\}$ | $\{O, XO, XXOO\}$ | .733614 |
| 6 | $\{XXXX, XXXOX\}$ | $\{O, XO, XXO, XXXOO\}$ | .804119 |
| ⋮ | ⋮ | ⋮ | ⋮ |

**Note**: The above $p$ used in phase I is NOT the same argument required for phase III

Sequences of $X$'s and $O$'s that trigger a stress level change may be relabeled $U$ or $D$. This gives the rule its name: the Up-Down-Transformed-Response rule, or UDTR rule for short.

Gonogo has a convention it follows for its $i_1$, $i_2$ entries:

If $i_2 = 0$, Phase I consists of an $L_{p|p \geq .5}$ test associated with $i_1$;

if $i_1 = 0$, Phase I consists of an $L_{p|p \leq .5}$ test associated with $i_2$;

if $i_1$ and $i_2$ are not both 0 and not both 1, then Phase I consists of two tests,

an $L_{p|p \geq .5}$ test followed by an $L_{p|p \leq .5}$ test associated with $i_1$ and $i_2$, respectively.

More information about the development of UDTR and OSTR rules and their applications may be found in Wetherill, Chen and Vasudeva (1966), Einbinder (1973) and MIL-STD-331D.

### 5.1.1 3pod Phase 1

Let's try recreating another test, the one that appeared in Table 1 (Wu and Tian 2014). Evidently, it was run using two resolutions: phases I and II with reso=.1 and phase III with reso=.0001. This will be a nice example to see how one can utilize batch mode's X option to recreate a legacy test you might later like to continue. To recreate this test, run the following three commands:

**yWT=c( 0, 1, 0, 1, 0, 1, 1, 1, 1, 0, 0, 0, 1, 0, 1, 1, 1, 0, 1, 1, 1, 1, 1, 1, 1, 1, 1, 1, 1, 0),**

**xWT=c( 5.5, 16.5, 11.0, 13.8, 10.1, 14.7, 10.4, 11.7, 9.7, 7.3, 7.8, 8.1, 12.2, 8.5, 11.8)** and

**wWT=gonogo(0,22,3,reso=.0001,Y=yWT,X=xWT)**.

The following graph is produced:

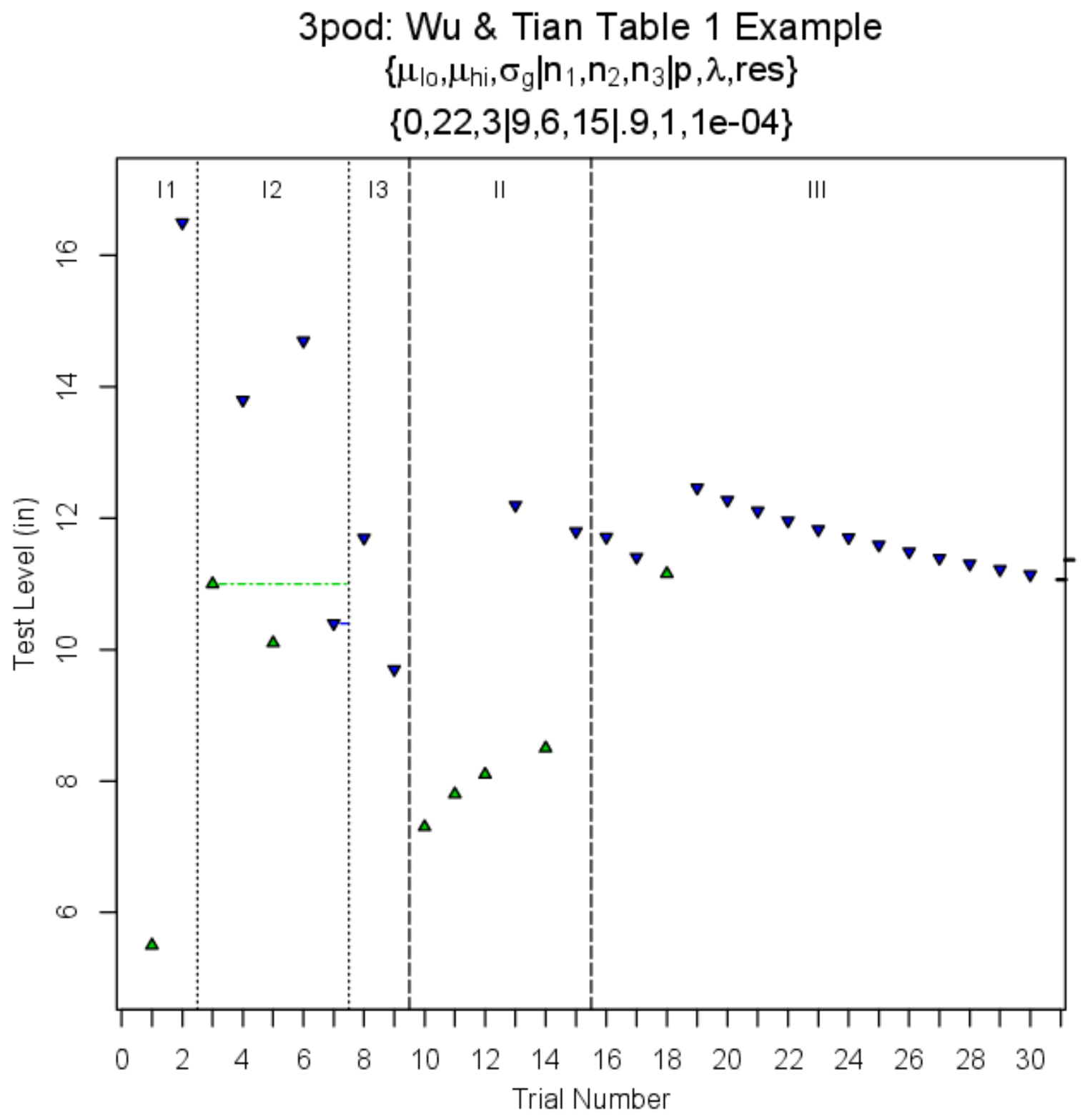

**Figure 3**. gonogo Rendition of Table 1 (Wu and Tian 2014)

The small black bar on the inside of the plot is the location of what would be the next recommended stress level, if the test had continued. The bar on the outside of the plot is $\hat{L}_p = \hat{\mu} + z_p\hat{\sigma}$, where $\hat{\mu}$ and $\hat{\sigma}$ are estimates based on the 30-run test. The values of $\hat{\mu}$ and $\hat{\sigma}$ (when they exist) are saved in the musig component of wWT, i.e., wWT$musig. In this case, you find that: $\hat{\mu} = 10.1707909$, $\hat{\sigma} = .9344091$ and $\hat{L}_p$=11.36828 (not 11.2816 as reported in Wu and Tian 2014). The values of $\hat{\mu}$ and $\hat{\sigma}$ may also be obtained by running **glmmle(wWT$d0).** They are also displayed in the console if you run **ptest(wWT,3)**.

Along with Figure 3, a text file is automatically created by gonogo in its working directory. It is called gonogo.txt. When it's made into an MS Word table, it provides the test's run details. For comparison, the contents of gonogo.txt are placed beside the original X column (taken from Wu and Tian 2013):

**Table 9**. Original version (Table 1, Wu and Tian (2013)) vs. gonogo version using Batch moded

| i | X | X | Y | COUNT | RX | EX | TX | ID |
|---|---|---|---|---|---|---|---|---|
| 1 | 5.5 | 5.5000 | 0 | 1 | 5.50000 | 5.500000 | 5.5000 | I1(iii) |
| 2 | 16.5 | 16.5000 | 1 | 1 | 16.50000 | 16.500000 | 16.5000 | I1(iii) |
| 3 | 11 | 11.0000 | 0 | 1 | 11.00000 | 11.000000 | 11.0000 | I2(ib) |
| 4 | 13.8 | 13.8000 | 1 | 1 | 13.75000 | 13.750000 | 13.8000 | I2(ib) |
| 5 | 10.1 | 10.1000 | 0 | 1 | 10.10000 | 10.100000 | 10.1000 | I2(id) |
| 6 | 14.7 | 14.7000 | 1 | 1 | 14.70000 | 14.700000 | 14.7000 | I2(id) |
| 7 | 10.4 | 10.4000 | 1 | 1 | 10.40000 | 10.400000 | 10.4000 | rI2(id) |
| 8 | 11.7 | 11.7000 | 1 | 1 | 11.70000 | 11.700000 | 11.7000 | I3 |
| 9 | 9.7 | 9.7000 | 1 | 1 | 9.70000 | 9.700000 | 9.7000 | I3 |
| 10 | 7.3 | 7.3000 | 0 | 1 | 7.26510 | 7.265078 | 7.3000 | II1 |
| 11 | 7.8 | 7.8000 | 0 | 1 | 7.75430 | 7.754301 | 7.8000 | II2 |
| 12 | 8.1 | 8.1000 | 0 | 1 | 8.08430 | 8.084262 | 8.1000 | II2 |
| 13 | 12.2 | 12.2000 | 1 | 1 | 12.16430 | 12.164304 | 12.2000 | II2 |
| 14 | 8.5 | 8.5000 | 0 | 1 | 8.51670 | 8.516679 | 8.5000 | II2 |
| 15 | 11.8 | 11.8000 | 1 | 1 | 11.82540 | 11.825443 | 11.8000 | II2 |
| 16 | 11.7106 | 11.7121 | 1 | 1 | 11.71210 | 11.712057 | 11.7121 | III1 |
| 17 | 11.4896 | 11.4083 | 1 | 1 | 11.40830 | 11.408272 | 11.4083 | III2 |
| 18 | 11.2980 | 11.1558 | 0 | 1 | 11.15580 | 11.155754 | 11.1558 | III2 |
| 19 | 12.3899 | 12.4633 | 1 | 1 | 12.46330 | 12.463306 | 12.4633 | III2 |
| 20 | 12.2393 | 12.2761 | 1 | 1 | 12.27610 | 12.276079 | 12.2761 | III2 |
| 21 | 12.1033 | 12.1107 | 1 | 1 | 12.11070 | 12.110741 | 12.1107 | III2 |
| 22 | 11.9796 | 11.9628 | 1 | 1 | 11.96280 | 11.962789 | 11.9628 | III2 |
| 23 | 11.8660 | 11.8291 | 1 | 1 | 11.82910 | 11.829108 | 11.8291 | III2 |
| 24 | 11.7612 | 11.7072 | 1 | 1 | 11.70720 | 11.707202 | 11.7072 | III2 |
| 25 | 11.6638 | 11.5952 | 1 | 1 | 11.59520 | 11.595231 | 11.5952 | III2 |
| 26 | 11.5730 | 11.4917 | 1 | 1 | 11.49170 | 11.491698 | 11.4917 | III2 |
| 27 | 11.4878 | 11.3955 | 1 | 1 | 11.39550 | 11.395498 | 11.3955 | III2 |
| 28 | 11.4077 | 11.3057 | 1 | 1 | 11.30570 | 11.305654 | 11.3057 | III2 |
| 29 | 11.3321 | 11.2214 | 1 | 1 | 11.22140 | 11.221436 | 11.2214 | III2 |
| 30 | 11.2605 | 11.1421 | 1 | 1 | 11.14210 | 11.142075 | 11.1421 | III2 |
| 301 | 11.1925 | 0.0000 | 0 | 0 | 11.06718 | 0.000000 | 0.0000 | III3 |

Notice how use of the X option limited the scope of the reso = .0001 option just to phase III. Readers may consider the gonogo X column to be an improvement over the original column, since the phase III differences are attributable to use of the recently proposed "equivariant" procedure described in Wang, Tian and Wu (2020).

More details about the test just run are contained in the list wWT. The contents of the 20 components are displayed by running the command **wWT,** which are presented below:

**Table 10**. 20 components of the list wWT

**$d0**

| i | X | Y | COUNT | RX | EX | TX | ID |
|---|---|---|---|---|---|---|---|
| 1 | 5.5000 | 0 | 1 | 5.50000 | 5.500000 | 5.5000 | I1(iii) |
| 2 | 16.5000 | 1 | 1 | 16.50000 | 16.500000 | 16.5000 | I1(iii) |
| 3 | 11.0000 | 0 | 1 | 11.00000 | 11.000000 | 11.0000 | I2(ib) |
| 4 | 13.8000 | 1 | 1 | 13.75000 | 13.750000 | 13.8000 | I2(ib) |
| 5 | 10.1000 | 0 | 1 | 10.10000 | 10.100000 | 10.1000 | I2(id) |
| 6 | 14.7000 | 1 | 1 | 14.70000 | 14.700000 | 14.7000 | I2(id) |
| 7 | 10.4000 | 1 | 1 | 10.40000 | 10.400000 | 10.4000 | rI2(id) |
| 8 | 11.7000 | 1 | 1 | 11.70000 | 11.700000 | 11.7000 | I3 |
| 9 | 9.7000 | 1 | 1 | 9.70000 | 9.700000 | 9.7000 | I3 |
| 10 | 7.3000 | 0 | 1 | 7.26510 | 7.265078 | 7.3000 | II1 |
| 11 | 7.8000 | 0 | 1 | 7.75430 | 7.754301 | 7.8000 | II2 |
| 12 | 8.1000 | 0 | 1 | 8.08430 | 8.084262 | 8.1000 | II2 |
| 13 | 12.2000 | 1 | 1 | 12.16430 | 12.164304 | 12.2000 | II2 |
| 14 | 8.5000 | 0 | 1 | 8.51670 | 8.516679 | 8.5000 | II2 |
| 15 | 11.8000 | 1 | 1 | 11.82540 | 11.825443 | 11.8000 | II2 |
| 16 | 11.7121 | 1 | 1 | 11.71210 | 11.712057 | 11.7121 | III1 |
| 17 | 11.4083 | 1 | 1 | 11.40830 | 11.408272 | 11.4083 | III2 |
| 18 | 11.1558 | 0 | 1 | 11.15580 | 11.155754 | 11.1558 | III2 |
| 19 | 12.4633 | 1 | 1 | 12.46330 | 12.463306 | 12.4633 | III2 |
| 20 | 12.2761 | 1 | 1 | 12.27610 | 12.276079 | 12.2761 | III2 |
| 21 | 12.1107 | 1 | 1 | 12.11070 | 12.110741 | 12.1107 | III2 |
| 22 | 11.9628 | 1 | 1 | 11.96280 | 11.962789 | 11.9628 | III2 |
| 23 | 11.8291 | 1 | 1 | 11.82910 | 11.829108 | 11.8291 | III2 |
| 24 | 11.7072 | 1 | 1 | 11.70720 | 11.707202 | 11.7072 | III2 |
| 25 | 11.5952 | 1 | 1 | 11.59520 | 11.595231 | 11.5952 | III2 |
| 26 | 11.4917 | 1 | 1 | 11.49170 | 11.491698 | 11.4917 | III2 |
| 27 | 11.3955 | 1 | 1 | 11.39550 | 11.395498 | 11.3955 | III2 |
| 28 | 11.3057 | 1 | 1 | 11.30570 | 11.305654 | 11.3057 | III2 |
| 29 | 11.2214 | 1 | 1 | 11.22140 | 11.221436 | 11.2214 | III2 |
| 30 | 11.1421 | 1 | 1 | 11.14210 | 11.142075 | 11.1421 | III2 |
| 301 | 0.0000 | 0 | 0 | 11.06718 | 0.000000 | 0.0000 | III3 |

| | |
|---|---|
| **$about** | "{0,22,3\|9,6,15\|.9,1,1e-04}" |
| **$title** | "3pod: Wu & Tian Table 1 Example" |
| **$ttl0** | "Wu & Tian Table 1 Example" |
| **$ttl1** | NULL |
| **$ttl2** | NULL |
| **$units** | "in" |
| **$en** | 9 6 15 |
| **$p** | 0.9 |
| **$reso** | 1e-04 |
| **$n2n3** | 7 |
| **$ln** | FALSE |
| **$init** | 0 22 3 |
| **$lam** | 1 |
| **$test** | 1 |
| **$savinit** | 0 22 3 |

**$jvec**

| i | j | k | v | u | a | tau2 | nu | b | x | y |
|---|---|---|---|---|---|---|---|---|---|---|
| 1 | 0.000000 | 0.000000 | 0.0000000 | 0.00000000 | 0.0000000 | 3.1630046 | 0 | 0.0000000 | 11.71206 | 1 |
| 2 | 1.281552 | 1.259256 | 0.8455910 | 0.25691424 | 1.9676839 | 2.6574786 | 0 | 0.8455910 | 11.40827 | 1 |
| 3 | 1.281552 | 1.221520 | 0.8529442 | 0.21540766 | 1.7173487 | 2.2875485 | 0 | 0.8529442 | 11.15575 | 0 |
| 4 | 1.281552 | 1.193150 | 0.8586089 | 0.18486973 | 1.5228195 | 2.0060253 | 0 | 0.8586089 | 12.46331 | 1 |
| 5 | 1.281552 | 1.171100 | 0.8630914 | 0.16158850 | 1.3674862 | 1.7850553 | 0 | 0.8630914 | 12.27608 | 1 |
| 6 | 1.281552 | 1.153497 | 0.8667188 | 0.14331974 | 1.2406774 | 1.6072417 | 0 | 0.8667188 | 12.11074 | 1 |
| 7 | 1.281552 | 1.139135 | 0.8697100 | 0.12863984 | 1.1352462 | 1.4612038 | 0 | 0.8697100 | 11.96279 | 1 |
| 8 | 1.281552 | 1.127203 | 0.8722164 | 0.11660810 | 1.0462355 | 1.3392043 | 0 | 0.8722164 | 11.82911 | 1 |
| 9 | 1.281552 | 1.117136 | 0.8743455 | 0.10658089 | 0.9701036 | 1.2358098 | 0 | 0.8743455 | 11.70720 | 1 |
| 10 | 1.281552 | 1.108534 | 0.8761754 | 0.09810449 | 0.9042548 | 1.1470983 | 0 | 0.8761754 | 11.59523 | 1 |
| 11 | 1.281552 | 1.101099 | 0.8777644 | 0.09085065 | 0.8467447 | 1.0701710 | 0 | 0.8777644 | 11.49170 | 1 |
| 12 | 1.281552 | 1.094611 | 0.8791568 | 0.08457652 | 0.7960885 | 1.0028406 | 0 | 0.8791568 | 11.39550 | 1 |
| 13 | 1.281552 | 1.088901 | 0.8803866 | 0.07909884 | 0.7511331 | 0.9434269 | 0 | 0.8803866 | 11.30565 | 1 |
| 14 | 1.281552 | 1.083838 | 0.8814805 | 0.07427688 | 0.7109699 | 0.8906183 | 0 | 0.8814805 | 11.22144 | 1 |
| 15 | 1.281552 | 1.079317 | 0.8824597 | 0.07000091 | 0.6748731 | 0.8433765 | 0 | 0.8824597 | 11.14208 | 1 |
| 16 | 1.281552 | 1.075256 | 0.8833413 | 0.06618415 | 0.6422561 | 0.8008694 | 0 | 0.8833413 | 11.06718 | NA |

| | |
|---|---|
| **$term1** | TRUE |
| **$en12** | 7 2 |
| **$musig** | 10.1707909 0.9344091 |

Most of the components are self-explanatory. The components about, title, titl0, titl1 and titl2 are titles available for various plots; term1 = T is the default that could possibly allow 3pod's stage I2 of Phase I to terminate (and enter Phase II) even when there is no interval overlap; en12 breaks up the phase I sample size into two parts (test=1 only), the second one being 3pod's stage 3 sample size; n2n3 is used internally in the code to control print statements; and en contains the sample sizes of the 3 phases. The user is asked to provide: a title; the units of the stress; n2 and n3, the phase II and III sample sizes; and p and lam, the phase III parameters. savinit is a vector placeholder for mlo, mhi and sg; and reso is the resolution.

More needs to be said about jvec, selecting p and lam, and term1. The rows of jvec provide details about how the phase III stress sequence is computed. Section 5.3 has more to say about jvec and the selection of p and lam. The term1 component is discussed presently.

Gonogo has a term1=F option that makes sure that 3pod's Stage I2 of Phase I (the "search for overlapping region" stage) ends with interval overlap. In console mode (keyboard entry), such an option wouldn't be necessary if the user's entered-in a stress sequence that remained faithful to the recommended values. However, if the user deviates enough, and/or the resolution setting (reso) is too big, 3pod's stage I2 could end without interval overlap being achieved. Here are two examples, a 3pod phase I test that ends badly, and another where the term1 = F option is used to force stage I2 to continue until interval overlap is achieved:

**Table 11**. gonogo's term1=F option – continue in phase I until overlap is achieved.

| gonogo(1.2, 1.6, .05) | gonogo(4,7,.3333,reso=.25,term1=F) |
|---|---|
| Enter title (without quotes): Test (with term1=T) | Enter title (without quotes): Test (with term1=F) |
| Enter units (without quotes): in | Enter units (without quotes): in |
| | |
| 1. Test at X ~ 1.3. Enter X & R: 1.3 0 | 1. Test at X ~ 4.75. Enter X & R: 4 1 |
| 2. Test at X ~ 1.5. Enter X & R: 1.5 1 | 2. Test at X ~ 6.25. Enter X & R: 6 1 |
| 3. Test at X ~ 1.4. Enter X & R: 1.4 0 | 3. Test at X ~ 3.5. Enter X & R: 3.5 0 |
| 4. Test at X ~ 1.45. Enter X & R: 1.45 1 | 4. Test at X ~ 3.75. Enter X & R: 4 1 |
| 5. Test at X ~ 1.385. Enter X & R: 1.38 0 | 5. Test at X ~ 3.75. Enter X & R: 4 1 |
| 6. Test at X ~ 1.465. Enter X & R: 1.46 1 | 6. Test at X ~ 3.75. Enter X & R: 4 1 |
| 7. Test at X ~ 1.425. Enter X & R: 1.42 1 | 7. Test at X ~ 3.75. Enter X & R: 3.5 1 |
| 8. Test at X ~ 1.39. Enter X & R: 1.39 0 | 8. Test at X ~ 3.5. Enter X & R: 3.5 0 |
| 9. Test at X ~ 1.43. Enter X & R: 1.43 1 | 9. Test at X ~ 3.5. Enter X & R: 3.5 1 |
| 10. Test at X ~ 1.39333. Enter X & R: 1.39 0 | 10. Test at X ~ 3.5. Enter X & R: 4 1 |
| 11. Test at X ~ 1.42667. Enter X & R: 1.43 1 | **** 3pod would normally enter I3 here **** |
| 12. Test at X ~ 1.39556. Enter X & R: 1.4 0 | 11. Test at X ~ 3.5. Enter X & R: 4 1 |
| 13. Test at X ~ 1.42444. Enter X & R: 1.42 1 | 12. Test at X ~ 3.5. Enter X & R: 3.75 1 |
| 14. Test at X ~ 1.41. Enter X & R: 1.41 1 | 13. Test at X ~ 3.5. Enter X & R: 3.55 0 |
| 15. Test at X ~ 1.39704. Enter X & R: 1.4 1 | 14. Test at X ~ 3.5. Enter X & R: 3.5 1 |
| 16. Test at X ~ 1.40494. Enter X & R: 1.4 1 | 15. Test at X ~ 3.5. Enter X & R: 3.5 1 |
| 17. Test at X ~ 1.39506. Enter X & R: 1.4 0 | 16. Test at X ~ 3.5. Enter X & R: 3.5 1 |
| | |
| Phase I complete, (Mu, Sig) = (1.4, **0**). | Phase I complete, (Mu, Sig) = (3.43427, 0.20521). |
| Enter Phase II (D-Optimal) size n2: -1 | Enter Phase II (D-Optimal) size n2: -1 |
| | |
| Test Suspended | Test Suspended |

**Note**. For the test=1 situation (3pod), this means – continue stage I2 of phase I

Further background material and specifics about the 3pod methodology, which forms the three-phase template of gonogo, may be found in Wu (1985), Wu and Tian (2014), Hung and Joseph (2014), Ray, Roediger and Neyer (2014), Steinberg and Dror (2014), Johnson, Freeman, Hester and Bell (2014), Wang, Tian and Wu (2015), Ray and Roediger (2018) and Wang, Tian and Wu (2020). For the case of one-dimensional predictors, the overlap condition reduces to $\min\left(X_{Y=1}\right) < \max\left(X_{Y=0}\right)$.

### 5.1.2 Neyer Phase 1

To recreate Neyer's Table 1 (Neyer 1994) using the batch mode, run the following commands: **yNY=c(rep(0,5),1,rep(0,6),1,0,1,0,1,1,1,1)**, **wNY=gonogo(.6,1.4,.1,test=2,reso=.01,Y=yNY).**

You get

**Table 12**. Reproduction of Neyer's Table 1 (Neyer 1994) using gonogo's Batch option

| Neyer Table 1 | | Gonogo Reproduction (wNY) | | | | | | |
|---|---|---|---|---|---|---|---|---|
| i | X | X | Y | COUNT | RX | EX | TX | ID |
| 1 | 1 | 1.00 | 0 | 1 | 1.00 | 1.000000 | 1.00 | B0 |
| 2 | 1.2 | 1.20 | 0 | 1 | 1.20 | 1.200000 | 1.20 | B1 |
| 3 | 1.4 | 1.40 | 0 | 1 | 1.40 | 1.400000 | 1.40 | B1 |
| 4 | 1.8 | 1.80 | 0 | 1 | 1.80 | 1.800000 | 1.80 | B1 |
| 5 | 2.6 | 2.60 | 0 | 1 | 2.60 | 2.600000 | 2.60 | B1 |
| 6 | 4.2 | 4.20 | 1 | 1 | 4.20 | 4.200000 | 4.20 | B1 |
| 7 | 3.4 | 3.40 | 0 | 1 | 3.40 | 3.400000 | 3.40 | B3 |
| 8 | 3.8 | 3.80 | 0 | 1 | 3.80 | 3.800000 | 3.80 | B3 |
| 9 | 4 | 4.00 | 0 | 1 | 4.00 | 4.000000 | 4.00 | B3 |
| 10 | 4.1 | 4.10 | 0 | 1 | 4.10 | 4.100000 | 4.10 | B3 |
| 11 | 4.28 | 4.28 | 0 | 1 | 4.28 | 4.280593 | 4.28 | B4 |
| 12 | 4.52 | 4.52 | 0 | 1 | 4.52 | 4.522707 | 4.52 | II1 |
| 13 | 5.55 | 5.55 | 1 | 1 | 5.55 | 5.546771 | 5.55 | II2 |
| 14 | 5.24 | 5.24 | 0 | 1 | 5.24 | 5.243292 | 5.24 | II2 |
| 15 | 6.37 | 6.37 | 1 | 1 | 6.37 | 6.371975 | 6.37 | II2 |
| 16 | 6.08 | 6.08 | 0 | 1 | 6.08 | 6.080515 | 6.08 | II2 |
| 17 | 7.38 | 7.38 | 1 | 1 | 7.38 | 7.384476 | 7.38 | II2 |
| 18 | 7.09 | 7.09 | 1 | 1 | 7.09 | 7.094232 | 7.09 | II2 |
| 19 | 6.89 | 6.89 | 1 | 1 | 6.89 | 6.893254 | 6.89 | II2 |
| 20 | 6.74 | 6.74 | 1 | 1 | 6.74 | 6.736082 | 6.74 | II2 |

The graph produced appears below as Figure 4.

### 5.1.3 Bruceton Phase I

Here is a simulated 3 Phase test with a Bruceton Phase I. The command we'll run won't specify BL, so by default $n\mathrm{Re}v$, $i_1$ and $i_2$ will be set to 4, 1 and 0, respectively. The command

**wb=gonogoSim(10,10,.25,6,6,.9,1,plt=1,test=3,reso=.01,iseed=62517)** produces Figure 5.

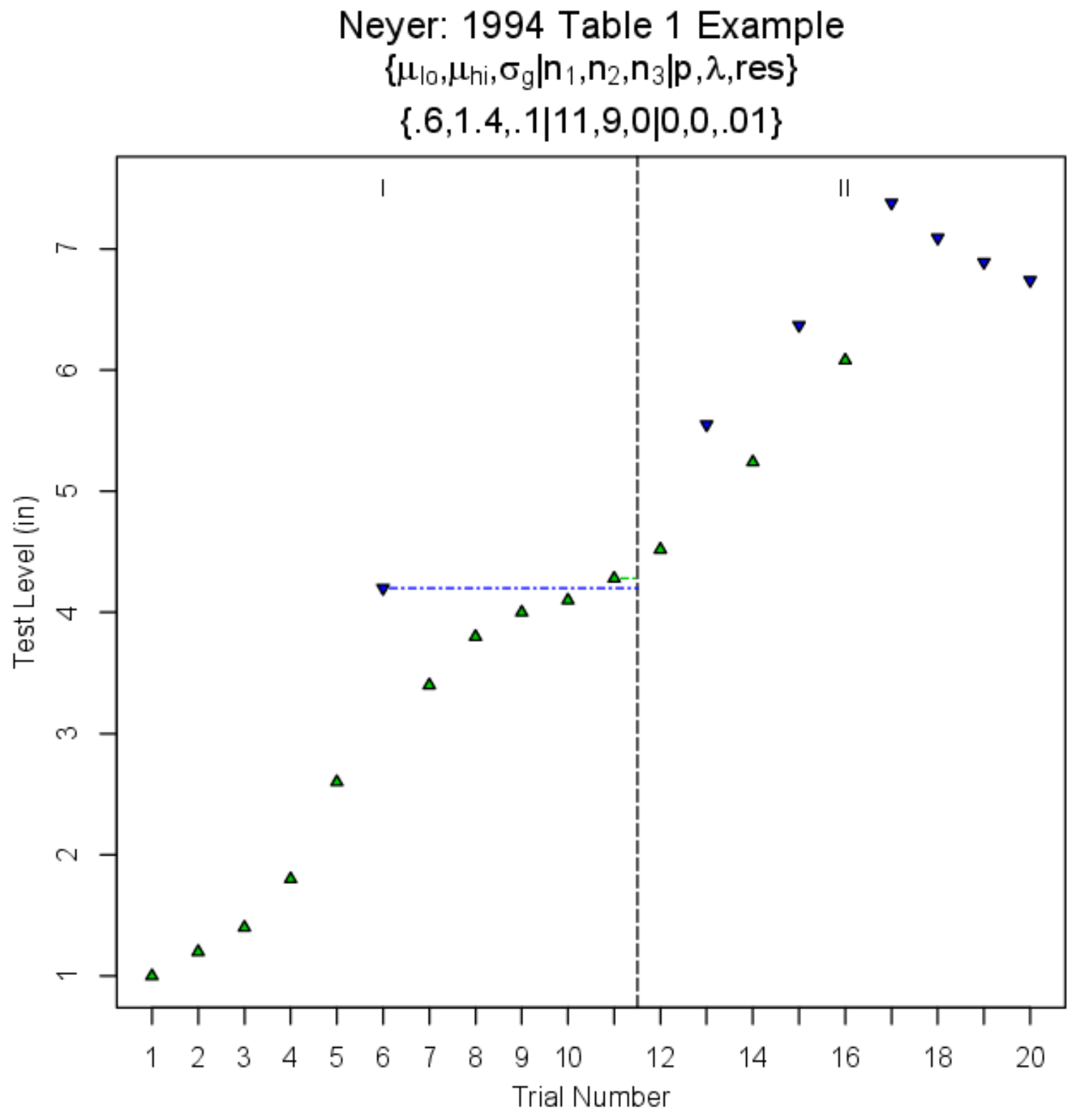

**Figure 4**. Neyer Table 1 History Plot

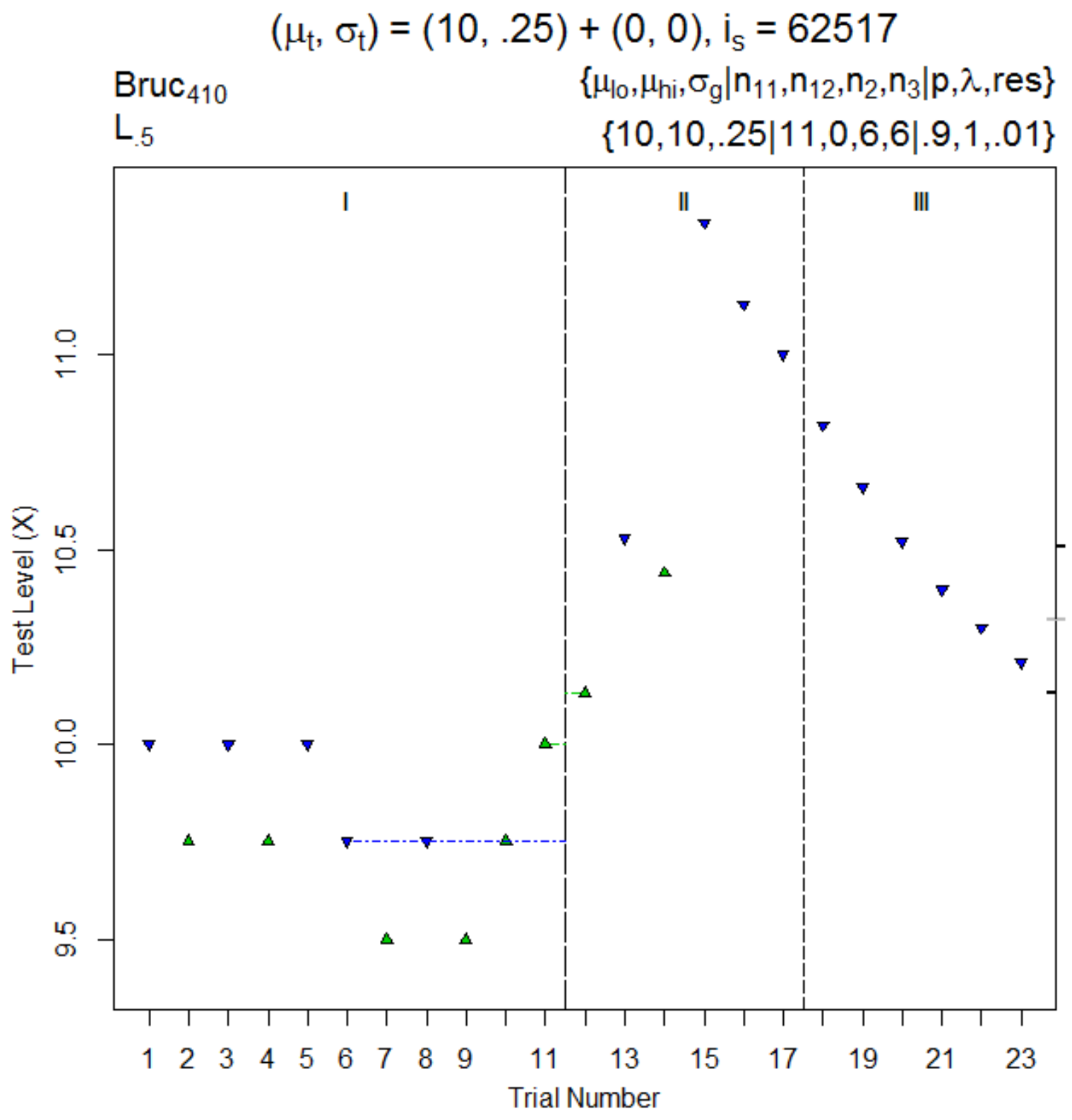

**Figure 5**. A randomly generated Bruceton Test

**Note**: if the iseed argument is left out it reverts to its default setting (-1) and a new (i.e., different) wb gets created each time the call is made. If iseed is specified (and not -1), rerunning the simulation will always give you the same result.

Another command, say with a user supplied BL selection of "4 2 2" for "$n\,\mathrm{Re}\,v \;\; i_1 \;\; i_2$" looks like

**ub=gonogoSim(10,10,.25,6,6,.9,1,plt=1,test=3,reso=.01,BL=c(4,2,2),iseed=62517)**.

Running this command produces a Phase I test consisting of two back-to-back Bruceton tests, as depicted below:

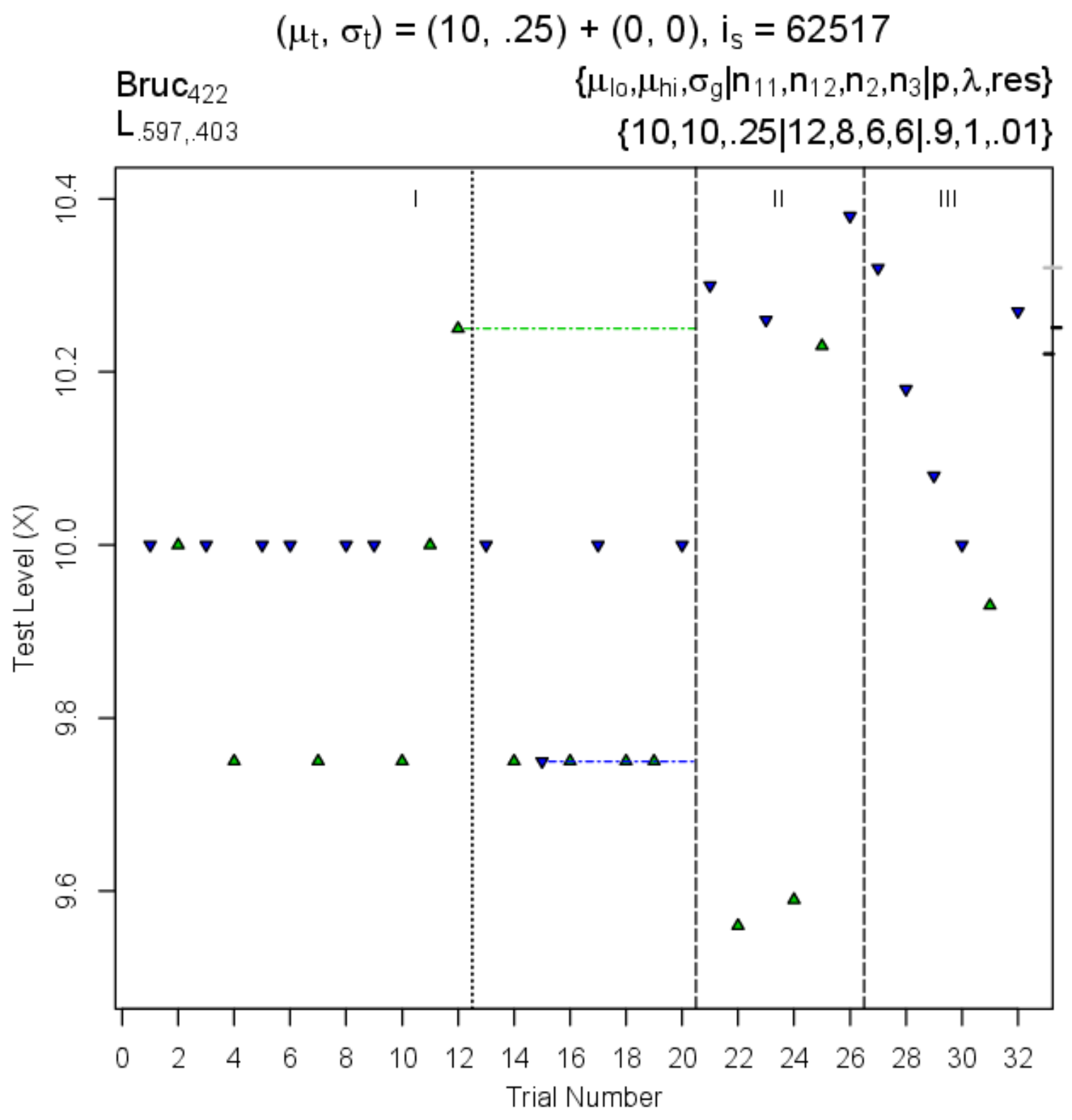

**Figure 6**. A simulated test (ub) with Phase I consisting of two Bruceton Tests

If you wanted to see the generalized U's and D's identified on the plot, you could replot the above graph using the pSdat1 function with the ud=T option (see Table 35 for alternate syntax for ptest(w,i=1)).

Running **pSdat1(ub,ud=T)** produces the following plot:

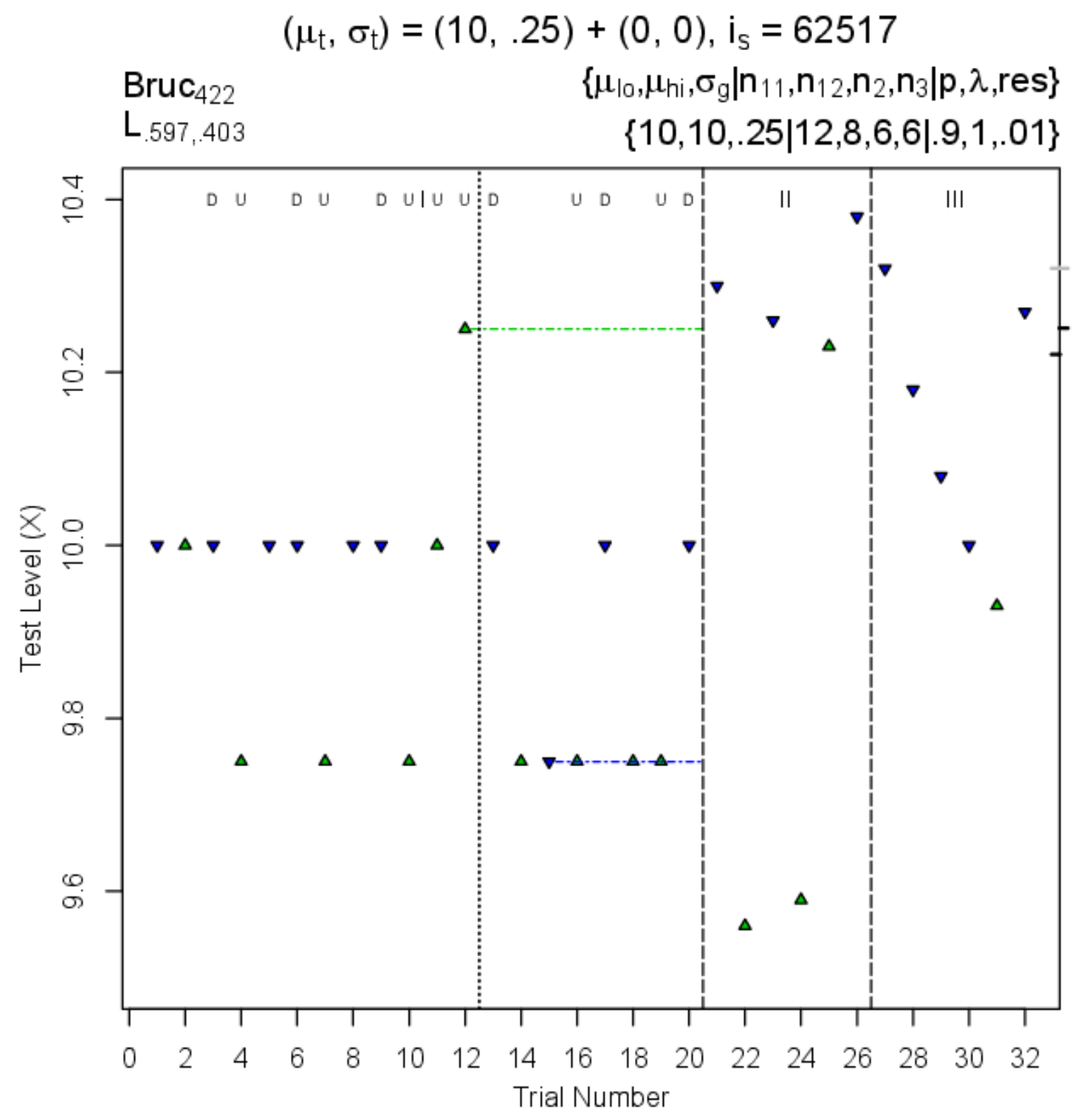

**Figure 7**. Same as Figure 6, but with annotated UDTR's

Further information about the Bruceton test may be found in Dixon and Mood (1948), Wetherill (1963) and MIL-STD-331D.

### 5.1.4 Langlie Phase 1

One way to to recreate, for example, the $L_p$ test cited in Table 6.2-3 on pp 202 of Einbinder (1973), initialized with $mlo = 0,\ mhi = 5,\ p = .2062995$, and stopping after 7 changes of response, would be to run it in console mode by entering in 7 0 5 at the "Enter BL (nRev and two i's (one 0 is OK): " prompt (see Table 8). You will then need to make sure the sequence starts correctly by entering in 2.5 for the first stress (because gonogo suggests $X_1 = 0 \cdot (1-p) + 5 \cdot p = 1.031497$). The console recreation is started by running the command **wLG1=gonogo(0,5,0,test=4)**. Details of the subsequent screen prompts and necessary user entries are described below:

**Table 13**. Console mode reproduction of Einbinder (1973, Table 6.2-3) using wLG1=gonogo(0,5,0,test=4)

| Prompt | Entry |
|---|---|
| `Enter title (without quotes):` | `Einbinder Example(Table 6.2-3)` |
| `Enter units (without quotes):` | `in` |
| `Three entries (separated by blanks) are required, namely` | |
| | |
| `(1) nRev: the number of reversals needed to exit Phase I` | |
| `(2) two I values (chosen from the following table 8)` | |

```
   I   Down(X=1,O=0)       Up(X=1,O=0)          p
 ---- --------------- --------------------- ----------
|  1 |              X |                  O  | .500000  |
|  3 |             XX |             {O, XO} | .707107  |
|  5 |            XXX |        {O, XO, XXO} | .793701  |
|  7 |           XXXX |  {O, XO, XXO, XXXO} | .840896  |
|  : |            :   |         :           |     :    |
 ---- --------------- --------------------- ---------
|  0 |              - |                   - |    -     |
|  2 |      {XX, XOX} |            {O, XOO} | .596968  |
|  4 |    {XXX, XXOX} |       {O, XO, XXOO} | .733614  |
|  6 | {XXXX, XXXOX} | {O, XO, XXO, XXXOO} | .804119  |
|  : |       :        |         :           |     :    |
 ---- --------------- --------------------- ----------
       Up(X=0,O=1)        Down(X=0,O=1)       1-p

Enter BL (nRev and two I's (one 0 is OK):                                    705
D = {1, 01, 001}, U = {000}, Lev = 0.206299
 1. Test at X ~ 1.0315. Enter X & R:                                       2.5 1
 2. Test at X ~ 1.25. Enter X & R:                                        1.25 1
 3. Test at X ~ 0.625. Enter X & R:                                       .625 1
 4. Test at X ~ 0.3125. Enter X & R:                                     .3125 0
 5. Test at X ~ 0.3125. Enter X & R:                                     .3125 0
 6. Test at X ~ 0.3125. Enter X & R:                                     .3125 0
 7. Test at X ~ 0.46875. Enter X & R:                                   .46875 0
 8. Test at X ~ 0.46875. Enter X & R:                                   .46875 0
 9. Test at X ~ 0.46875. Enter X & R:                                   .46875 1
10. Test at X ~ 0.39062. Enter X & R:                                   .39062 0
11. Test at X ~ 0.39062. Enter X & R:                                   .39062 0
12. Test at X ~ 0.39062. Enter X & R:                                   .39062 0
13. Test at X ~ 0.42968. Enter X & R:                                   .42968 0
14. Test at X ~ 0.42968. Enter X & R:                                   .42968 0
15. Test at X ~ 0.42968. Enter X & R:                                   .42968 0
16. Test at X ~ 0.83984. Enter X & R:                                   .83984 0
17. Test at X ~ 0.83984. Enter X & R:                                   .83984 0
18. Test at X ~ 0.83984. Enter X & R:                                   .83984 1
19. Test at X ~ 0.63476. Enter X & R:                                   .63476 0
20. Test at X ~ 0.63476. Enter X & R:                                   .63476 0
21. Test at X ~ 0.63476. Enter X & R:                                   .63476 0
22. Test at X ~ 0.7373. Enter X & R:                                     .7373 1
23. Test at X ~ 0.68603. Enter X & R:                                   .68603 0
24. Test at X ~ 0.68603. Enter X & R:                                   .68603 0
25. Test at X ~ 0.68603. Enter X & R:                                   .68603 0

Phase I complete, (Mu, Sig) = (0.8625, 0.29107).
Enter Phase II (D-Optimal) size n2:                                            0

Phase II skipped, (Mu, Sig) = (0.8625, 0.29107).
Enter Phase III (S-RMJ) size n3:                                               0

Phase III complete, (Mu, Sig) = (0.8625, 0.29107).
```

**Note**. Recommended first x was 1.031497, but actual first x was 2.5

A quicker way would be to use batch mode. It's invoked by defining Y to be the entire vector of responses in the call to gonogo. If all of the stresses were the ones suggested by gonogo, then the default X=NULL would suffice. When one or more stresses differ, X has to be defined within the gonogo call to be the vector of all stresses *UP TO AND INCLUDING THE LAST ONE THAT DIFFERS*.

To recreate the Einbinder example this way, define **ywLG1=wLG1$d0$Y**, the response sequence for the example. Then run **wLG2=gonogo(0,5,0,test=4,Y=ywLG1,X=2.5).** This command reproduces wLG1 and its history plot, which is presented below:

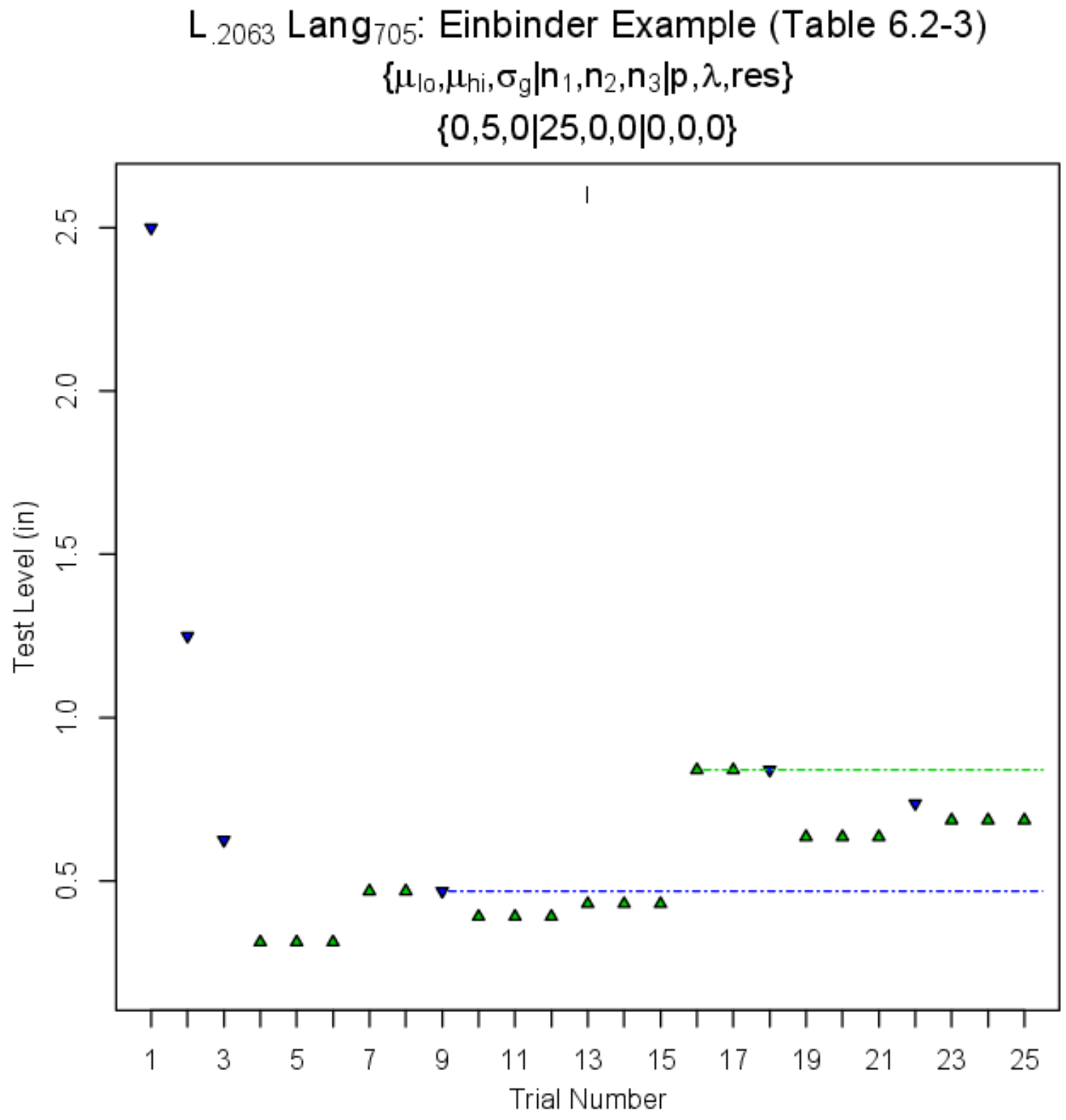

**Figure 8**. Langlie Test (wLG2) reproduction using gonogo's batch mode

For a final comparison, run **wLG3=gonogo(0,5,0,test=4,Y=ywLG1)**. This test begins with X1 = 1.031497 (instead of 2.5). This has the effect of substantially reducing the subsequent X's. Comparisons of Einbinder's original example with wLG1, wLG2 and wLG3 are summarized below:

**Table 14**. Comparison of console and batch versions

| Original Table 6.2-3 | | | | Batch (wLG1) & Console (wLG2) | | | Batch (wLG3) | |
|---|---|---|---|---|---|---|---|---|
| i | Y | X | ID | X | RX | ID | $X_1 = RX_1$ | ID |
| 1 | 1 | 2.5 | D | 2.5 | 2.5 | D | 1.03150 | D |
| 2 | 1 | 1.25 | D | 1.25 | 1.25 | D | .51575 | D |
| 3 | 1 | .625 | D | .625 | .625 | D | .25788 | D |
| 4 | 0 | .3125 | | .3125 | .3125 | | .12894 | |
| 5 | 0 | .3125 | | .3125 | .3125 | | .12894 | |
| 6 | 0 | .3125 | U | .3125 | .3125 | U | .12894 | U |
| 7 | 0 | .46875 | | .46875 | .46875 | | .19341 | |
| 8 | 0 | .46875 | | .46875 | .46875 | | .19341 | |
| 9 | 1 | .46875 | D | .46875 | .46875 | D | .19341 | D |
| 10 | 0 | .3906 | | .39062 | .39062 | | .16118 | |
| 11 | 0 | .3906 | | .39062 | .39062 | | .16118 | |
| 12 | 0 | .3906 | U | .39062 | .39062 | U | .16118 | U |
| 13 | 0 | .4297 | | .42968 | .42968 | | .17729 | |
| 14 | 0 | .4297 | | .42968 | .42968 | | .17729 | |
| 15 | 0 | .4297 | U | .42968 | .42968 | U | .17729 | U |
| 16 | 0 | .83984 | | .83984 | .83984 | | .34652 | |
| 17 | 0 | .83984 | | .83984 | .83984 | | .34652 | |
| 18 | 1 | .83984 | D | .83984 | .83984 | D | .34652 | D |
| 19 | 0 | .63477 | | .63476 | .63476 | | .26190 | |
| 20 | 0 | .63477 | | .63476 | .63476 | | .26190 | |
| 21 | 0 | .63477 | U | .63476 | .63476 | U | .26190 | U |
| 22 | 1 | .7373 | D | .7373 | .7373 | D | .30421 | D |
| 23 | 0 | .68603 | | .68603 | .68603 | | .28306 | |
| 24 | 0 | .68603 | | .68603 | .68603 | | .28306 | |
| 25 | 0 | .68603 | U | .68603 | .68603 | U | .28306 | U |

Further details about the Langlie procedure may be obtained in Langlie (1962), Einbinder (1973) and MIL-STD-331D.

## 5.2 Phase II

Upon successful completion of phase I, we're ready to begin refining our MLE's of mu and sigma. A D-optimal methodology is employed to accomplish this – which is described as follows:

> For large sample sizes, the area of the standard confidence ellipsoid for the parameters is inversely proportional to the determinant of the information matrix. Since a D-optimal result will be obtained when the determinant of the information matrix is maximized, a D-optimal design gives the smallest confidence ellipsoid for the parameters. Since the off-diagonal terms of the matrix are typically small compared to the diagonal terms, a D-optimal test will also approximately minimize the product of the asymptotic variances of $\mu$ and $\sigma$. This condition is achieved (for a normal distribution) by testing near $\mu \pm 1.138 \cdot \sigma$ (Neyer 1994).

This tendancy to test at two levels in phase II can be illustrated by considering the history plot depicted in Figure 9. It is simulated Neyer test given by **ny=gonogoSim(.4,1.6,.1,20,test=2,plt=1,iseed=7865)**, having overlaid horizontal lines at stresses $\mu_t \pm 1.138 \cdot \sigma_t$ obtained via **abline(h=1+1.138/10,lty=2)** and **abline(1-1.138/10,lty=2)**.

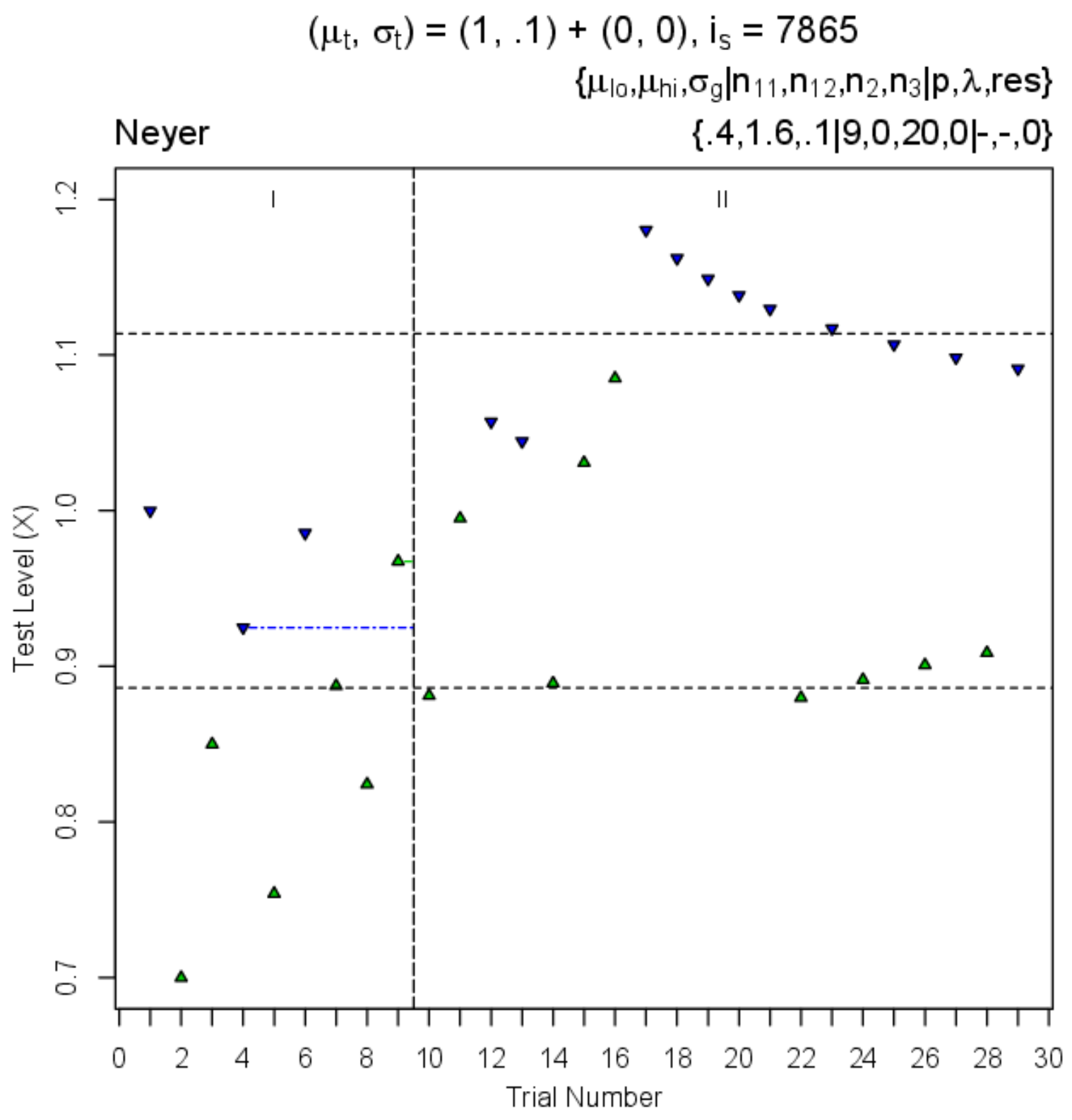

**Figure 9**. Phase II testing tends to be conducted at two levels (horizontal lines)

The steps to compute D-Optimal stresses are described succinctly in the Neyer procedure diagram (Phase II block) in Appendix B, and are expanded upon below. Define the information matrix

$$b = \begin{pmatrix} \sum_{i=1}^{n} G_i^2 & \sum_{i=1}^{n} G_i^2 k_i \\ \sum_{i=1}^{n} G_i^2 k_i & \sum_{i=1}^{n} G_i^2 k_i^2 \end{pmatrix}, \text{ where } k_i = \frac{X_i - \tilde{\mu}}{\tilde{\sigma}} \text{ and } G_i = \frac{\varphi(k_i)}{\sqrt{\Phi(k_i)(1-\Phi(k_i))}}.$$

With $\tilde{\mu} = median\left(X_{min}, X_{\max}, \hat{\mu}\right)$ and $\tilde{\sigma} = min\left(X_{\max} - X_{\min}, \hat{\sigma}\right)$, the next stress computed by the D-Optimal methodology is given by $X_{n+1} = \tilde{\mu} + k^* \tilde{\sigma}$, where $G^2(k)\left(b_{11}k^2 - 2b_{12}k + b_{22}\right)$ attains its maximum at $k = k^*$. The steps to perform the maximization are presented below:

**Table 15**. D-optimal steps to compute $X_{next} \mid d0$

| Compute | R Code |
|---|---|
| $\hat{\mu}$ | nq=glmmle(d0), mu=nq$mu |
| $\hat{\sigma}$ | sig=nq$sig (***when there is overlap***) |
| $\tilde{\mu}$ | xl=min(d0$X); xu=max(d0$X), mut=max(xl,min(mu,xu)) |
| $\tilde{\sigma}$ | sigt=min(sig,xu-xl) |
| $k_i$ | ki=(d0$X-mut)/sigt |
| $G_i$ | p=pnorm(ki), g=dnorm(ki)/sqrt(p*(1-p)) |
| $b$ | b=yinfomat(d0,mut,sigt)$infm |
| $k^*$ | ks=kstar(b) |
| $X_{next}$ | xnext=mut + ks · sigt |

## 5.3 Phase III

This optional phase utilizes the skewed version of the Robbins-Monro-Joseph (RMJ) procedure described in Wang, Tian and Wu (2015). Background on the development of the procedure is provided in Robbins and Monro (1952) and Joseph (2004).

The purpose of phase III is to accelerate convergence to the unknown quantile $L_p$ once the region of interest has been identified in phases I and II. Phase III requires the user to specify two parameters, $p$ (a probability between 0 and 1) and $\lambda$ (a positive skewness coefficient).

For $p \gg .5$, picking $\lambda < 1$ lets the user force bigger downward steps than upward steps, and for $p \ll .5$, picking $\lambda > 1$ lets the user force bigger upward steps than downward steps. Both choices are recommended in these cases, as they increase the chance of getting the less probable response. The source materials unfortunately don't have much more to say about choosing $p$ and $\lambda$, or how to allocate sample sizes for phase II and III testing. However, in regards to $p$, Neyer has said (personal communication):

> I am more than ever convinced that (sensitivity) *testing is mainly useful for finding process variation*, and any prediction of extreme levels must be treated with a grain of salt.

To sequentially compute the phase III stresses, consider the jvec component of a returned gonogo test, e.g., **wWT** (discussed earlier). It can be displayed by the command **wWT$jvec**:

**Table 16**. The jvec component of wWT

| i | j | k | v | u | a | tau2 | nu | b | X | Y |
|---|---|---|---|---|---|---|---|---|---|---|
| 1 | 0.000000 | 0.000000 | 0.0000000 | 0.00000000 | 0.0000000 | 3.1630046 | 0 | 0.0000000 | 11.71206 | 1 |
| 2 | 1.281552 | 1.259256 | 0.8455910 | 0.25691424 | 1.9676839 | 2.6574786 | 0 | 0.8455910 | 11.40827 | 1 |
| 3 | 1.281552 | 1.221520 | 0.8529442 | 0.21540766 | 1.7173487 | 2.2875485 | 0 | 0.8529442 | 11.15575 | 0 |
| 4 | 1.281552 | 1.193150 | 0.8586089 | 0.18486973 | 1.5228195 | 2.0060253 | 0 | 0.8586089 | 12.46331 | 1 |
| 5 | 1.281552 | 1.171100 | 0.8630914 | 0.16158850 | 1.3674862 | 1.7850553 | 0 | 0.8630914 | 12.27608 | 1 |
| 6 | 1.281552 | 1.153497 | 0.8667188 | 0.14331974 | 1.2406774 | 1.6072417 | 0 | 0.8667188 | 12.11074 | 1 |
| 7 | 1.281552 | 1.139135 | 0.8697100 | 0.12863984 | 1.1352462 | 1.4612038 | 0 | 0.8697100 | 11.96279 | 1 |
| 8 | 1.281552 | 1.127203 | 0.8722164 | 0.11660810 | 1.0462355 | 1.3392043 | 0 | 0.8722164 | 11.82911 | 1 |
| 9 | 1.281552 | 1.117136 | 0.8743455 | 0.10658089 | 0.9701036 | 1.2358098 | 0 | 0.8743455 | 11.70720 | 1 |
| 10 | 1.281552 | 1.108534 | 0.8761754 | 0.09810449 | 0.9042548 | 1.1470983 | 0 | 0.8761754 | 11.59523 | 1 |
| 11 | 1.281552 | 1.101099 | 0.8777644 | 0.09085065 | 0.8467447 | 1.0701710 | 0 | 0.8777644 | 11.49170 | 1 |
| 12 | 1.281552 | 1.094611 | 0.8791568 | 0.08457652 | 0.7960885 | 1.0028406 | 0 | 0.8791568 | 11.39550 | 1 |
| 13 | 1.281552 | 1.088901 | 0.8803866 | 0.07909884 | 0.7511331 | 0.9434269 | 0 | 0.8803866 | 11.30565 | 1 |
| 14 | 1.281552 | 1.083838 | 0.8814805 | 0.07427688 | 0.7109699 | 0.8906183 | 0 | 0.8814805 | 11.22144 | 1 |
| 15 | 1.281552 | 1.079317 | 0.8824597 | 0.07000091 | 0.6748731 | 0.8433765 | 0 | 0.8824597 | 11.14208 | 1 |
| 16 | 1.281552 | 1.075256 | 0.8833413 | 0.06618415 | 0.6422561 | 0.8008694 | 0 | 0.8833413 | 11.06718 | NA |

jvec was computed and returned by gonogo's function **skewL**. Stresses beyond the first run were calculated iteratively from preceding rows of jvec. Precise details are described below:

**Table 17**. Phase III Skewed RMJ Computations

| | Steps | Code |
|---|---|---|
| | Calculate $\hat{\mu}_{I+II}$, $\hat{\sigma}_{I+II}$ | g=glmmle(d0), mu=g$muhat, sig=g$sighat |
| | Calculate $\tau_1^2$ | ww=yinfomat(d0,mu,sig)<br>tau2=sum(t(c(1,qnorm(p)^2))*diag(ww$vcov1)) |
| | Truncate $\tau_1^2$ | ti=round((c(3,5)/qnorm(.975))^2,4)*sig^2<br>***if(ln) round((c(3,5)/qlnorm(.975))^2,4)*sig^2***<br>tau2=min(max(tau2,ti[1]),ti[2]) |
| | Calculate $\tilde{\mu}$, $\tilde{\sigma}$ | m1=min(d0$X,na.rm=T) ;<br>m2=max(d0$X,na.rm=T);<br>m2=min(c(mu,m2),na.rm=T);<br>mut=max(c(m1,m2),na.rm=T);<br>sigt=min(sig,diff(range(d0$X)),na.rm=T); |
| | Calculate $\beta$ | be=1/(2*sigt)<br>***if(ln) be=plnorm(qlnorm(p))/(pnorm(qnorm(p))*sigt)*** |
| | Calculate $\upsilon_1$ | c1=f3point8(lam);<br>nu=sqrt(tau2)*c1; |
| | Calculate $X_1$, Obtain $Y_1$ | xx=mut+qnorm(p)*sigt+nu |
| for(i in 1:n3) | Compute jvec<br>(using skewL) | j=qnorm(p)+be*nu<br>k=sqrt(1+be^2*tau2)<br>v=pnorm(j/k)<br>u=be*tau2*dnorm(j/k)/k+nu*v<br>a=(u-nu*v)/(v*(1-v))<br>ntau2=a^2*v*(1-v)-2*a*(u-nu*v)+tau2<br>nnu=sqrt(ntau2)*c1<br>b=v-(nu-nnu)/a |
| | Compute<br>$X_{i+1} = X_i - a_i\left(Y_i - b_i\right)\hat{\sigma}_{I+II}$ | tau2=ntau2<br>nu=nnu |
| | If(i<n3) Obtain $Y_{i+1}$ | xx=d0$X[nrow(d0)]-a*(d0$Y[nrow(d0)]-b) |

## 6. Confidence

Gonogo offers a **lims** function that implements three computational methods to compute confidence intervals: Fisher Matrix (FM), General Linear Model (GLM) and Likelihood Ratio (LR).

The call to the lims function is **lims(ctyp,dat,conf,P=numeric(0),Q=numeric(0))**. The arguments to lims are described in the following table:

**Table 18**. Five arguments to the lims function

| Argument | Description |
|---|---|
| ctyp | An integer: 1 for FM; 2 for LR; and 3 for GLM. |
| dat | The d0 component of a named list generated by gonogo (or gonogoSim).<br>For example, for the previously defined list wWT, dat would be wWT$d0. |
| conf $(C)$ | A confidence associated with the interval, e.g., .95. Think of conf as ***2-sided*** |
| P | A vector of probabilities, e.g., .95, c(.95,.99), al15 or al49 (see below), etc. |
| Q | A vector of quantiles (i.e., stresses) |

**Note**. gonogo users are advised that confidence level inputs to lims and ptest are always 2-Sided. Thus, if you're primarily interested in 1-sided limits, say at 95% confidence, you'd need to do your work in gonogo with a 90% 2-sided confidence limit entry.

There are two handy vector objects defined by gonogo containing 15 and 49 values of alpha that span the (0,1) range of possible interest. They are:

**al15** =c(.000001, .00001, .0001, .001, .01, .1, .25, .5, .75, .9, .99, .999, .9999, .99999, .999999), and

**al49** =c(.000001, .00001, .0001, .001, .01,seq(.025,.975,by=.025),.99, .999, .9999, .99999, .999999).

The two vector constants are particularly useful to compute confidence interval tables with the lims function.

For example, here's how you would compute a 95% FM confidence interval (CI) about the stress 8.5 for the gonogo test wWT: **tbl=lims(1,wWT$d0,.95,Q=8.5)**. The vector output is presented below:

**Table 19**. 95% CI's for gonogo's wWT test. The predicted $p \mid q = 8.5$ is based on MLE's $\hat{\mu}$ and $\hat{\sigma}$

| FM 95% CI about $q = 8.5$ | | | FM 95% CI about $p = .036882$ | | |
|---|---|---|---|---|---|
| $q_l$ | $q$ | $q_u$ | $p_l$ | $p$ | $p_u$ |
| 6.519019 | 8.5 | 10.48098 | 0 | 0.036882 | 0.20788 |

The command **tbl=lims(1,wWT$d0,.95,P=al15,Q=8.5)** returns the following matrix (as a component of tbl):

**Table 20**. Output of lims using the vector al15

| FM 95% CI about $q$ | | | FM 95% CI about $p$ | | | Notes |
|---|---|---|---|---|---|---|
| $q_l$ | $q$ | $q_u$ | $p_l$ | $p$ | $p_u$ | |
| 1.713070 | 5.729148 | 9.745225 | 0.000000 | 0.000001 | 0.000022 | |
| 2.509827 | 6.185638 | 9.861449 | 0.000000 | 0.000010 | 0.000186 | |
| 3.398713 | 6.695708 | 9.992703 | 0.000000 | 0.000100 | 0.001497 | |
| 4.420058 | 7.283250 | 10.146441 | 0.000000 | 0.001000 | 0.011317 | |
| 5.655082 | 7.997030 | 10.338979 | 0.000000 | 0.010000 | 0.076799 | |
| 7.323686 | 8.973297 | 10.622909 | 0.000000 | 0.100000 | 0.409826 | |
| **8.268435** | 9.540542 | **10.812648** | 0.000000 | **0.250000** | 0.682622 | **See Figure 10** |
| 9.261870 | 10.170791 | 11.079711 | 0.111940 | 0.500000 | 0.888060 | |
| **10.100920** | 10.801040 | **11.501160** | 0.511901 | **0.750000** | 0.988099 | **See Figure 10** |
| 10.608143 | 11.368284 | 12.128426 | 0.757232 | 0.900000 | 1.000000 | |
| 11.082057 | 12.344552 | 13.607046 | 0.953990 | 0.990000 | 1.000000 | |
| 11.317805 | 13.058332 | 14.798859 | 0.992728 | 0.999000 | 1.000000 | |
| 11.489465 | 13.645874 | 15.802283 | 0.998986 | 0.999900 | 1.000000 | |
| 11.630368 | 14.155944 | 16.681520 | 0.999869 | 0.999990 | 1.000000 | |
| 11.752583 | 14.612434 | 17.472285 | 0.999984 | 0.999999 | 1.000000 | |
| **6.519019** | **8.500000** | **10.480981** | **0.000000** | **0.036882** | **0.207880** | **See Table 19** |

An important plotting function of confidence is **ptest** with the plt=3 option. Here is a gonogo screen menu that contains all of the available options when running the command **ptest(wWT,3)**:

**Table 21**. Fifteen possible entries for J can be supplied to ptest(w,3)

This function requires two inputs, conf & J. Choose J from the following ...

| | To Plot Confidence Interval(s) about: | | | Via the Method(s) | | |
|---|---|---|---|---|---|---|
| | Probability (p) | Quantile (q) | p&q | FM | LR | GLM |
| Enter this value for J | 1 | 2 | 3 | X | | |
| | | | 4 | | X | |
| | 5 | 6 | 7 | | | X |
| | 8 | 9 | | X | X | |
| | 10 | 11 | | X | | X |
| | 12 | 13 | | | X | X |
| | 14 | 15 | | X | X | X |

By entering .95 for conf and 15 for J at the **" Enter conf and J:"** prompt, we get the following graph:

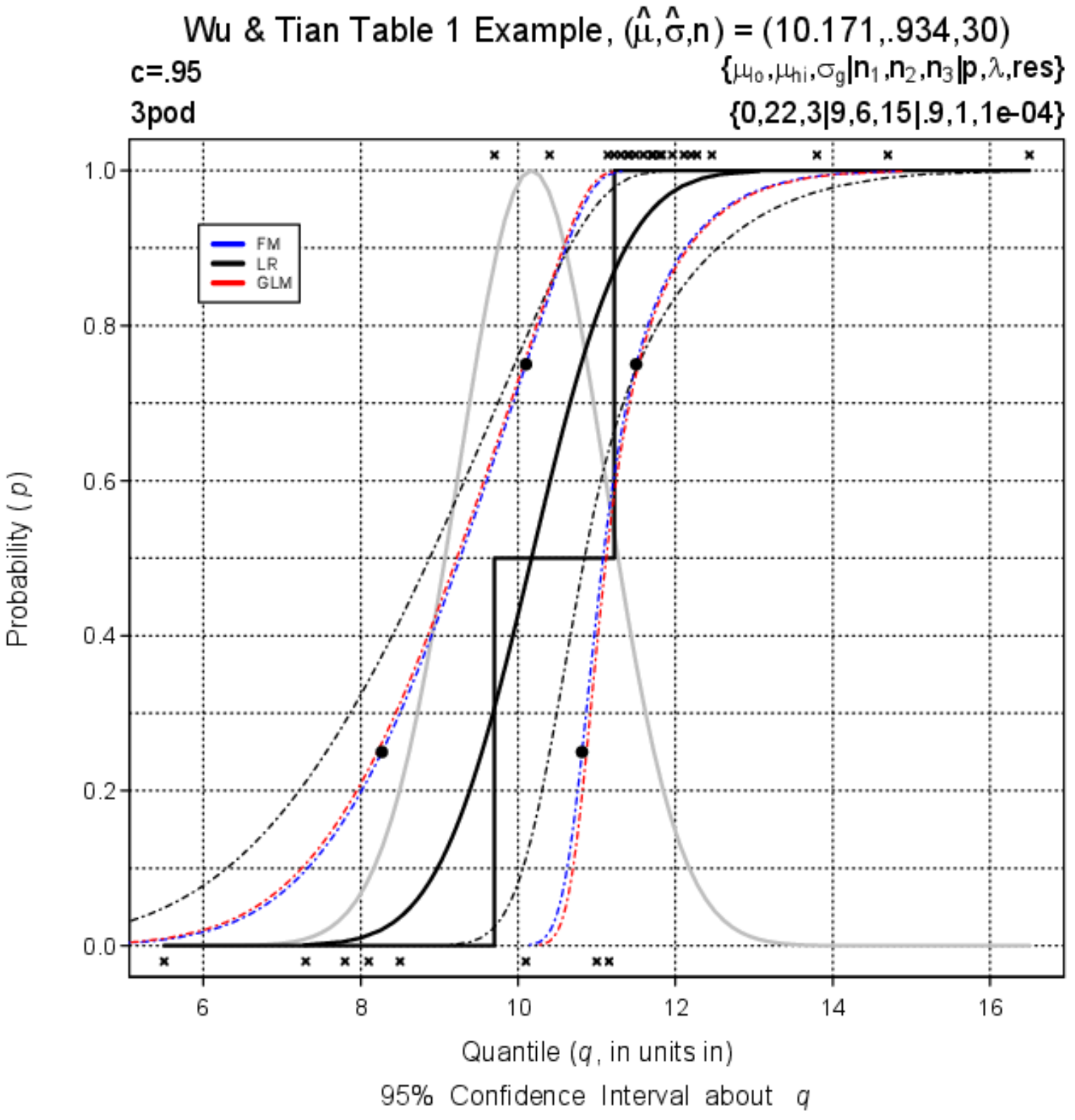

**Figure 10**. 95% CI's about $q$ , with Pooled Adjacent Violators solution.
See Table 20 for the values of the four bold points (calculated with lims)

According to maximum likelihood criterion, the so-called pooled adjacent violators (PAV) solution depicted in the above graph is the best piecewise-constant, non-decreasing CDF representation. It appears in all ptest(w,3) graphs (for all values of J). A flat PAV provides a strong indication that one minus a CDF (i.e., a survivor function) provides a better model of the data. The CDF model is better when $\overline{X}_{Y=1} > \overline{X}_{Y=0}$, whereas the 1-CDF model is better when $\overline{X}_{Y=1} < \overline{X}_{Y=0}$. More details about the PAV methodology are provided in Ayer, Brunk, Ewing, Reid and Silverman (1955). Note that all of the data is plotted in the above Figure 10: the $X_{Y=1}$ data appears above the $p=1$ line, and the $X_{Y=0}$ data appears below the $p=0$ line.

## 6.1 FM Method

The Fisher Matrix (FM) methodology has been widely used in DoD and industrial applications for many years. Brief details about FM confidence limit computations are presented below.

Suppose w is the saved list from a gonogo (or gonogoSim test). Today we can use the R function glm to compute the maximum likelihood estimates of $\mu$ and $\sigma$, $\hat{\mu}$ and $\hat{\sigma}$, pertaining to the test. This we do with just a few lines of code:

xglm=glm(y ~ x, family = binomial(link = probit)), where y is the vector of responses (i.e., y=w$d0$Y), and x is the vector of stresses (i.e., x=w$d0$X).

With ab=as.vector(xglm$coef), we get muhat=-ab[1]/ab[2] and sighat=1/ab[2].

Prior to formulating the maximum likelihood estimation problem in terms of a general linear model (glm), muhat and sighat were typically found by a brute force method (i.e., search over a crude grid of possible values and continue refining the estimates).

Gonogo computes a two by two covariance matrix $v$ with its function yinfomat, as follows:

v= yinfomat(w$d0,muhat,sighat)$vcov1*sighat^2.

Using $se=\sqrt{v_{11}+2v_{12}z_p+v_{22}z_p^2}$ to estimate the standard error about the predicted quantile $q$, the following lower and upper limits may be computed:

(a) $q_{lo}$, and $q_{hi}$, about $q=\hat{\mu}+z_p\hat{\sigma}$, to be $q \mp z_{(1+conf)/2} \times se$, respectively; and

(b) $p_{lo}$ and $p_{hi}$, about $p$, to be $p \mp \phi\left(z_p\right) z_{(1+conf)/2} \times se/\hat{\sigma}$

Again, in the above computations, $conf$ is understood to be **2-sided** and apply to a confidence interval.

## 6.2 GLM Method

For gonogo's glm method, two standard error estimates are calculated:

$se_1$ =mdose.p(xglm,p)$se

$se_2$ =predict(xglm,list(muhat+qnorm(p)*sighat), se.fit=T)$se.fit

The lower and upper limits are given by

(a) $q_{lo}$ and $q_{hi}$, about $q=\hat{\mu}+z_p\hat{\sigma}$, are $q \mp t_{(1+conf)/2,df} \times se_1$, respectively; and

(b) $p_{lo}$ and $p_{hi}$, about $p$, are $pnorm\left(q \mp t_{(1+conf)/2,df} \times se_2\right)$, respectively.

Again, in the above computations, $conf$ is understood to be **2-sided**.

## 6.3 LR Method

In this method, -2 times the log likelihood divided by its maximum is calibrated to a critical value of the chi square distribution with 2 degrees of freedom. Over a grid of mu, sig values, a contour curve about $\hat{\mu}$, $\hat{\sigma}$ is constructed having the desired critical value. It's termed a 100*Jconf% **joint** confidence region about $\hat{\mu}$, $\hat{\sigma}$. From its boundary points, (m, s), **individual** 100*conf% confidence interval curves are derived. Jconf (or $C_J$) and conf (or $C$) are related by qchisq($C_J$,2) = qchisq($C$,1). If $M_{FULL}$ is the maximum of the log-likelihood achieved by a full model, then the level $L$ of the contour is related to Jconf by L=(1-$C_J$)*$M_{FULL}$ if there's overlap, and L=(1-$C_J$)*$M_{FULL}$/4, if there isn't. Needless to say, computing LR confidence curves is more time consuming than computing the other two. Again, let's remember conf needs to be thought of as being **2-sided**. Users are encouraged to familiarize themselves with gonogo's LR graphical capability by examining the 27 data sets included in the function wxdat(i) using the versatile plot function ptest (see section 11). More is said about the LR methodology on pages 43-44 and wxdat(i) in section 12 on page 48.

## 7. Tips on Usage

The user of gonogo should become familiar with some of the basic features provided. They are briefly summarized in the following sections.

## 7.1 Console

Recall that the general form of a console-initiated test is:

**w=gonogo(mlo=0,mhi=0,sg=0,newz=T,reso=0,ln=F,test=1,term1=T,BL=NULL,Y=NULL,X=NULL)**

**Table 22**. How-to tips for operating the console

| How To | Details |
|---|---|
| Start a Test | w=gonogo(mlo, mhi, sg) starts a 3pod test (test=1 is the default).<br>The name (here it is w) assigned to the output of gonogo is NOT important.<br>For a Neyer, Bruceton or Langlie, specify test equal to 2, 3 or 4, respectively.<br>The newz=T option (the default) starts a new test by creating a list w.<br>A list called w retains test details. Whatever w was previously will be lost. |
| Suspend a Test | A test may be suspended any time upon entry of<br>(1) an invalid Y response (e.g., a blank or -1);<br>(2) a negative sample size; or (3) p <= 0 (or >= 1) or lam < 0. |
| Resume Testing | The newz = F option allows one to resume a suspended test saved in the list z.<br>**gonogo assumes z is the name of the list that you want to continue with where you left off**. Thus, resuming a test saved in a list named U can ONLY be accomplished via commands: z = U followed by w=gonogo(newz=F). |
| Skip a Phase | Phase II or III may be skipped by entering 0 (zero) for n2 or n3, respectively. |
| Correct an Entry | fixw is a function that lops off the last n entries of a test. **If a test is saved in a list V, then z=fixw(V,n) allows you to go back and correct an entry.** |
| Set the Resolution | reso = 0 (the default) means recommended stresses (x) are rounded to 5 decimal places.<br>reso = .1 (for example) means suggested X's will be rounded to the nearest tenth. |
| Tabulate Output | Upon suspension or completion of a test, a text file is generated (gonogo.txt).<br>The file is table ready: (1) copy contents into an MS Word document; and<br>(2) highlight and select "convert text to table" (from the Insert tab). |

It would be worthwhile for the new user to try out the gonogo's **fixw** function to see how (and that) it really works. The **fixw** function is meant only for tests being created in console mode, i.e., with gonogo, not with gonogoSim. It makes sense that you don't want to fiddle with simulated tests. To convert, e.g., the simulated test we saved as ub (Figure 6), define it (if you don't already have it) by running **ub=gonogoSim(10,10,.25,6,6,.9,1,plt=1,test=3,reso=.01,BL=c(4,2,2),iseed=62517)**. Then convert ub into a gonogo test by running **UB=gonogo(10,10,.25,test=3,reso=.01,Y=ub$d0$Y)**. To verify that the test sequences are the same, look at **ub$d0$X** and **UB$d0$X** and compare them.

Now, suppose you wanted to redo UB's run# 17. This means we want to first create a test, called z, consisting of UB's runs 1 through 16, and then continue that test with gonogo(newz=F).

To create such a z, examine UB$d0 in R's console widow and count back through the number of reads gonogo would have to backtrack to get to run 16: this author counts 32-16, plus 1 (for the p lam read), plus 1 (for the n3 read), plus 1 (for the n2 read) , which equals 19 reads total. Therefore, if you execute the command **z=fixw(UB,19)**, you should find that **z$d0** consists of runs# 1 through #16. To resume testing at run #17, run **UB1=gonogo(newz=F)**. gonogo will grab z and allow you to pick up testing beginning at the $17^{th}$ run.

**Note**: instead of backtracking 19 reads, we could have backtracked 7, 8 or 15 to read in a new p lam line, a different n3, or a different n2, respectively.

### 7.2 Batch

The function gonogo has an optional Y argument that invokes a batch mode of operation. This recent addition has been invaluable in the preparation of this document. It was designed to quickly recreate gonogo tests by passing in the defining arguments along with the entire sequence of responses (Y). When the number of responses exceeds the length of phase I, the user is prompted to enter n2. When the number of responses exceeds the length of the first two phases, the next set of prompts for user input are issued. Variants of the 3pod test depicted in Figure 3 are possible, e.g., **gonogo(0,22,3,reso=.0001,Y=yWT)**. You could also lop off any number of responses from the end of yWT. It is recommended to decrement the n2 entry accordingly (and possibly the n3 entry), so as to be consistent with the length of Y. In this example, you wouldn't want to remove the $1^{st}$ response. If you did, the resulting 3pod sequence will consist of a four-run stage I1, followed by a zero-run stage I2, and a one run stage I3, which ultimately exits phase I with an infinite $\hat{\sigma}$ estimate. In such rare cases, gonogo 3pod's or Neyer's will terminate with the message:

**Table 23**. Infinite Sigma message from gonogo

| |
|---|
| Entry into Phase II requires that a positive and finite sigma exists. Thus, M0 > m1 (for overlap) & delta = Avg(X[Y==1]) – Avg(X[Y==0]) > 0. The second condition ensures that the regression slope coefficient is positive. Since your completed 3pod or Neyer Phase I did not meet both conditions, it has been suspended for your further review. Bruceton and Langlie Phase I's have been programmed to continue on until both conditions are met. |

The output of **gonogo(0,22,3,reso=.0001,Y=yWT[-1])** produces the following history plot of phase I:

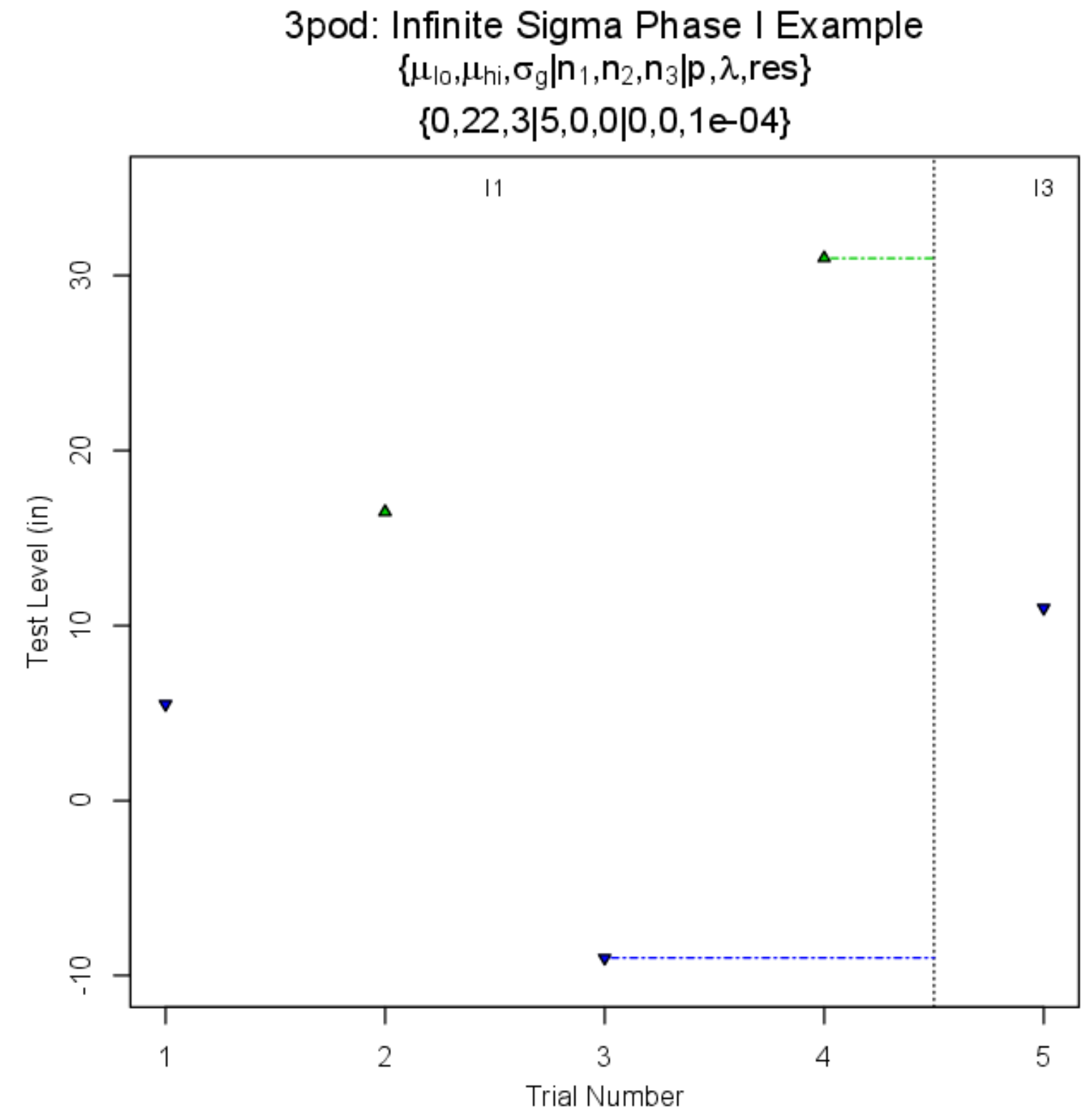

**Figure 11**. Infinite Sigma 3pod Phase I inadvertently produced in batch mode

The above example shows how easy it can be to do something in batch mode you just shouldn't do. Of course, the same can be said of console mode, e.g., don't reverse the order of 3pod's $X_1$ and $X_2$, and watch out for keyboard typos. It would be possible to build better safeguards against misusing the batch mode. For now, this mode comes very much recommended, but for **careful** use only.

## 7.3 Random

The following table contains some tips on how to specify details of the test(s) to be simulated by gonogoSim:

**Table 24**. How-to tips to conduct simulations with gonogoSim

| Argument | Usage |
|---|---|
| test | test = 1: 3pod, test = 2: Neyer<br>test = 3: Bruceton, test = 4: Langlie |
| n2 | **To fix the combined size of Phases I and II**<br>Enter -n2 if n11+n12+abs(-n2) is to be capped at abs(-n2)<br>**To fix the combined size of Phases I, II and III** |
| n3 | Enter -n3 if n11+n12+n2+abs(-n3) is to be capped at abs(-n3) |
| lIgo | lIgo = F <--> Stop Simulation at End of<br>(1) 3pod Phase I2; (2) Neyer Phase I |
| dm, ds | Deviations about Target Mean (tm) and Target Standard Deviation (ts) |
| iseed | 0 <= iseed < Inf if you want repeatability in the X's and Y's<br>if iseed is NOT the default (-1), then set.seed is initialized in gd0 as nd0+iseed, where nd0 is the current trial number within gd0. |
| M | To verify the Phase III is really scale-free, run gonogoSim twice with different M's (with same fixed iseed option). For example:<br>y1=gonogoSim(0,22,3,5,5,.9,1,plt=1,dm=1,ds=1,M=1,iseed=5,reso=.0001)<br>y10=gonogoSim(0,22,3,5,5,.9,1,plt=1,dm=1,ds=1,M=10,iseed=5,reso=.001)<br>Examine graphs: ptest(y1,i) v ptest(y10,i), for i=1,2 and 3.<br>Examine y1$d0$X and y10$d0$X (or, see section 8). |
| reso | The reso option has been enabled in gonogoSim. |

gonogoSim iteratively generates random strengths (from an underlying normal distribution), stresses (via one the sequential procedures described), and responses (Y) according to the formulae

$$Y_{stress \geq strength} = 1, \text{ and } Y_{stress < strength} = 0$$

To start a simulation, gonogoSim needs a true mean and true standard deviation for the distribution of strengths. To get them, a target mean and a target standard deviation (tm and ts) are computed (based on mlo, mhi and sg), to which deviations (dm and ds) are then added. On graphical outputs, the true mean ($\sigma_t$) and the true standard deviation ($\sigma_t$), are represented as sums of their target components and deviation components (in that order). The list returned by gonogoSim contains dm and ds components, along with tmu and tsig components, for the true mean and the true standard deviation, respectively.

The following table describes the target mean (tm) and target standard deviation (ts) computations and provides details how gonogoSim generates random responses:

**Table 25**. Strengths (S or log(S)) are simulated from a latent $N\left(\mu_{true}, \sigma_{true}\right)$

| ln | test | $tm = \mu_{true} - dm$ | $ts = \sigma_{true} - ds$ | Notes |
|---|---|---|---|---|
| FALSE | 1 | $(\mu_{lo} + \mu_{hi})/2$ | $\sigma_g$ | - |
| | 2 | $(\mu_{lo} + \mu_{hi})/2$ | $\sigma_g$ | - |
| | 3 | $(\mu_{lo} + \mu_{hi})/2$ | $\sigma_g$ | $x_1 = (1-p)\cdot\mu_{lo} + p\cdot\mu_{hi}$ |
| | 4 | $(\mu_{lo} + \mu_{hi})/2$ | $(\mu_{hi} - \mu_{lo})/6$ | $x_1 = (1-p)\cdot\mu_{lo} + p\cdot\mu_{hi}$ |
| TRUE | 1 | $(u[1]+u[2])/2$ | $u[3]$ | $u = fgs(\mu_{lo}, \mu_{hi}, \sigma_g)$ |
| | 2 | $(u[1]+u[2])/2$ | $u[3]$ | $u = fgs(\mu_{lo}, \mu_{hi}, \sigma_g)$ |
| | 3 | $\log(\sqrt{\mu_{lo}\cdot\mu_{hi}})$ | $\log(1+\sigma_g/\log(\mu_{lo}))$ | $x_1 = (1-p)\cdot\log(\mu_{lo}) + p\cdot\log(\mu_{hi})$ |
| | 4 | $\log(\sqrt{\mu_{lo}\cdot\mu_{hi}})$ | $\log(\mu_{hi}/\mu_{lo})/6$ | $x_1 = (1-p)\cdot\log(\mu_{lo}) + p\cdot\log(\mu_{hi})$ |
| | | | $p$ comes from $L_p$ defined in Table 8 | *fgs* picks $\mu_{lo}$, $\mu_{hi}$ on a log scale to yield an $x_1$ and $x_2$ provided by 3pod |

## 8. Unit Scaling

It's fairly obvious that Phases I and II will produce comparable X's (for the same Y's) if stresses were measured in inches rather than, say, centimeters. However, it's not clear if this "equivariance" property of unit-independence holds in Phase III. To check this, let's redo the 3pod of Table 1 in Wu and Tian (2014) with starting values 10 times the original ones and compare the two X sequences. Also, let's do simulations with identical starting conditions, one with M=1 (the default multiplier) and the other with M=10 . Here are the side-by-side sequences for two sets of comparisons, each with $p = .9$ and $\lambda = 1$. The first, second, third and fourth X columns were obtained via **gonogo(0,22,3,reso=.0001,Y=yWT,X=xWT), gonogo(0,220,30,reso=.001,Y=yWT,X=xWT)**, **gonogoSim(0,22,3,6,15,.9,1,reso=.0001,iseed=10)** and **gonogoSim(0,220,30,reso=.001,iseed=10,M=10)**, respectively.

**Tables 26, 27, 28 and 29**. Demonstration of the equivariance property phase III (with $\lambda = 1$)

| | Verify Phase III Equivariance with: $n_2 = 6,\ n_3 = 15,\ p = .9,\ \lambda = 1$ | | | | | | | |
|---|---|---|---|---|---|---|---|---|
| | Console & Batch (gonogo) | | | | Simulation (gonogoSim, $i_{seed} = 10$ ) | | | |
| i | Y | $X_{0,22,3,reso=.0001}$ | $X_{0,220,30,reso=.001}$ | ID | Y | $X_{0,22,3,reso=.0001}$ | $X_{0,22,3,reso=.001,M=10}$ | ID |
| 1 | 0 | 5.5000 | 55.000 | I1(iii) | 0 | 5.5000 | 55.000 | I1(iii) |
| 2 | 1 | 16.5000 | 165.000 | I1(iii) | 1 | 16.5000 | 165.000 | I1(iii) |
| 3 | 0 | 11.0000 | 110.000 | I2(ib) | 0 | 11.0000 | 110.000 | I2(ib) |
| 4 | 1 | 13.8000 | 138.000 | I2(ib) | 1 | 13.7500 | 137.500 | I2(ib) |
| 5 | 0 | 10.1000 | 101.000 | I2(id) | 0 | 10.1000 | 101.000 | I2(id) |
| 6 | 1 | 14.7000 | 147.000 | I2(id) | 1 | 14.6500 | 146.500 | I2(id) |
| 7 | 1 | 10.4000 | 104.000 | rI2(id) | 1 | 10.4000 | 104.000 | rI2(id) |
| 8 | 1 | 11.7000 | 117.000 | I3 | 0 | 11.7000 | 117.000 | I3 |
| 9 | 1 | 9.7000 | 97.000 | I3 | 1 | 9.7000 | 97.000 | I3 |
| 10 | 0 | 7.3000 | 73.000 | II1 | 0 | 7.0544 | 70.544 | II1 |
| 11 | 0 | 7.8000 | 78.000 | II2 | 1 | 14.4491 | 144.491 | II2 |
| 12 | 0 | 8.1000 | 81.000 | II2 | 0 | 7.8403 | 78.403 | II2 |
| 13 | 1 | 12.2000 | 122.000 | II2 | 1 | 13.8033 | 138.033 | II2 |
| 14 | 0 | 8.5000 | 85.000 | II2 | 0 | 8.3442 | 83.442 | II2 |
| 15 | 1 | 11.8000 | 118.000 | II2 | 1 | 13.3558 | 133.558 | II2 |
| 16 | 1 | 11.7121 | 117.121 | III1 | 1 | 13.2531 | 132.531 | III1 |
| 17 | 1 | 11.4083 | 114.083 | III2 | 0 | 12.7738 | 127.738 | III2 |
| 18 | 0 | 11.1558 | 111.558 | III2 | 1 | 15.0847 | 150.847 | III2 |
| 19 | 1 | 12.4633 | 124.633 | III2 | 1 | 14.7450 | 147.450 | III2 |
| 20 | 1 | 12.2761 | 122.761 | III2 | 1 | 14.4496 | 144.496 | III2 |
| 21 | 1 | 12.1107 | 121.107 | III2 | 1 | 14.1887 | 141.887 | III2 |
| 22 | 1 | 11.9628 | 119.628 | III2 | 1 | 13.9554 | 139.554 | III2 |
| 23 | 1 | 11.8291 | 118.291 | III2 | 1 | 13.7445 | 137.445 | III2 |
| 24 | 1 | 11.7072 | 117.072 | III2 | 1 | 13.5522 | 135.522 | III2 |
| 25 | 1 | 11.5952 | 115.952 | III2 | 0 | 13.3756 | 133.756 | III2 |
| 26 | 1 | 11.4917 | 114.917 | III2 | 1 | 14.5481 | 145.481 | III2 |
| 27 | 1 | 11.3955 | 113.955 | III2 | 1 | 14.3963 | 143.963 | III2 |
| 28 | 1 | 11.3057 | 113.057 | III2 | 1 | 14.2546 | 142.546 | III2 |
| 29 | 1 | 11.2214 | 112.214 | III2 | 1 | 14.1217 | 141.217 | III2 |
| 30 | 1 | 11.1421 | 111.421 | III2 | 1 | 13.9966 | 139.966 | III2 |
| 31 | 0 | 0.0000 | 0.000 | III3 | 0 | 0.0000 | 0.000 | III3 |

Here is a similar set of comparisons demonstrating equivariance in a case when $\lambda \neq 1$:

**Tables 30, 31, 32 and 33**. Demonstration of the equivariance property phase III (with $\lambda \neq 1$ )

| | Verify Phase III Equivariance with: $n_2 = 6,\ n_3 = 15,\ p = .9,\ \lambda = .8$ | | | | | | | |
|---|---|---|---|---|---|---|---|---|
| | Console & Batch (gonogo) | | | | Simulation (gonogoSim, $i_{seed} = 10$ ) | | | |
| i | Y | $X_{0,22,3,reso=.0001}$ | $X_{0,220,30,reso=.001}$ | ID | Y | $X_{0,22,3,reso=.0001}$ | $X_{0,22,3,reso=.001,M=10}$ | ID |
| 1 | 0 | 5.5000 | 55.000 | I1(iii) | 0 | 5.5000 | 55.000 | I1(iii) |
| 2 | 1 | 16.5000 | 165.000 | I1(iii) | 1 | 16.5000 | 165.000 | I1(iii) |
| 3 | 0 | 11.0000 | 110.000 | I2(ib) | 0 | 11.0000 | 110.000 | I2(ib) |
| 4 | 1 | 13.8000 | 138.000 | I2(ib) | 1 | 13.7500 | 137.500 | I2(ib) |
| 5 | 0 | 10.1000 | 101.000 | I2(id) | 0 | 10.1000 | 101.000 | I2(id) |
| 6 | 1 | 14.7000 | 147.000 | I2(id) | 1 | 14.6500 | 146.500 | I2(id) |
| 7 | 1 | 10.4000 | 104.000 | rI2(id) | 1 | 10.4000 | 104.000 | rI2(id) |
| 8 | 1 | 11.7000 | 117.000 | I3 | 0 | 11.7000 | 117.000 | I3 |
| 9 | 1 | 9.7000 | 97.000 | I3 | 1 | 9.7000 | 97.000 | I3 |
| 10 | 0 | 7.3000 | 73.000 | II1 | 0 | 7.0544 | 70.544 | II1 |
| 11 | 0 | 7.8000 | 78.000 | II2 | 1 | 14.4491 | 144.491 | II2 |
| 12 | 0 | 8.1000 | 81.000 | II2 | 0 | 7.8403 | 78.403 | II2 |
| 13 | 1 | 12.2000 | 122.000 | II2 | 1 | 13.8033 | 138.033 | II2 |
| 14 | 0 | 8.5000 | 85.000 | II2 | 0 | 8.3442 | 83.442 | II2 |
| 15 | 1 | 11.8000 | 118.000 | II2 | 1 | 13.3558 | 133.558 | II2 |
| 16 | 1 | 11.5538 | 115.538 | III1 | 1 | 13.0034 | 130.034 | III1 |
| 17 | 1 | 11.2419 | 112.419 | III2 | 0 | 12.5114 | 125.114 | III2 |
| 18 | 0 | 10.9851 | 109.851 | III2 | 1 | 14.7655 | 147.655 | III2 |
| 19 | 1 | 12.2560 | 122.560 | III2 | 1 | 14.4224 | 144.224 | III2 |
| 20 | 1 | 12.0678 | 120.678 | III2 | 1 | 14.1255 | 141.255 | III2 |
| 21 | 1 | 11.9022 | 119.022 | III2 | 1 | 13.8643 | 138.643 | III2 |
| 22 | 1 | 11.7545 | 117.545 | III2 | 1 | 13.6313 | 136.313 | III2 |
| 23 | 1 | 11.6213 | 116.213 | III2 | 1 | 13.4212 | 134.212 | III2 |
| 24 | 1 | 11.5001 | 115.001 | III2 | 1 | 13.2300 | 132.300 | III2 |
| 25 | 1 | 11.3889 | 113.889 | III2 | 0 | 13.0546 | 130.546 | III2 |
| 26 | 1 | 11.2863 | 112.863 | III2 | 1 | 14.1844 | 141.844 | III2 |
| 27 | 1 | 11.1910 | 111.910 | III2 | 1 | 14.0341 | 140.341 | III2 |
| 28 | 1 | 11.1021 | 111.021 | III2 | 1 | 13.8938 | 138.938 | III2 |
| 29 | 1 | 11.0188 | 110.188 | III2 | 1 | 13.7623 | 137.623 | III2 |
| 30 | 1 | 10.9404 | 109.404 | III2 | 1 | 13.6386 | 136.386 | III2 |
| 31 | 0 | 0.0000 | 0.000 | III3 | 0 | 0.0000 | 0.000 | III3 |

The interested reader may also want to check that the equivariance property includes translation of units by a constant amount. This property was not included in the equivariance discussion Wang, Tian and Wu (2020), but it holds nevertheless.

## 9. Log Transform

Gonogo (and gonogoSim) offer a ln=T option. This allows the gonogo test methods to operate on the log(X) scale while it maintains all dialogue with the user on the X scale. Gonogo is able to do this by computing new mlo, mhi and sg with its fgs function (described in Table 25). To the user, the test appears to be running in X scale. This option may be useful when stresses are required to be positive, or when the log-normal distribution is deemed an appropriate latent stress distribution. Here's how it might have worked if the gonogo version of Table 1 (Wu and Tian 2014) was run with the log=T option (with the same vector of responses $Y$ ). Recall we ran the batch command wWT=gonogo(0,22,3,reso=.0001,Y=yWT,X=xWT) to get Figure 3 and Table 10. This time, run **wWTL=gonogo(0,22,3,reso=.0001,Y=yWT,X=xWT,ln=T)** to get the log counterparts of these outputs.

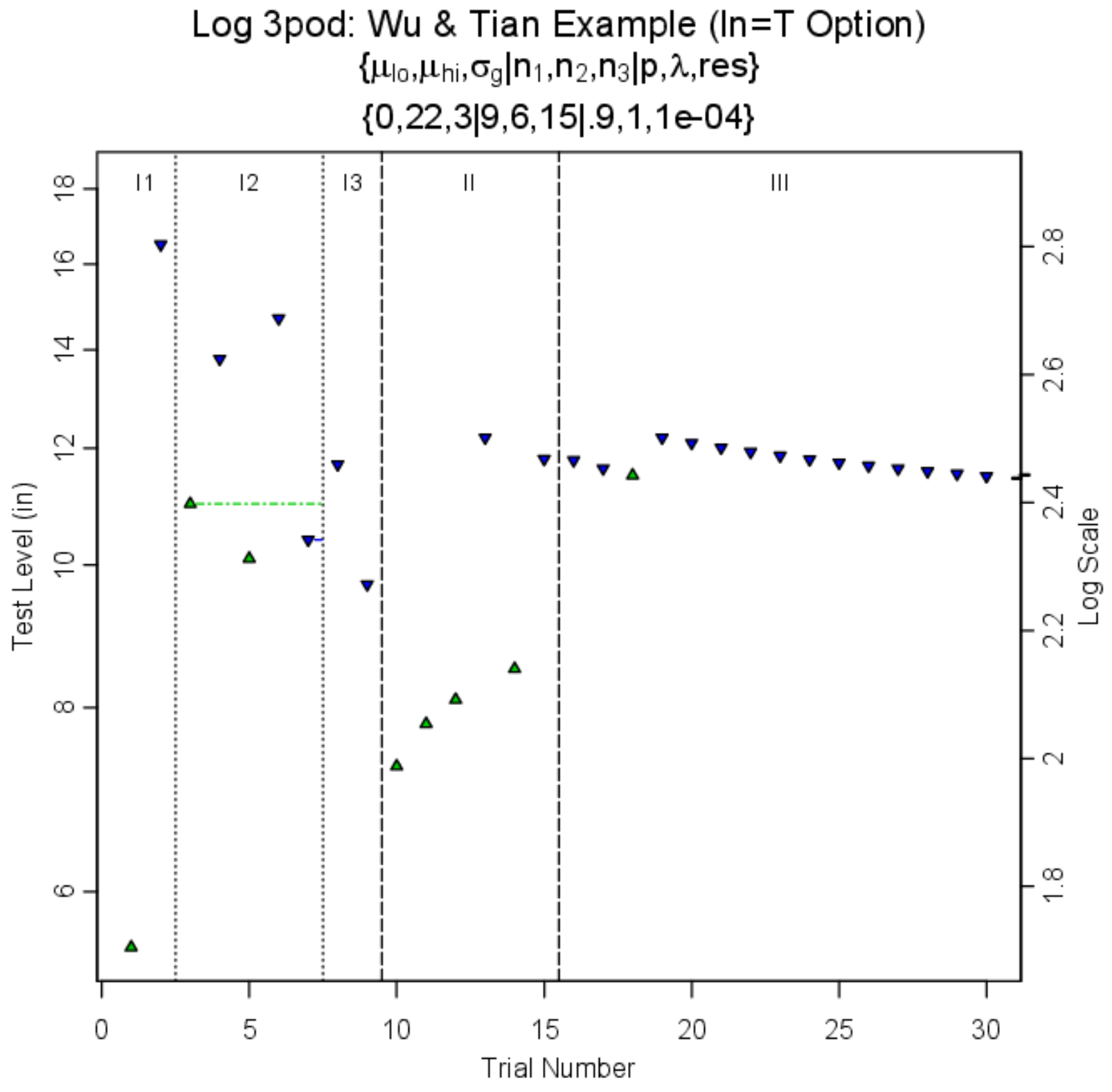

**Figure 12**. Wu Tian Example run on the log scale (wWTL, Log counterpart to wWT)

After the user enters n2, the following alert appears in the console (just as a reminder):

***** Starting values (tau2[1] & be) for Phase III, ln=T may need tweaking.***

The gonogo.txt file automatically produced in the above Log 3pod example is presented below:

**Table 34**. Log 3pod example (wWTL) produced in batch mode

| i | X | Y | COUNT | RX | EX | TX | ID |
|---|---|---|---|---|---|---|---|
| 1 | 1.70475 | 0 | 1 | 5.5 | 1.70475 | 5.5 | I1(iii) |
| 2 | 2.80336 | 1 | 1 | 16.5 | 2.80336 | 16.5 | I1(iii) |
| 3 | 2.3979 | 0 | 1 | 9.5263 | 2.25406 | 11 | I2(ib) |
| 4 | 2.62467 | 1 | 1 | 18.1293 | 2.89753 | 13.8 | I2(ic) |
| 5 | 2.31254 | 0 | 1 | 10.0115 | 2.30373 | 10.1 | I2(ic) |
| 6 | 2.68785 | 1 | 1 | 14.6941 | 2.68745 | 14.7 | rI2(ic) |
| 7 | 2.34181 | 1 | 1 | 10.3307 | 2.33512 | 10.4 | rI2(ic) |
| 8 | 2.45959 | 1 | 1 | 11.8756 | 2.47448 | 11.7 | I3 |
| 9 | 2.27213 | 1 | 1 | 9.6333 | 2.26523 | 9.7 | I3 |
| 10 | 1.98787 | 0 | 1 | 7.6627 | 2.03636 | 7.3 | II1 |
| 11 | 2.05412 | 0 | 1 | 8.0148 | 2.08129 | 7.8 | II2 |
| 12 | 2.09186 | 0 | 1 | 8.2731 | 2.11301 | 8.1 | II2 |
| 13 | 2.50144 | 1 | 1 | 8.4705 | 2.13659 | 12.2 | II2 |
| 14 | 2.14007 | 0 | 1 | 8.594 | 2.15107 | 8.5 | II2 |
| 15 | 2.4681 | 1 | 1 | 8.7617 | 2.1704 | 11.8 | II2 |
| 16 | 2.46604 | 1 | 1 | 11.7757 | 2.46604 | 11.7757 | III1 |
| 17 | 2.45339 | 1 | 1 | 11.6277 | 2.45339 | 11.6277 | III2 |
| 18 | 2.44262 | 0 | 1 | 11.5031 | 2.44262 | 11.5031 | III2 |
| 19 | 2.50151 | 1 | 1 | 12.2009 | 2.50151 | 12.2009 | III2 |
| 20 | 2.49326 | 1 | 1 | 12.1006 | 2.49325 | 12.1006 | III2 |
| 21 | 2.48587 | 1 | 1 | 12.0116 | 2.48588 | 12.0116 | III2 |
| 22 | 2.4792 | 1 | 1 | 11.9317 | 2.4792 | 11.9317 | III2 |
| 23 | 2.47312 | 1 | 1 | 11.8594 | 2.47312 | 11.8594 | III2 |
| 24 | 2.46754 | 1 | 1 | 11.7934 | 2.46754 | 11.7934 | III2 |
| 25 | 2.46238 | 1 | 1 | 11.7327 | 2.46238 | 11.7327 | III2 |
| 26 | 2.45759 | 1 | 1 | 11.6766 | 2.45759 | 11.6766 | III2 |
| 27 | 2.45311 | 1 | 1 | 11.6245 | 2.45311 | 11.6245 | III2 |
| 28 | 2.44892 | 1 | 1 | 11.5758 | 2.44891 | 11.5758 | III2 |
| 29 | 2.44497 | 1 | 1 | 11.5302 | 2.44497 | 11.5302 | III2 |
| 30 | 2.44124 | 1 | 1 | 11.4873 | 2.44124 | 11.4873 | III2 |
| 301 | 0 | 0 | 0 | 2.43771 | 0 | 0 | III3 |

To repeat this log example in console mode, the user would have to enter in X's from the TX column. The user's first 15 entries would differ from gonogo's recommended X's – which is the sequence of untransformed X's appearing in the RX column. Thus, all manual X entries are made in the units familiar to the user, whereas all of the log transformations take place in the background.

The following graph depicts the latent lognormal PDF and CDF and 90% confidence intervals about the quantiles:

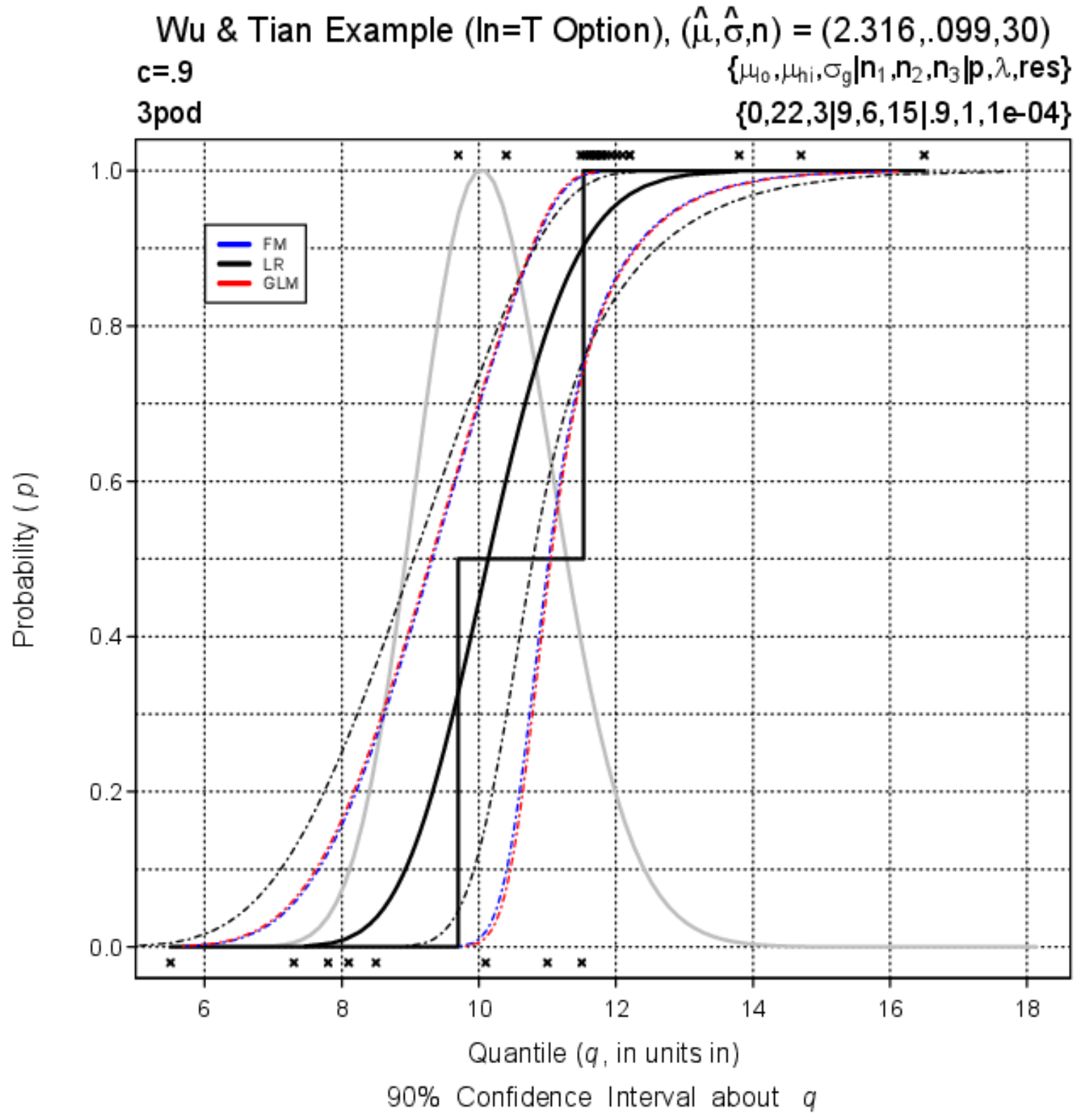

**Figure 13**. Sample ptest(wWTL,3) graph with conf=.9 and J=15

## 10. Simulation

You may want to use the gonogoSim function to perform a simulation study. For that, the user will have to write an R script within the gonogo workspace. Simulations can be rather time consuming since the R environment is not the best setting for this task. Also, gonogo was not designed to do large scale studies. With that said, here's a modest and sanitized study using gonogoSim reported in Ray and Roediger (2018). Two R functions prepared for the study are included (nmel and plotmm). First, we scaled the papers $V_{nom}$ of 700 and $\sigma_{nom}$ of 29.2 by dividing each by 10.

We then ran **nmel3(70,2.92,.9,30,2000,83)**. One point on the resulting graph was then picked out and identified to be the 1431st (MNFV,MASS) pair. The 3pod test of size 30 corresponding to that pair was reproduced by gonogoSim with seed = isd0+nt*(i-1) where isd0=83, nt=30 and i=1431. Figure 3 of Ray and Roediger (2018) is obtained by running **w1431=gonogoSim(mlo,mhi,sg,iseed=42983,plt=1)** with mlo=10*(70-4*2.92), mhi=10*(70+4*2.92), sg=10*2.92 and its Table 1 was then obtained by running **lims(3,w1431$d0,.9,P=c(.000001,.001))**; and its Figure 4 was obtained by running

**nmel3(70,2.92,.9,30,2000,83,icirc=1431)**.

The latter command took 260.62 seconds to complete and its result is presented below:

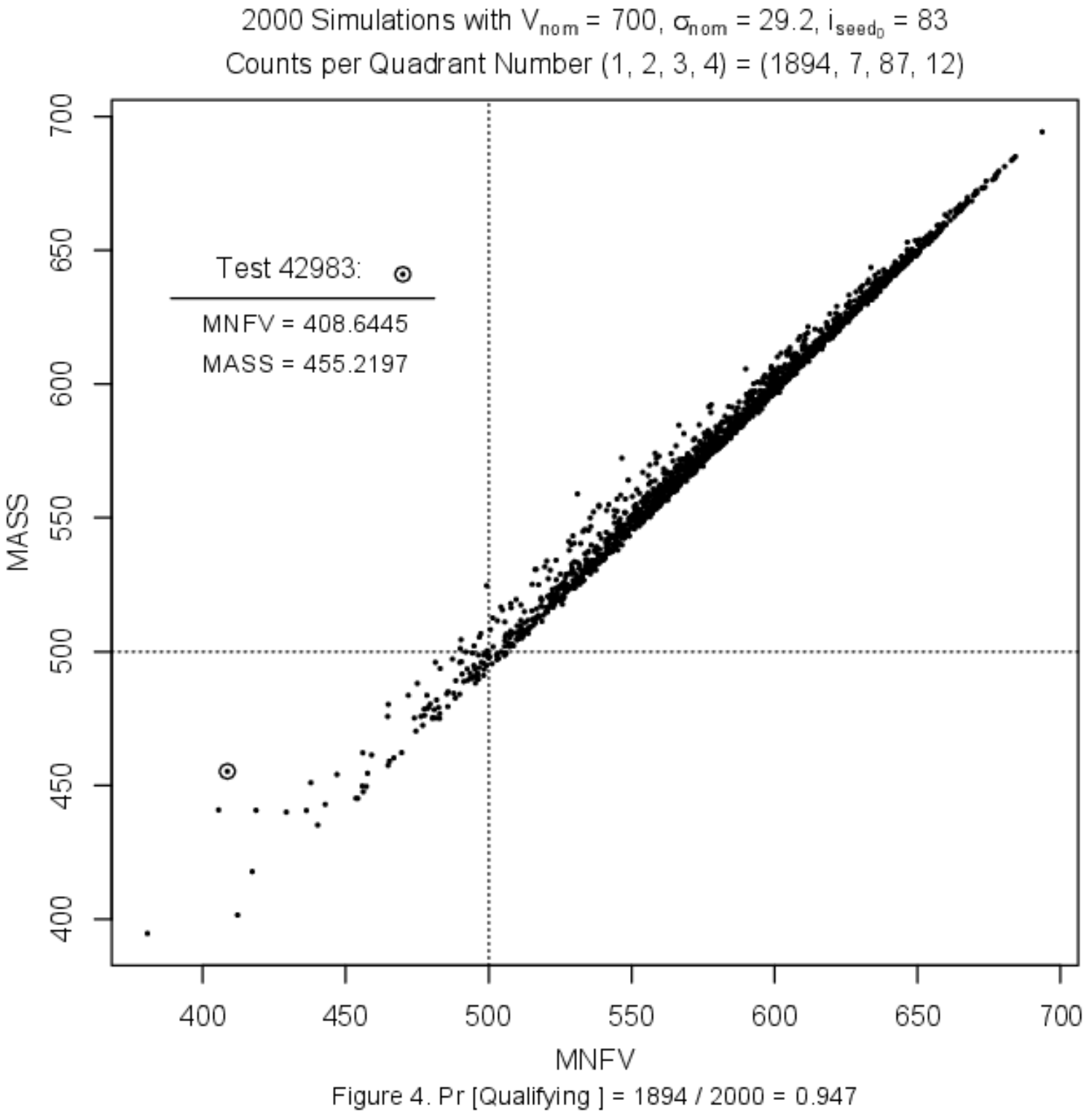

**Figure 14**. Reproduction of Chance Article's Figure 4 (Ray and Roediger 2018)

One of the tricky things about running a simulation like this is to ensure that the seeds passed into the random number generator didn't overlap or repeat themselves during the looped calls to gonogoSim.

## 11. Graphs

Eight plots are included in gonogo. A history plot of X's and Y's is produced at (practically) every console entry or upon completion of a simulation (with plt=1 option). The history plot can be generated from the list associated with the test, as can its remaining 7 plots. The command formats are v=ptest(w,i,notitle=F) for various i. Defining v in this way captures any tabular output generated by the call. Details are provided as follows:

**Tables 35**. Eight plots are provided by gonogo.R

| Syntax | i | Graphic | Alternate Syntax |
|---|---|---|---|
| ptest(w,i,notitle=F) | 1 | History plot | pdat1(w, notitle=F, ud=F) or pSdat1(w, notitle=F, ud=F) |
| notitle=F (for no title) | 2 | MLE's of mu and sigma | pdat2(w, notitle=F) or pSdat2(w,notitle=F) |
| | 3 | Response curve, with data, Pooled Adjacent violators solution, and 100*conf 2-Sided confidence bounds | pdat3(w, notitle=F) or pSdat3(w, notitle=F) |
| | 4 | A simple visual of the data | picdat(w) |
| | 5 | Joint likelihood ratio (LR) multi-confidence bounds | jlrcb(w, notitle=F) |
| | 6 | Joint & Individual LR multi-*confidence bounds* | lrcb(w, notitle=F) |
| | 7 | Joint and/or Individual LR confidence bounds | cbs(w,1, notitle=F) or cbs(w,7, notitle=F) |
| | 8 | Confidence bounds on Probability (p) and Quantile (q) computed via 3 methods: Likelihood Ratio (LR), Fisher Matrix (FM) and General Linear Model (GLM) | cbs(w,2, notitle=F) or cbs(w,8, notitle=F) |

The LR tools (graphics and tabulations) focus on confidence bounds as featured in Neyer's SenTest™ program. Confidence bounds computed via the other more commonly used methods (FM & GLM) are also included in cbs(w,2)**.** LR confidence bounds graphed with ptest(w,7) and ptest(w,8) (cbs(w,1) and cbs(w,2), respectively) are restricted to data sets having interval overlap where conf < cmax. Less restrictive (including point overlap, no overlap, conf > cmax) LR confidence bounds may be graphed with ptest(w,5) (same as jlrcb(w)) and ptest(w,6) (same as lrcb(w)).

**An aside**: Critical decisions are often based upon confidence bound estimates. Better guidance on their use for practitioners and decision makers would be welcomed. Some experts (see Neyer's comment on page 25) hold the view that confidence bounds in the sensitivity testing setting can't be taken too seriously. On the othe hand, using confidence bounds as in choosing a D-Optimal phase II test sequence is a useful application of them.

### 11.1 ptest(w,1)

Let's do a history plot of a simulated 3 phase test that consists of a Neyer phase I followed by a phase II of size 6 and a phase III of size 15 having p and lambda equal to .9 and 1, respectively. gonogoSim will only generate a history plot when its plt argument is set equal to 1. To generate one simulated test that is saved and plotted, run the following command:

**un=gonogoSim(.6,1.4,.1,6,15,.9,1,plt=1,reso=.01,test=2,iseed=4257)**

The output is presented below:

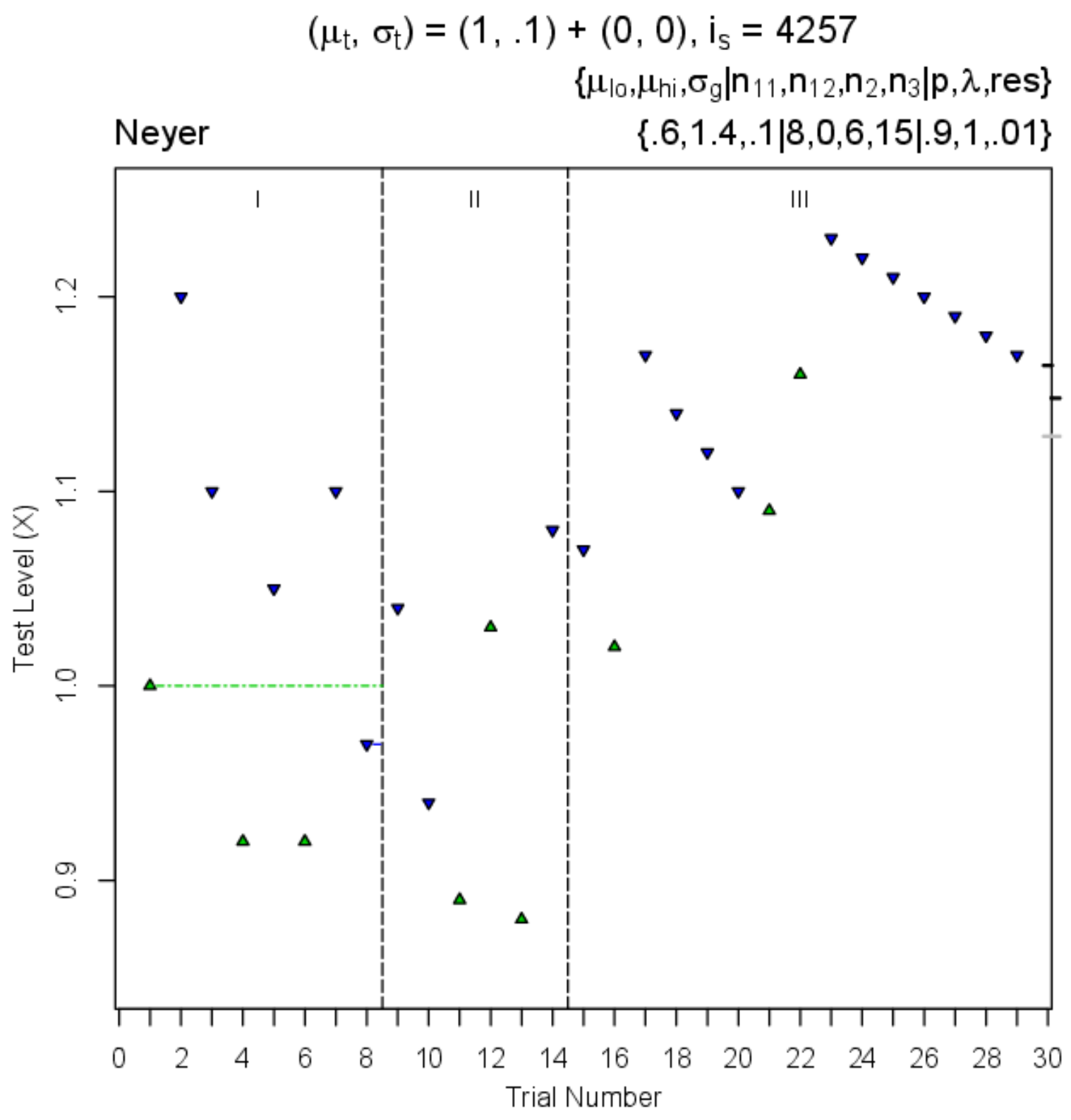

**Figure 15**. Sample ptest(w,1) plot for w=un

**un=gonogoSim(.6,1.4,.1,6,15,.9,1,reso=.01,test=2,iseed=4257)** followed by **ptest(un,1)** would have produced the same result. Saving the test (as un) enables us to produce different graphs of it later. On the graph, the interior black bar indicates the location of the next stress level – had the test continued. The exterior black bar represents the predicted $L_{.90}$ stress level. The grey bar represents the actual $L_{.90}$ point based upon the true underlying distribution, assumed here to have **true values**

$$(\mu_t, \sigma_t) = (1, .1).$$

GonogoSim adds these points (bars) to the plot only upon completion of phase III.

## 11.2 ptest(w,2)

Continuing with the Neyer sample test (un), by running **ptest(un,2)** you get the following plot:

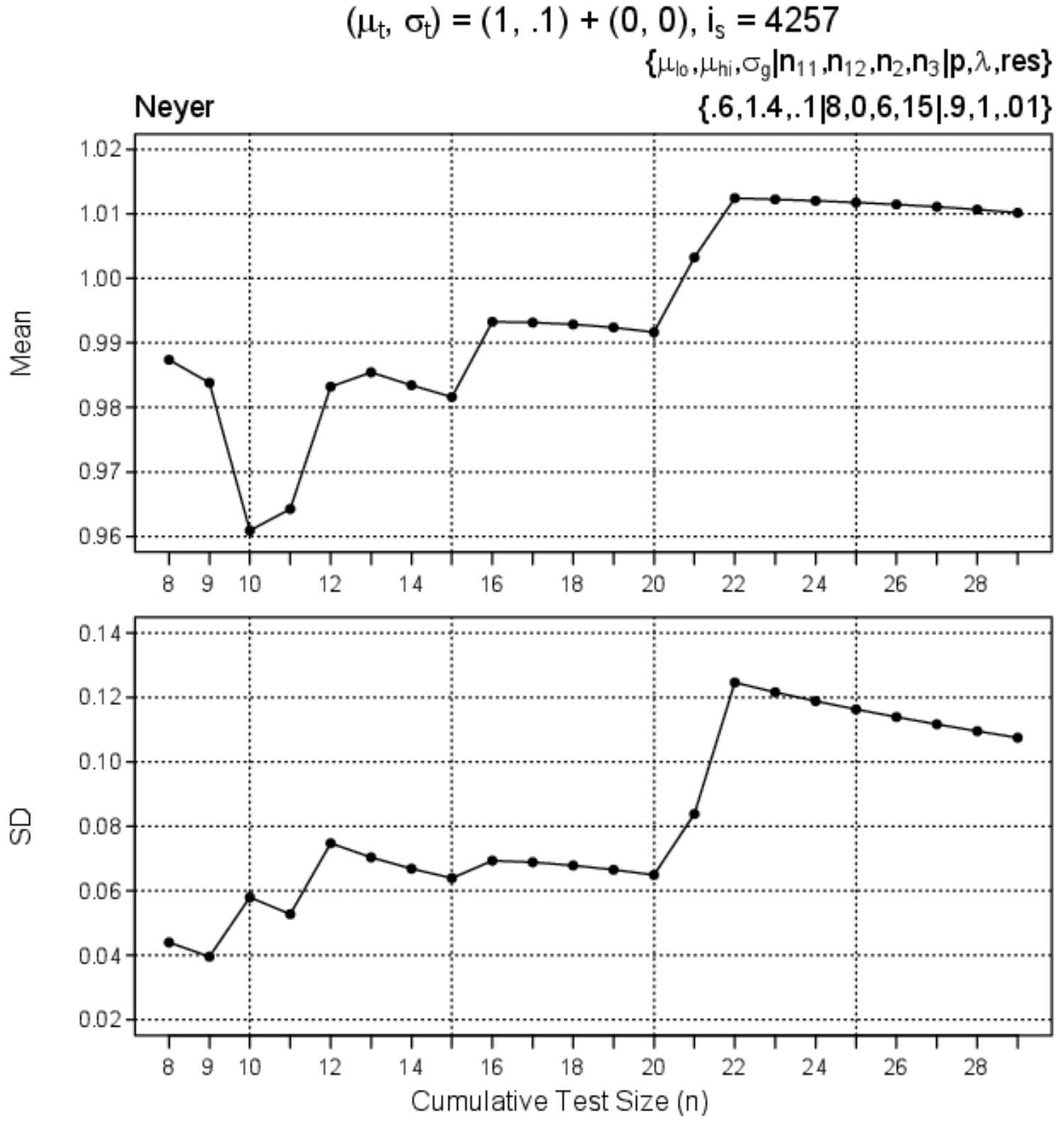

**Figure 16**. Sample ptest(w,2) for w=un

A table of $\hat{\mu}$, $\hat{\sigma}$ and $\hat{\mu} + z_p\hat{\sigma}$ (if there is a phase III) is automatically generated in R's console window. It is presented below:

**Table 36**. Estimated Means, Standard Deviations and $\hat{L}_p = \hat{\mu} + z_p\hat{\sigma}$

| run# | $\hat{\mu}$ | $\hat{\sigma}$ | $\hat{L}_p$ | run# | $\hat{\mu}$ | $\hat{\sigma}$ | $\hat{L}_p$ |
|---|---|---|---|---|---|---|---|
| 8 | 0.9873797 | 0.04399377 | 1.043760 | 19 | 0.9923936 | 0.06653730 | 1.077665 |
| 9 | 0.9838174 | 0.03959259 | 1.034557 | 20 | 0.9916517 | 0.06492514 | 1.074857 |
| 10 | 0.9609069 | 0.05798142 | 1.035213 | 21 | 1.0032469 | 0.08383680 | 1.110688 |
| 11 | 0.9642461 | 0.05275399 | 1.031853 | 22 | 1.0124456 | 0.12460515 | 1.172133 |
| 12 | 0.9832098 | 0.07475745 | 1.079015 | 23 | 1.0122534 | 0.12158827 | 1.168075 |
| 13 | 0.9854295 | 0.07038439 | 1.075631 | 24 | 1.0120305 | 0.11883798 | 1.164328 |
| 14 | 0.9834347 | 0.06686423 | 1.069125 | 25 | 1.0117695 | 0.11629103 | 1.160802 |
| 15 | 0.9816069 | 0.06393483 | 1.063543 | 26 | 1.0114630 | 0.11390552 | 1.157439 |
| 16 | 0.9932668 | 0.06935069 | 1.082143 | 27 | 1.0111026 | 0.11165430 | 1.154193 |
| 17 | 0.9931582 | 0.06885981 | 1.081406 | 28 | 1.0106785 | 0.10952007 | 1.151034 |
| 18 | 0.9928694 | 0.06786451 | 1.079841 | 29 | 1.0101797 | 0.10749281 | 1.147937 |

### 11.3 ptest(w,3)

Running the command **ptest(un,3)** triggers a screen dump of the J menu (Table 21) and the following prompt for two inputs:

Enter conf and J: .**9 14**

The user's response (in bold) was conf=.9 and J=14. This plot will produce three upper and three lower 1-sided (1+.9)/2 = 95% confidence limit curves **about the percentile** p, one pair for FM, and two other pairs for LR and GLM. The plot is presented below:

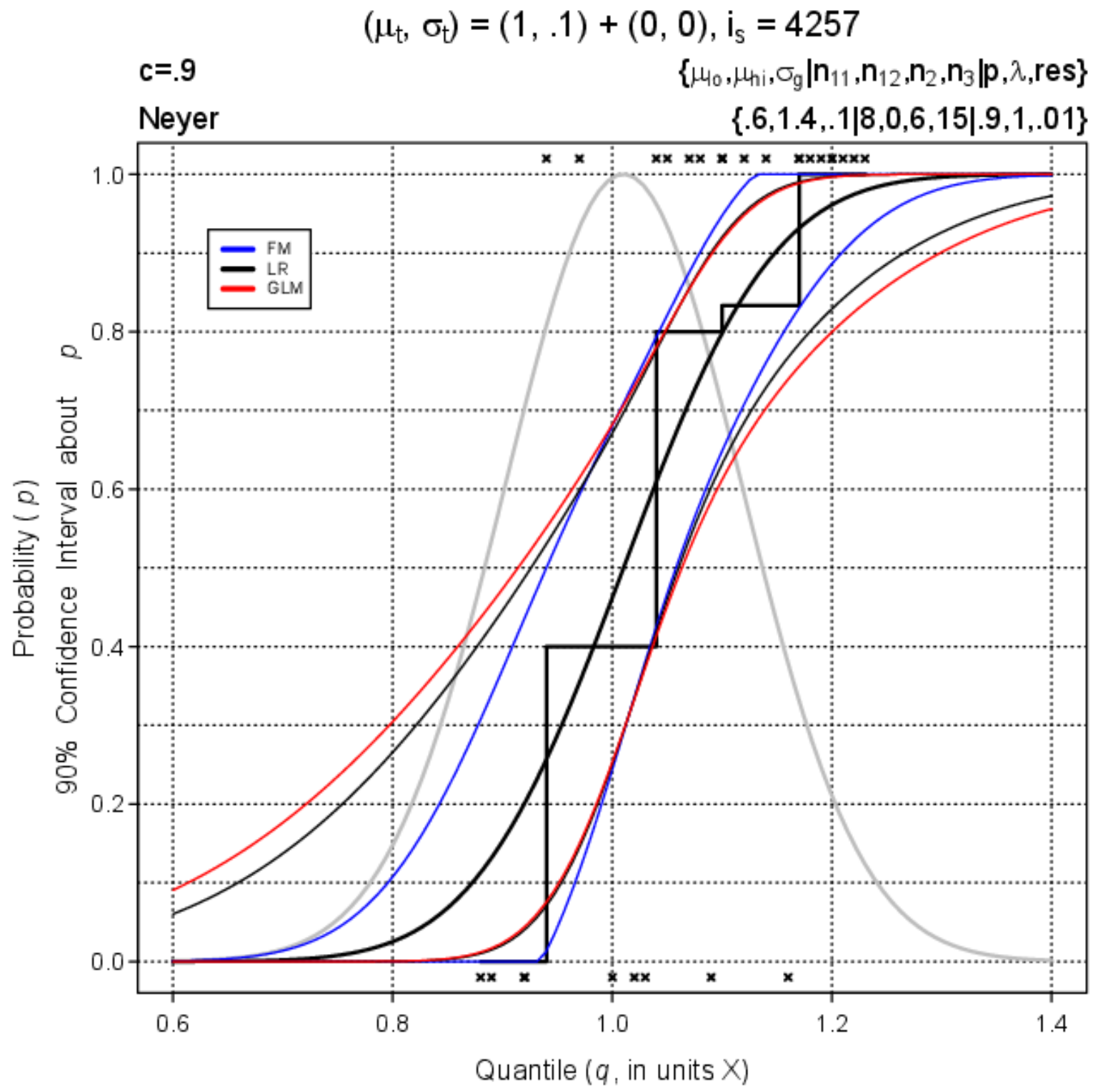

**Figure 17**. Sample ptest(w,3) for w=un, conf=.9 and J=14

The J=14 option requests that confidence intervals be placed about p|q (i.e., about p given q). These intervals occur (and should be read) in the up/down direction of the graph. The p axis label hints how to read the graph properly. Also, the curves are solid lines whenever this interpretation applies.

In the next graph, we'll redo the graph with the J=15 option, which requests confidence intervals to be placed about q|p (i.e., about q given p). Intervals about q|p are plotted with dashed lines (to distinguish the two types of confidence intervals whenever they happen to be plotted together, e.g., when J=7).

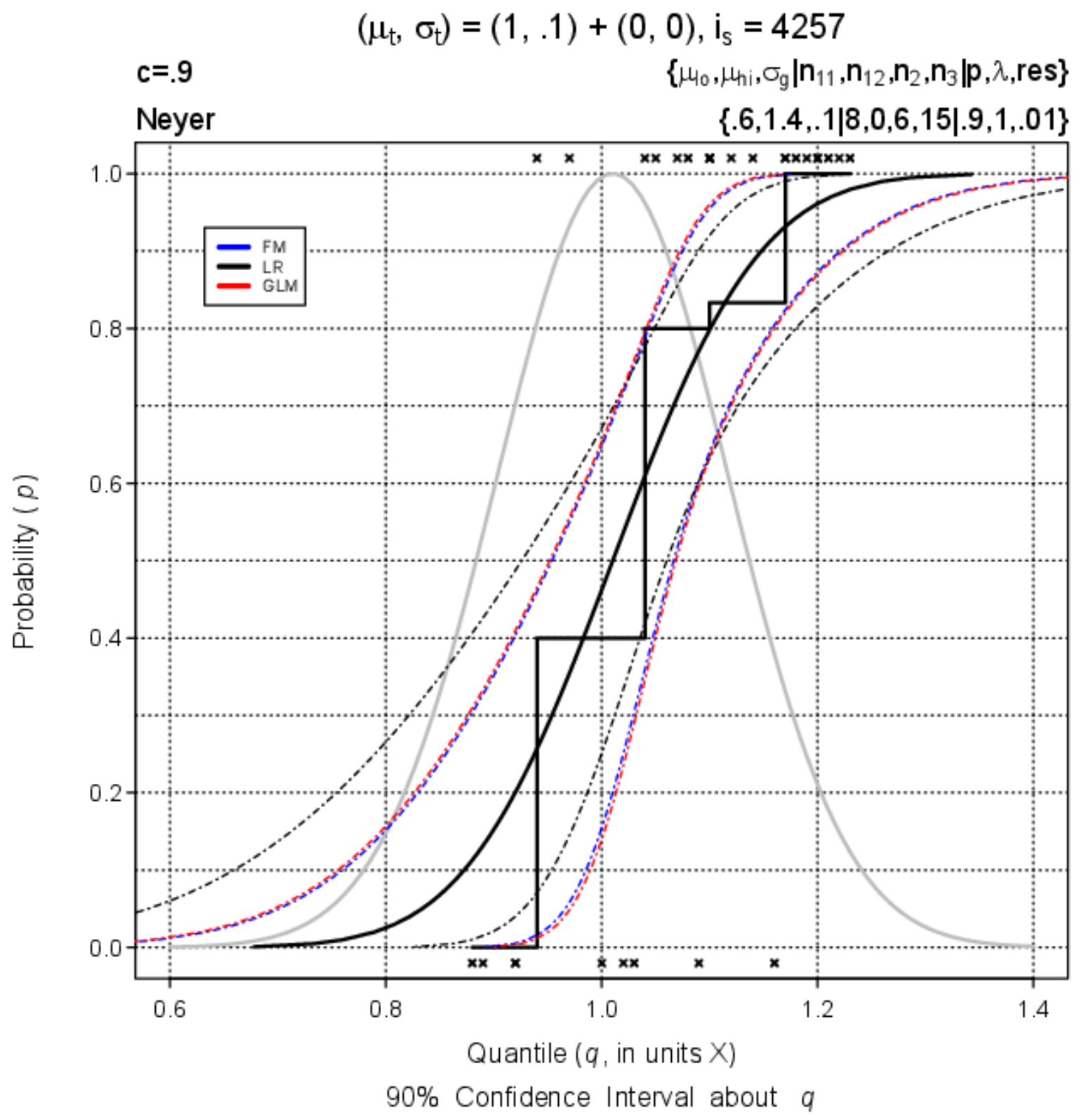

**Figure 18**. Sample ptest(w,3) for w=un, conf=.9 and J=15

The q axis label hints how to read the graph properly, which in this case is the left/right direction. Note, the blue LR curves for p|q (Figure 17) and q|p (Figure 18) are identical. The symmetry property that holds for two types of LR curves does not hold for its FM or GLM counterparts.

### 11.4 ptest(w,4)

Running **ptest(un,4)** produces a simplistic visual of the responses (Y=1) on the top horizontal line and the non-responses (Y=0) on the bottom horizontal line. The two vertical lines delineate the zone of mixed results (ZMR) bounded by m1 and M0. This plot is presented below:

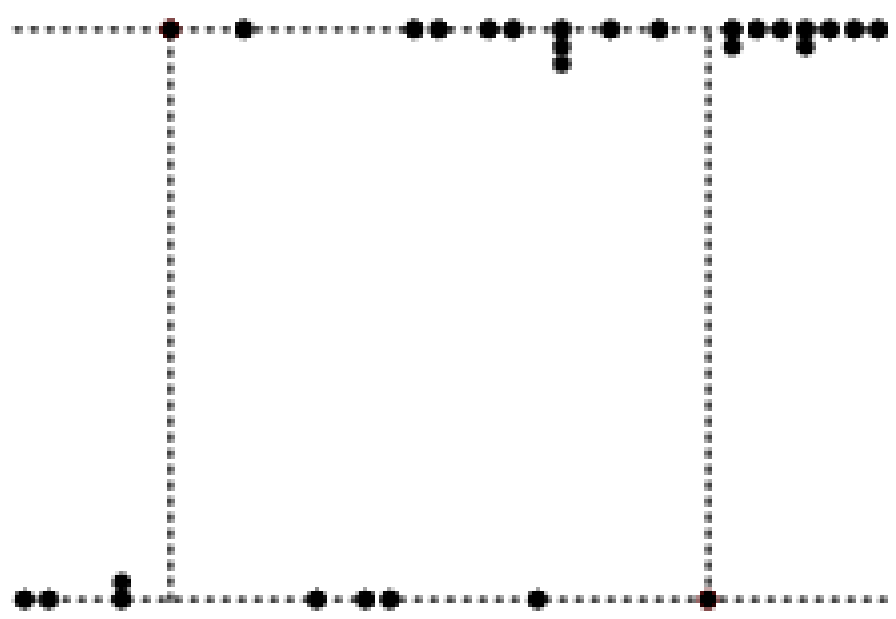

**Figure 19**. Sample ptest(w,4) for w=un

The next table attempts to summarize additional graphics related to confidence intervals and tabular outputs provided by gonogo's ptest(w,i), for i= 5, 6,7, or 8.

**Table 37**. Capabilities provided by ptest(w,i) for i=5,6,7 and 8

| input | method | joint conf region | overlap | output | i=5 jlrcb(w) | i=6 lrcb(w) | i=7 cbs1(w,1) | i=8 cbs2(w,2) |
|---|---|---|---|---|---|---|---|---|
| conf | | | | | | | | X |
| conf | | | | | X | X | X | |
| conf's | | | | | X | X | | |
| p 0 or 0 q | | | | | | X | X | |
| p 0 or 0 q or 0 0 | | | | | | | X | |
| | LR | bounded | Y | | X | X | X* | X* |
| | | | N | | X | X | | |
| | | unbounded | Y | | X | X | X* | X* |
| | | | N | | X | X | | |
| | FM | | Y | | | | | X* |
| | GLM | | Y | | | | | X* |
| | | | | joint LRCB(s) | X | X | X | |
| | | | | individual LRCB's | | X | X | |
| | | | | (x,qnorm(p)), x=q, qlo & qhi | | X | | X |
| | | | | (q,x), x=q, qlo & qhi | | X | | X |
| | | | | (q,qnorm(x)), x=p, plo & phi | | X | | X |
| | | | | 100*(p,x), x=p, plo & phi | | X | | X |
| | | | | list returned | X | | | |
| | | | | text file | | X | X | X |

**Note**. * interval overlap only

The LR confidence bound computations are quite complex. For small to moderate confidence entries (conf, or $C$ ), the joint LR confidence boundary about $(\hat{\mu},\hat{\sigma})$ forms a closed loop. When $C$ is greater than or equal to a critical value, $C_{\max}$, the closed loop opens and becomes unbounded.

$C_{\max}$ depends upon two quantities, the maximum of the log likelihood function achieved by a FULL model (normal or lognormal with response depending on $\mu$ and $\sigma$), and the maximum achieved by a NULL model (response ≡ $p_{con}$, a constant). Calling these maxima $M_{FULL}$ and $M_{NULL}$, one finds that $C_{\max} = pchisq\left(2(M_{FULL} - M_{NULL}),1\right)$ (equivalently, $C_{J\max} = pchisq\left(2(M_{FULL} - M_{NULL}),2\right)$). The Null model maximum occurs when $p_{con} = \overline{Y}$, where $Y$ is the vector of $n$ responses.

In the joint likelihood graphs produced by ptest(w, i=5, 6 or 7), a dashed line is graphed. It represents an axis about which (or to which) the closed LR confidence bounds tend to get stretched out and approach as $C$ nears $C_{\max}$ from below. The slope and intercept of this line are given by $m_0 = -1/\,qnorm(p_{con})$, and $b_0 = \hat{\sigma} - m_0$, respectively.

Two individual LR confidence bounds may be depicted with ptest (w, i=6 or 7), one of standard deviation (s) versus quantile (q) for a given probability (p), the other of standard deviation (s) versus probability (p) for a given quantile (q). The dashed line in the first case has slope and intercept given by $m_1 = 1/\left(qnorm(p) + 1/m_0\right)$ and $b_1 = \hat{\sigma} - m_1 \cdot (\hat{\mu} + qnorm(p)\cdot\hat{\sigma})$, respectively. In the second case, the dashed line is vertical at $p = p_{con}$.

Interestingly, joint and individual LR confidence bounds exist when technical overlap does not. In these cases (of complete and quasi complete separation) gonogo recalibrates the log likelihood function the same way that SenTest™ does.

Details of the joint LR curve calculations are described below:

**Table 38**. How gonogo computes a Joint LR confidence curve (given d0 and $C$ )

| Step | Description | Detail |
|---|---|---|
| 1 | Compute $M_{FULL}$ (Max Log Likelihood, Full Model) | $M_{FULL} = glmmle(d_0)\$maxll$ |
| 2 | Compute $M_{NULL}$ (Max Log Likelihood, Null Model) | $M_{NULL} = glmmle(d_0)\$maxlc$ |
| 3 | Recalibrate (if there's no overlap) | If( $M_{FULL} == 0$ ) $M_{FULL} = M_{NULL}$ |
| 4 | Compute Joint Confidence Level | $C_J$ = pchisq(qchisq( $C$ ,1),2) |
| 5 | Compute Contour Level (of Likelihood function) | lev = (1- $C_J$ )*exp( $M_{FULL}$ ) |
| 6 | Compute $C_{\max}$ ( $C < C_{\max}$ yields bounded joint regions) | $C_{\max}$ = pchisq(2*( $M_{FULL}$ - $M_{NULL}$ ),1) |

In terms of points $(s,\ m)$ on the joint LR curve computed with associated confidence $C_J$, and points $(p,\ q)$ on the MLE response curve, two other relationships are examined: $s$ versus $q$ and $s$ versus $p$. This is accomplished by setting $q = m + qnorm(p) \cdot s$ and $p = pnorm\big((q-m)/s\big)$, respectively.

When the LR curve is closed and bounded ( $C_J < C_{J\max}$ , or equivalently $C < C_{\max}$ ), upper and lower 2-sided 100*C confidence limits for q and p may be determined individually. From the latter two individual LR curves one may eliminate the variable $s$ and graph (or tabulate) four curves:

> $p_u$ versus $q$ , $p_l$ versus $q$ , $p$ versus $q_l$ and $p$ versus $q_u$ . But these 4 curves reduce to two, since $p_u$ versus $q$ is the same as $p$ versus $q_l$, and $p_l$ versus $q$ is the same as $p$ versus $q_u$ .

That the four curves reduce to two is a symmetry property inherited by having been derived from a Joint LR curve. Unfortunately, FM and GLM confidence limits do not share this property of symmetry.

Note that the $p_u$ and $p_l$ versus $q$ curves, and the $p$ versus $q_l$ and $q_u$ curves are not available in the unbounded LR case, i.e., when $C_j \geq C_{J\max}$ , or equivalently $C \geq C_{\max}$ .

### 11.5 ptest(w,5)

The next three ptest graphs (i=5, 6 and 7) create LR curves like the ones featured in SenTest™. Run the command **ul=gonogoSim(10,20,0,3,4,.5,1,plt=0,test=4,reso=.01,dm=-1, ds=.2,iseed=217)**, followed by **ptest(ul,5)**. The latter command invokes the following prompt for conf inputs (separated by blanks).

Enter conf's (separated by blanks): **.5 .6 .7 .8 .9 .95**

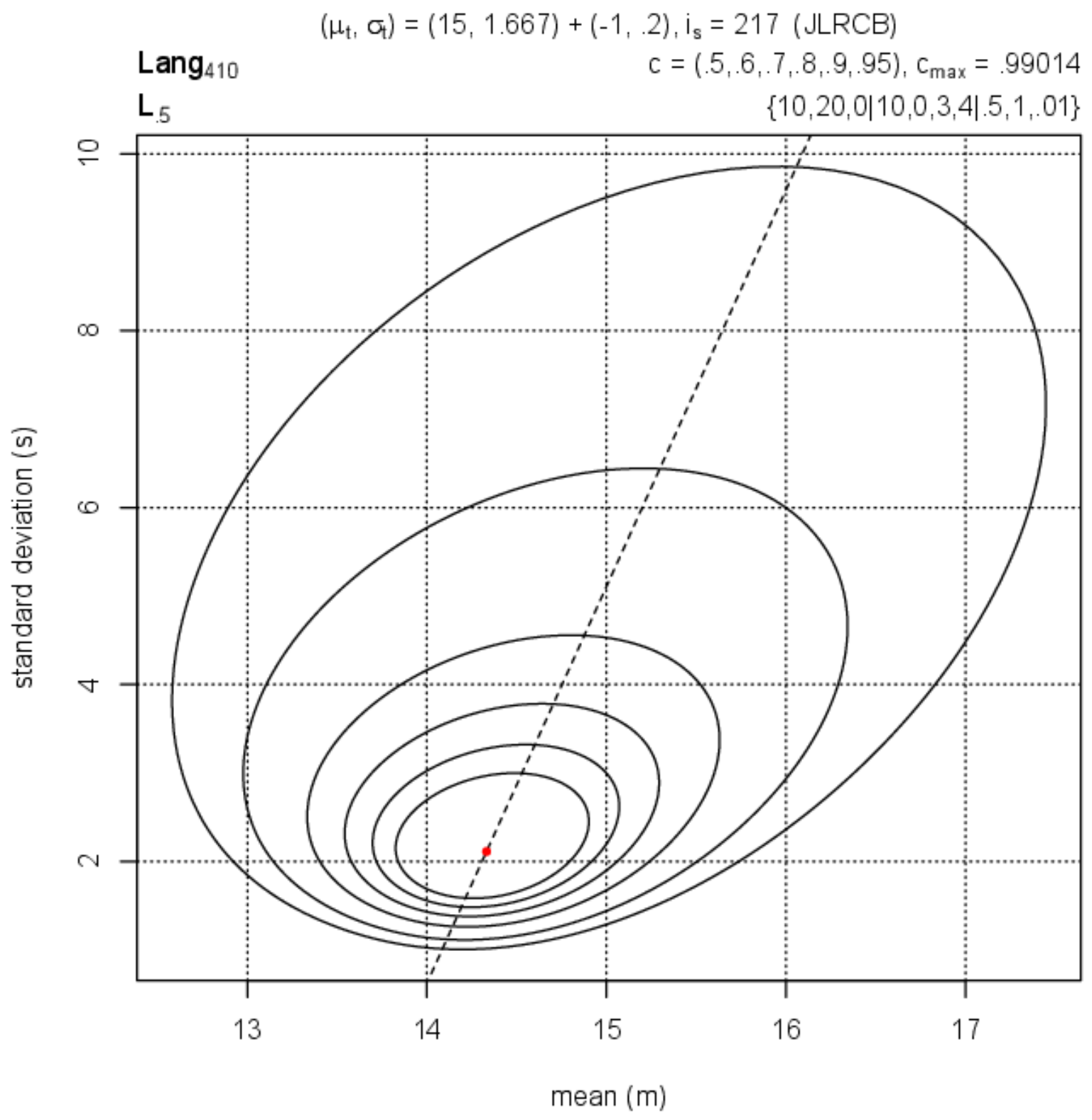

**Figure 20**. Sample ptest(w,5) for w=ul and conf=.5 .6 .7 .8 .9 .95

Since this is our first example with non-zero dm and ds, let's verify that the true mu and sigma reported by gonogoSim is as advertised. Typing ul$tmu and ul$tsig yields14 and 1.866667, respectively, which checks out perfectly.

## 11.6 ptest(w,6)

All of the important details about a test are captured in the list argument, w, to ptest. Sometimes, it may seem that there's not much room in the title area to capture every detail. This is especially true in this graph - where there's only one line available. The user is reminded that the ptest notitle = T option suppresses the default titles and allows you to create your own. Running the command **ptest(ul,6)** invokes two prompts, one for conf inputs (separated by blanks), followed by either p and 0, or 0 and q (**Note**: each non-zero entry determines the other one). An example follows:

Enter conf's (separated by blanks): **.5 .6 .7 .8 .9 .95**

Enter p and q (one must be 0): **.8 0**

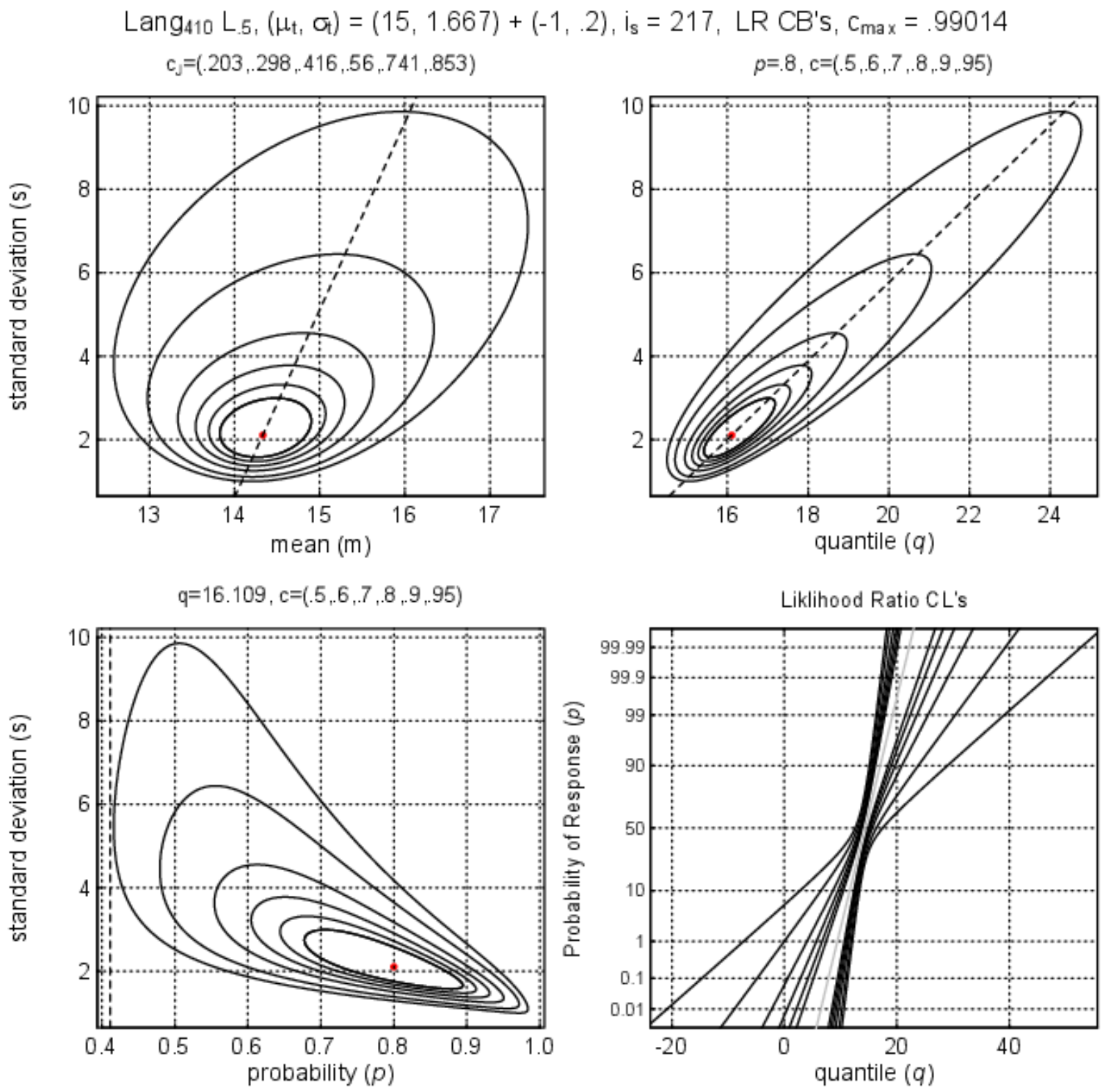

**Figure 21**. Sample ptest(w,6), for conf=.5 .6 .7 .8 .9 .95 and p=.8, q=0 with w=ul.

## 11.7 ptest(w,7)

Running the command **ptest(ul,7)** invokes two prompts, the first for two values, conf and Jconf (separated by blanks, one of which must be 0), the second for two more values, p and q (separated by blanks, at least one must be 0). **Note**: in both cases, the non-zero entry determines the other one.

Here's an example:

Enter conf and Jconf (one must be 0): **.9 0**

Enter p and q (at least one must be 0): **.8 0**

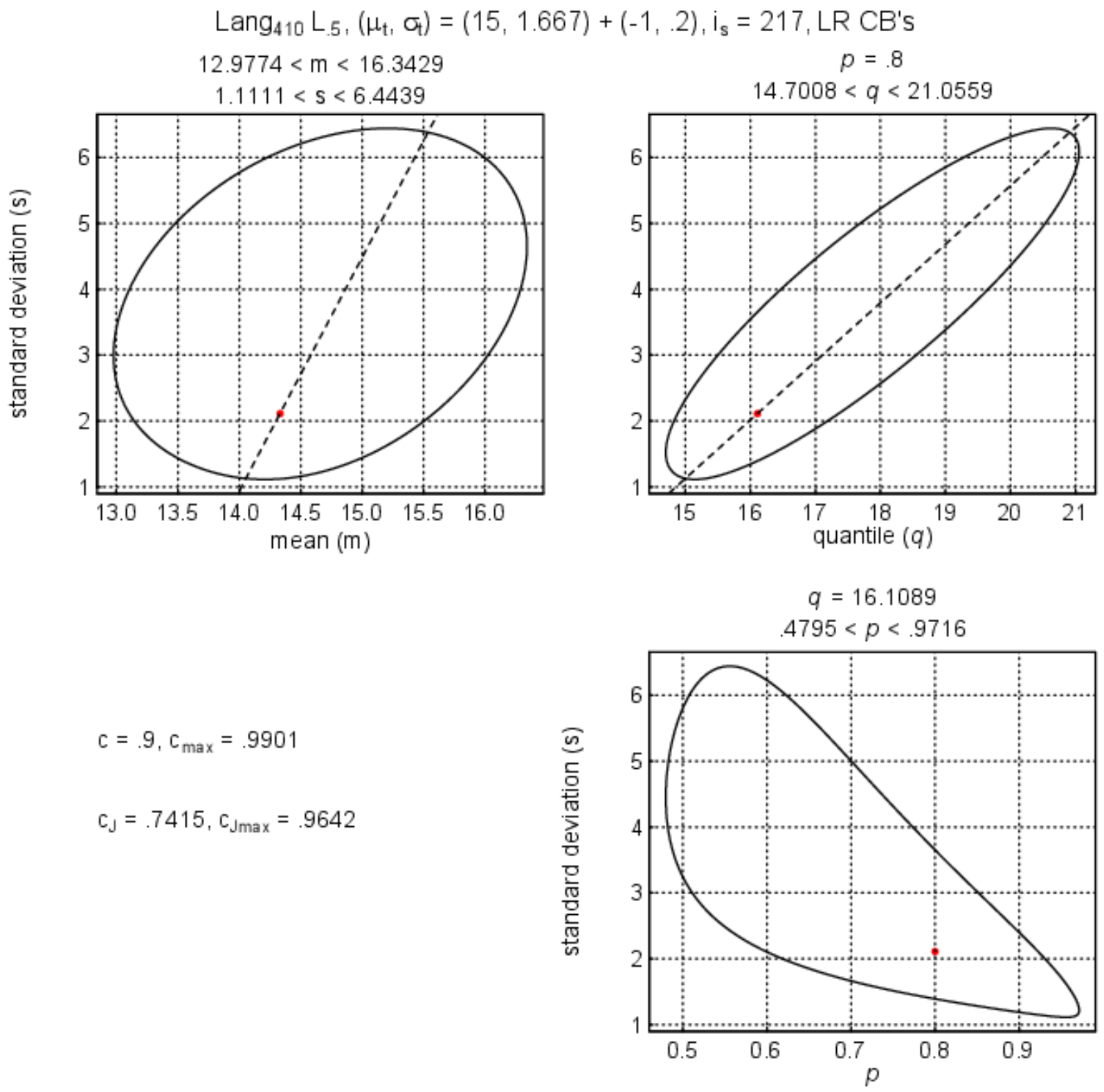

**Figure 22**. Sample ptest(w,7) for conf=.9 and p=.8 (with q=0) with w=ul

## 11.8 ptest(w,8)

Linearized confidence bounds for all three methodologies, FM, LR and GLM, are computed with this graphical function. Here's a sample output for the simulated Langlie test ul, obtained by running the command

**ptest(ul,8)** (which elicits the following prompt)

Enter conf: **.9**

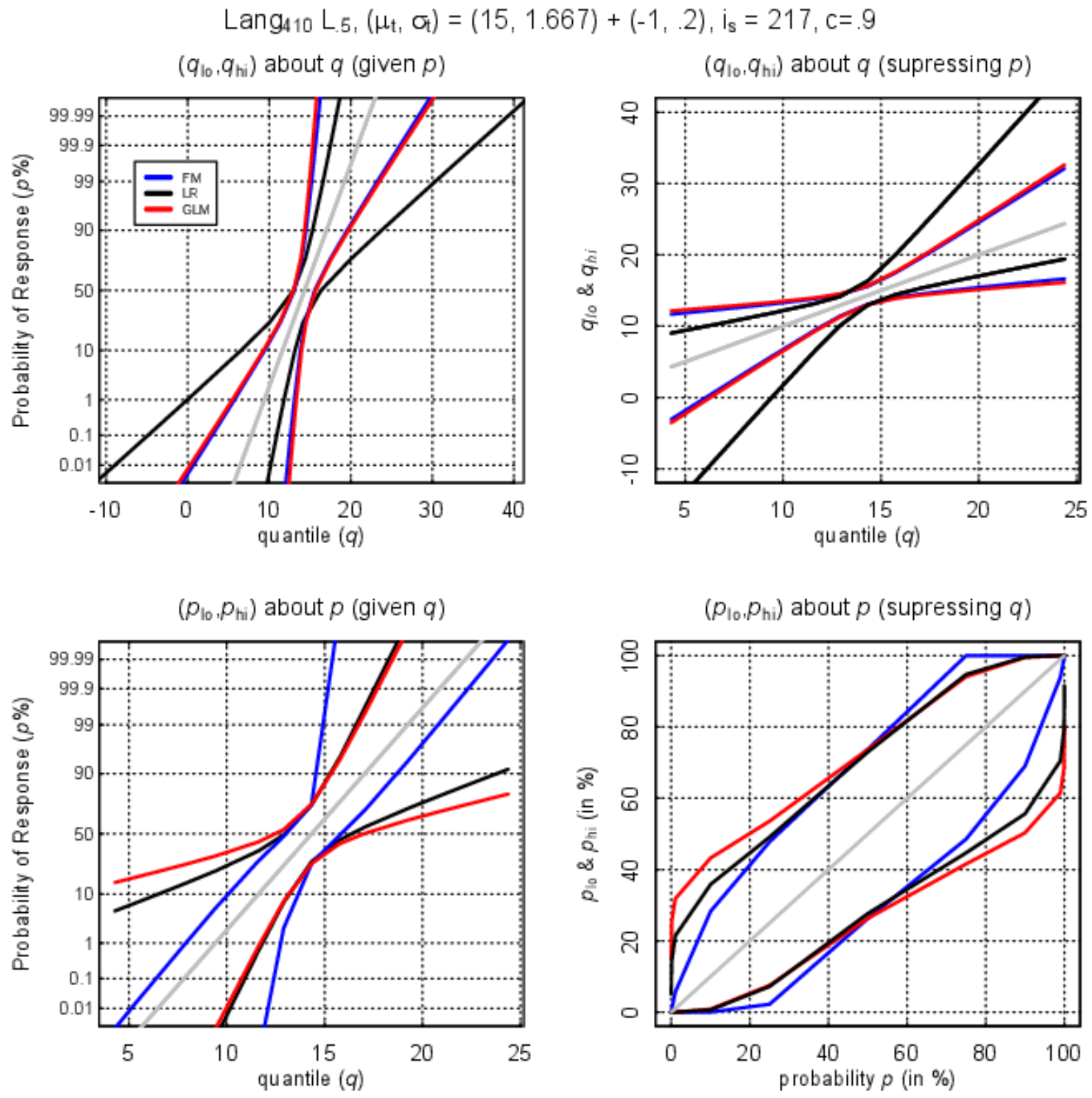

**Figure 23**. Sample ptest(w,8) with w=ul

The above graph is styled like graphs produced by SenTest™ and features linearized versions of ptest(w,3) with J=14 and 15.

## 12. wxdat

The function wxdat contains 27 sample tests, most of which were NOT obtained via gonogo. The call to get the $i^{th}$ data set is w=wxdat(i). You may plot any of these tests with ptest(w,j), for j=4,...,8. If you try ptest(w,j) for j= 1, 2 or 3, you'll get the message: "This function only works for lists created by gonogo".

wxdat contains many degenerate tests (no overlap, single point overlap, infinite sigma). They provide interesting examples which can be examined with the applicable ptest functions. Neyer's recalibration of the likelihood enables "worst case" confidence limits to be calculated for tests having no-overlap. Here's a summary of the data sets (tests) that are included in wxdat:

**Table 39**. Twenty-Seven data sets contained in wxdat(i), for $1 \le i \le 27$

| ic | Description |
|---|---|
| 1 | My Neyer Demo Data, n=8 (SenTest input - My8Shot.sen) |
| 2 | Neyer Data from MIL-STD-331D, n=20 (Appendix G, page 273) |
| 3 | Neyer No ZMR Data, n=8 (SenTest input: nozmr.sen) |
| 4 | Neyer Data, n=30 |
| 5 | No ZMR, n=17 |
| 6 | Infinite Sigma Data, n=4 (SenTest input - InfSig.sen) |
| 7 | Velocity, n=15 |
| 8 | VariDensity, n=24 |
| 9 | VariGap, n=21 |
| 10 | NO ZMR Example |
| 11 | Dror & Steinberg, RP SOR 0607 |
| 12 | Eli Data n=73 |
| 13 | A Neyer Data Set |
| 14 | JF's Data |
| 15 | An n=3 Ex. |
| 16 | An n=4, con=.5 Ex. |
| 17-19 | No Overlap Ex's, (n=3, 2, 4). |
| 20-24 | One Point of Overlap Examples |
| 25 | A Simulated Test |
| 26-27 | 2 point tests: n=4 (overlap, Infinite Sigma), n=6 (overlap) |

Using data catalogued in the wxdat function, some joint LR confidence bounds for four extremely small data sets are illustrated in the next set of graphs:

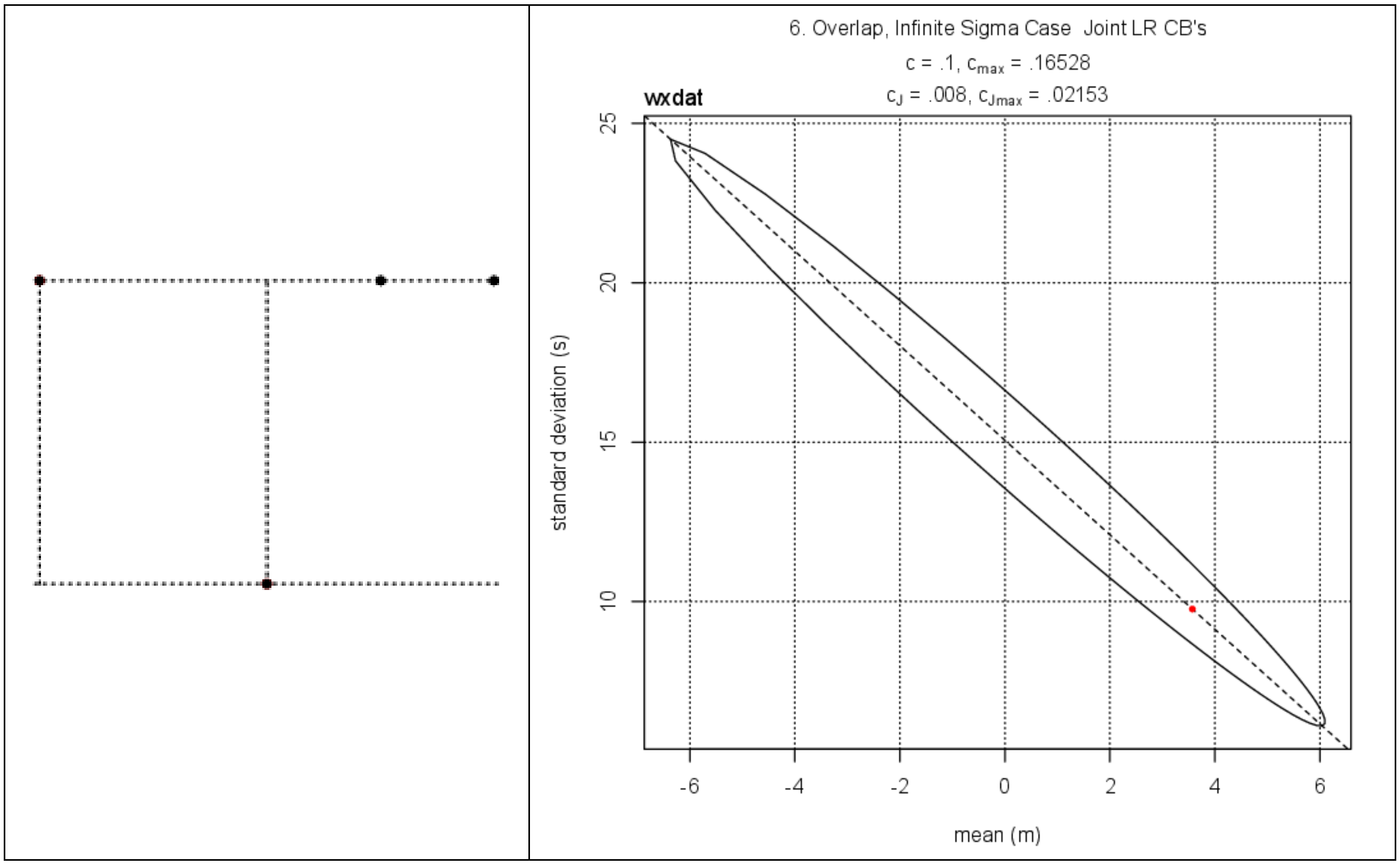

**Figures 24 and 25**. Results of w=wxdat(6), on left and ptest(w,5), on right ( *conf* = *.1* )

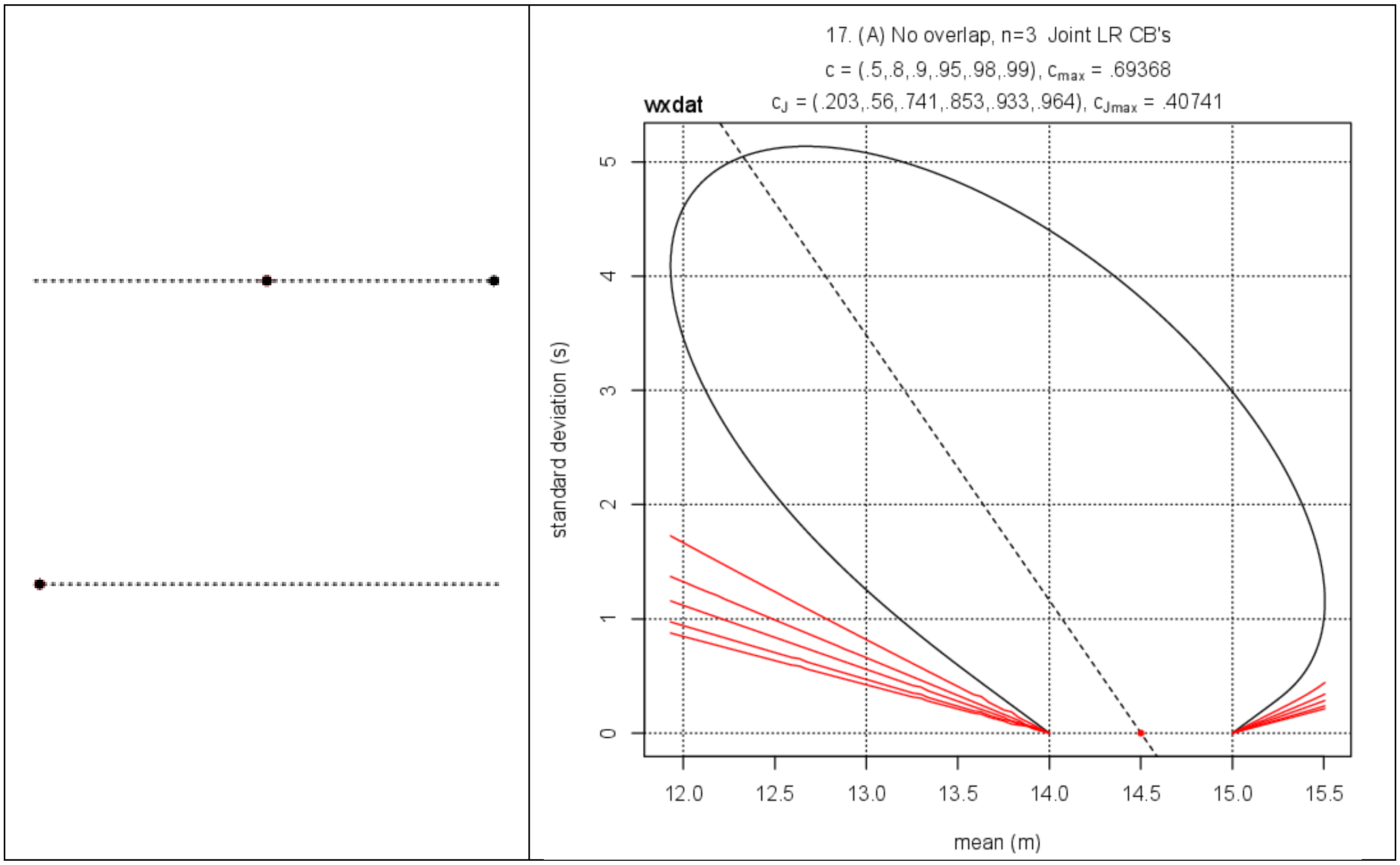

**Figures 26 and 27**. Results of w=wxdat(17), on left and ptest(w,5), on right

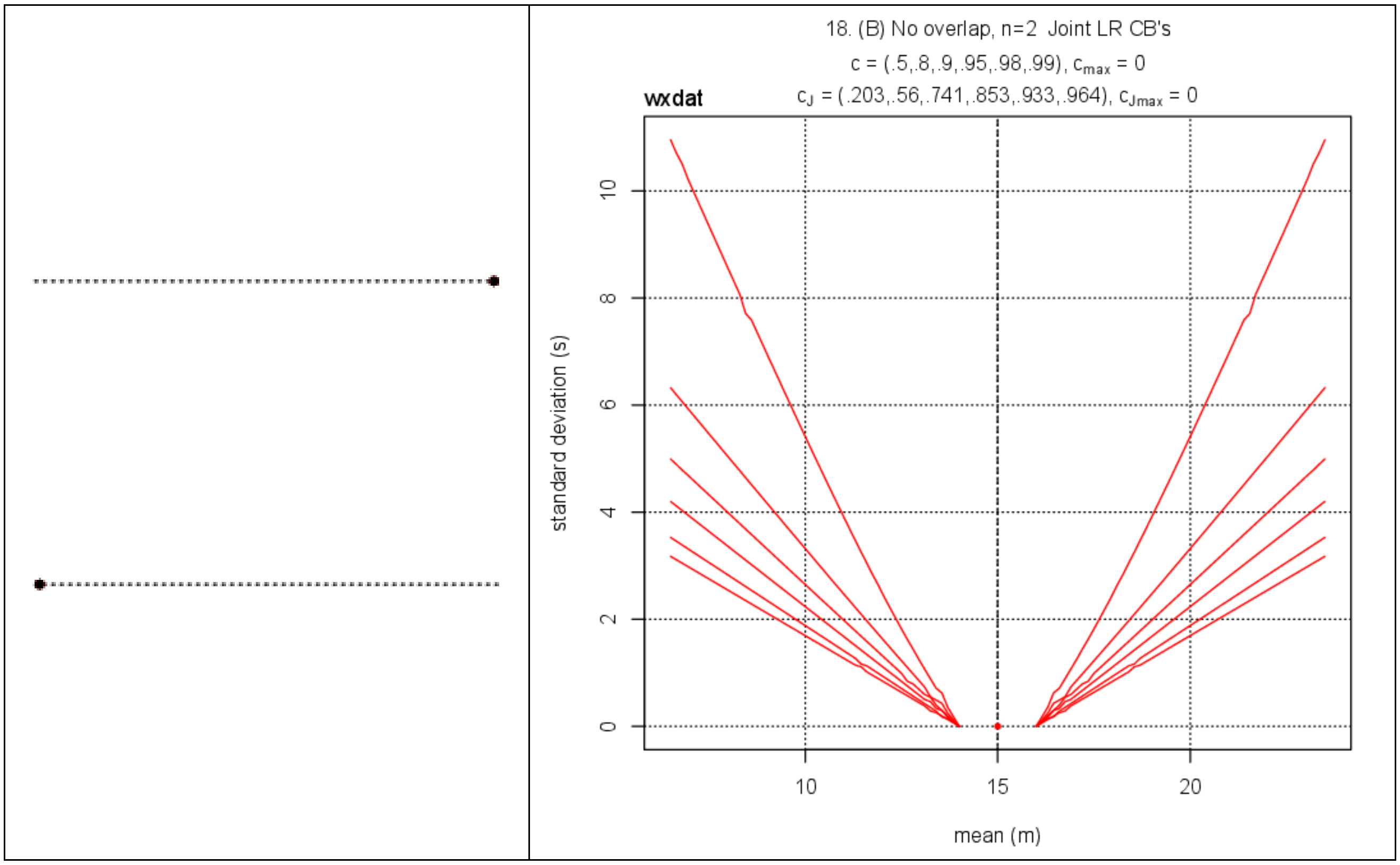

**Figures 28 and 29**. Results of w=wxdat(18), on left and ptest(w,5), on right

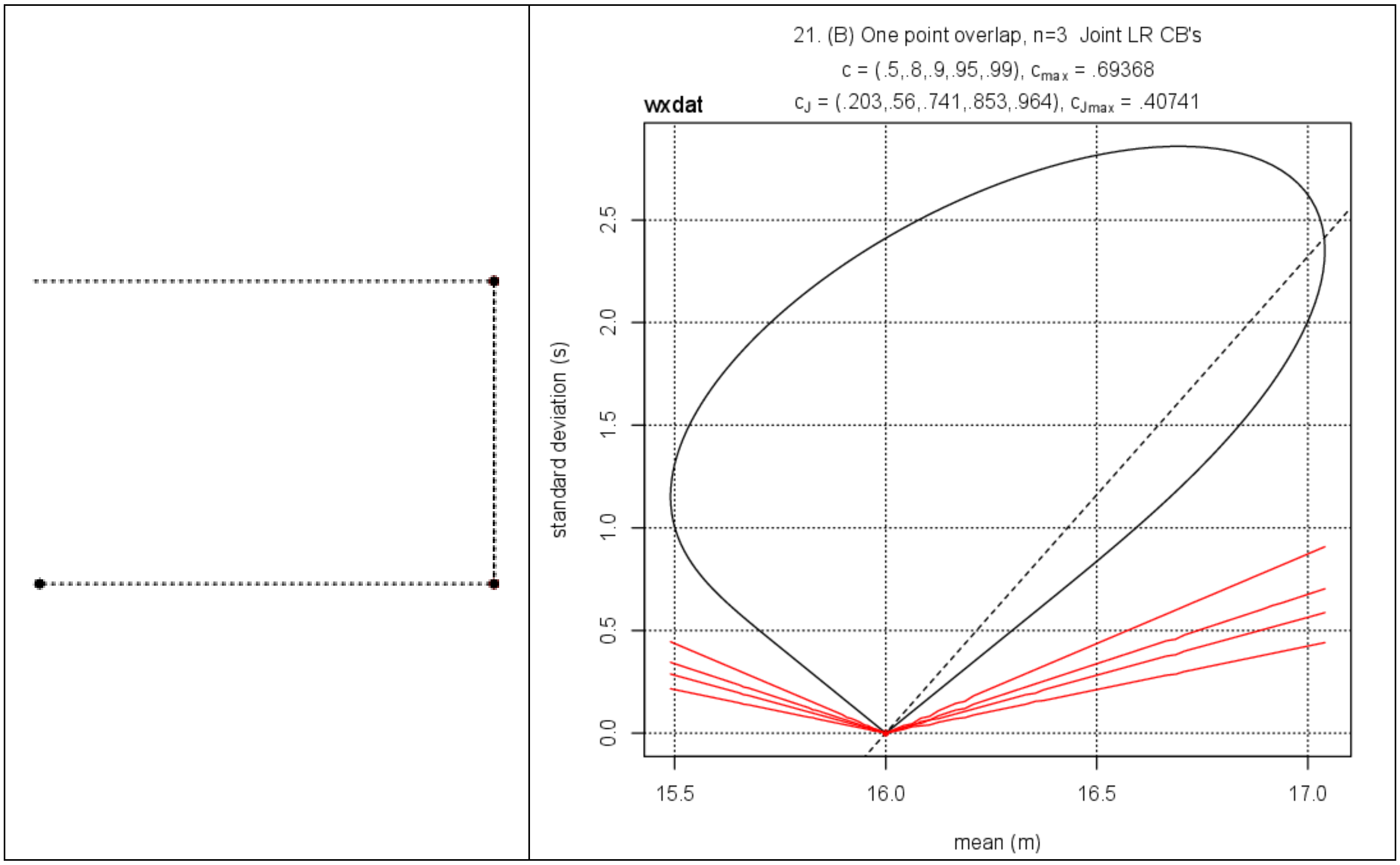

**Figures 30 and 31.** Results of w=wxdat(21), on left and ptest(w,5), on right

## 13. Dimension

All procedures described in this document operate on a single stress factor. According Jeff Wu ( https://casd.wordpress.ncsu.edu/program/)

> As far as we know, there is *no* known procedure for sensitivity testing with two or more stress factors. But the problems are encountered in practice. A good procedure is sorely needed!

There has been considerable interest in developing similar binary response strategies when there are two (or more) stress factors involved. Progress on this front has been slow going.

## 14. Overlap

The existence of maximum likelihood estimates for binary regression problems is a subtle and technical issue. In the one-dimensional case, e.g., all of the gonogo applications described here, the overlap criterion is simple enough: the stress having the largest non-response (M0) must exceed the stress having the smallest response (m1). Since the definitions of response and non-response are reversable, overlap really means there's an **interval** zone of mixed results (ZMR), and either a CDF or a survivor function (one minus a CDF) properly fits the data. Occasionally, when the proper fit is a survivor function, an infinite estimate of sigma is returned for the gonogo procedures discussed here. This subject, related to the sign of the slope coefficient, has been discussed in detail in Owen and Roediger (2014).

A proper infinite sigma exists only in the rare case when the slope coefficient is exactly zero. More frequently, an Infinite sigma improperly occurs because the slope coefficient is negative, which violates the procedural assumption that $\sigma$ is positive.

The first version of this work included in this section the topic of n-dimensional overlap determination (based upon empirical likelihood) and the beginnings of a taxonomy of minimal overlap configurations in dimensions fewer than four. That discussion has been separated from this document and greatly expanded in Roediger (2020).

## 15. Infinite Sigma

Occasionally an infinite sigma estimate can arise. This can happen if the underlying sigma is very small or the responses (Y) were incorrectly labelled (and should have been 1-Y). Also, one erroneous test entry can flip the sign of this estimate.

The assumption for all tests considered here has been that Pr[Y=1|X] is a non-decreasing CDF.

This forces the ML estimate of Sigma to be infinite whenever the sign of $\bar{X}_{Y=1} - \bar{X}_{Y=0}$ is not positive, since the "best" CDF, according to MLE, assigns a constant probability of response, namely $\bar{Y}$, over the entire range of the X's.

Thus, for $pnorm\left((x-\hat{\mu})/\hat{\sigma}\right)=\bar{Y}$ to be true, it follows that $\hat{\mu}$ and $\hat{\sigma}$ satisfy $\lim_{\hat{\sigma}\to\infty}\left(\hat{\mu}+qnorm(\bar{Y})\cdot\hat{\sigma}\right)=0$.

Although the Infinite Sigma case seems far-fetched, it's important to note that overlap can be present. The sign of $\bar{X}_{Y=1} - \bar{X}_{Y=0}$ also has no bearing on overlap. However, $\hat{\sigma}=0$ would be entirely inconsistent with overlap.

Finally, it seems obvious that test procedures considered here could be modified to accommodate both kinds of response models:

a CDF and a Survivor function (one minus a CDF).

If formulated this way, the modified test procedures could adjust the tilt of the model to match the current sign of the logistic regression.

## Appendix A. Gonogo Functions and Constants

**Table A1**. GONOGO.R Functions and Constants (*), in the file's order

| i | gonogo.R |
|---|---|
| 1 | **gonogo(mlo=0,mhi=0,sg=0,newz=T,reso=0,ln=F,test=1,term1=T,BL=NULL,Y=NULL,X=NULL)** |
| 2 | phaseI1(dat0,mlo,mhi,sg,reso,about,titl,unit,ln) |
| 3 | phaseI2(d0,dat0,sg,reso,about,titl,unit,ln,term1) |
| 4 | phaseI3(d0,dat0,sg,reso,about,titl,unit,ln,term1) |
| 5 | nphaseI(dat0,mlo,mhi,sg,reso,about,titl,unit,ln,term1) |
| 6 | bphaseI(dat0,mlo,mhi,sg,reso,about,titl,unit,ln,BL) |
| 7 | lphaseI(dat0,mlo,mhi,sg,reso,about,titl,unit,ln,BL) |
| 8 | phaseII(d0,dat0,n2,reso,about,titl,unit,ln,term1) |
| 9 | sphaseIII(d0,dat0,n3,p,reso,about,titl,unit,ln,lam=0) |
| 10 | getd0(xx,d0,dat0,ID,reso,about,titl,unit,ln,cab=F) |
| 11 | getxr(x,nd0,reso,ln) |
| 12 | phaseBI1(dat0,mlo,mhi,sg,reso,about,titl,unit,ln,Y,X) |
| 13 | phaseBI2(d0,dat0,sg,reso,about,titl,unit,ln,term1,Y,X) |
| 14 | phaseBI3(d0,dat0,sg,reso,about,titl,unit,ln,term1,Y,X) |
| 15 | nphaseBI(dat0,mlo,mhi,sg,reso,about,titl,unit,ln,term1,Y,X) |
| 16 | bphaseBI(dat0,mlo,mhi,sg,reso,about,titl,unit,ln,BL,Y,Xx) |
| 17 | lphaseBI(dat0,mlo,mhi,sg,reso,about,titl,unit,ln,BL,Y,Xx) |
| 18 | phaseBII(d0,dat0,n2,reso,about,titl,unit,ln,term1,Y,X) |
| 19 | sphaseBIII(d0,dat0,n3,p,reso,about,titl,unit,ln,Y,X,lam=0) |
| 20 | getBd0(xx,d0,dat0,ID,reso,about,titl,unit,ln,Y,X,cab=F) |
| 21 | getBxr(x,nd0,reso,ln,Y,X) |
| 22 | prd0(z) |
| 23 | d.update(dat) |
| 24 | ok1(dat,y=2) |
| 25 | chabout(about,s47,loc) |
| 26 | wabout(vv) |
| 27 | blrb7() |
| 28 | blrb8() |
| 29 | about4(x) |
| 30 | **fixw(w,k=1)** |
| 31 | fixw1(w) |
| 32 | **pdat1(dat,notitle=F,ud=F)** |
| 33 | pdat2(dat,notitle=F) |
| 34 | pdat3(w,notitle=F) |
| 35 | intToBitVect(x) |
| 36 | **gonogoSim(mlo,mhi,sg,n2=0,n3=0,p=0,lam=0,dm=0,ds=0,reso=0,ln=F,plt=0, iseed=-1,IIgo=T,M=1,test=1,BL=c(4,1,0))** |
| 37 | pI1(mlo,mhi,sg,tmu,tsig,reso,ln,iseed,dat0=data.frame(numeric(0))) |
| 38 | pI2(d0,dat0,sg,tmu,tsig,reso,ln,iseed) |
| 39 | pI3(d0,dat0,sg,tmu,tsig,reso,ln,iseed) |
| 40 | npI(mlo,mhi,sg,tmu,tsig,reso,ln,iseed,dat0=data.frame(numeric(0))) |
| 41 | bpI(mlo,mhi,sg,tmu,tsig,reso,ln,iseed,BL,dat0=data.frame(numeric(0))) |
| 42 | lpI(mlo,mhi,sg,tmu,tsig,reso,ln,iseed,BL,dat0=data.frame(numeric(0))) |
| 43 | pII(d0,dat0,tmu,tsig,n2,reso,ln,iseed) |
| 44 | spIIIsim(d0,dat0,tmu,tsig,n3,p,lam,reso,ln,iseed) |
| 45 | gd0(xx,d0,dat0,ID,mu,sig,reso,ln,iseed=-1) |
| 46 | gxr(x,mu,sig,reso,ln,iset) |
| 47 | ntau(dat,response=1) |
| 48 | wabout13(M,cmlo,cmhi,csg,p,n11,n12,n2,n3,lam,reso) |
| 49 | **pSdat1(dat,notitle=F,ud=F)** |
| 50 | pSdat2(dat,notitle=F) |
| 51 | pSdat3(dat,notitle=F) |
| 52 | **nmel3(mu,sg,conf,nt,ns,isd0,test=1,targp=.5,meth=3,fixx=T,inum=-1,icirc=numeric(0))** |
| 53 | **plotmm=function(mm,icirc=numeric(0))** |
| 54 | skewL(c1,nu,tau2,p,be) |
| 55 | n.update(dat) |
| 56 | m.update(dat) |
| 57 | **glmmle(mydata)** |
| 58 | llik(mydata, mu, sig) |
| 59 | tauf(x,y) |
| 60 | yinfomat(dat, mu, sig) |
| 61 | Sk(k,b) |
| 62 | Gk(k) |
| 63 | dgs(k,b) |
| 64 | kstar(b) |
| 65 | pavdf(data.df,ln,plotit=F,lineit=F,labx="STIMULUS",laby="PROBABILITY OF RESPONSE",titl="PAV SOLUTION") |
| 66 | blrb1() |
| 67 | blrb2() |
| 68 | blrb3() |
| 69 | blrb4() |
| 70 | blrb5() |
| 71 | blrb6() |
| 72 | f38(x,l) |
| 73 | f3point8(l) |
| 74 | fgs(mlo,mhi,sg) |
| 75 | ifg(fg0,fg1) |
| 76 | addneyr(dtt,ylm,sim=F) |
| 77 | add3pod(dtt,ylm,sim=F) |
| 78 | addBorL(dtt,ylm,ud) |
| 79 | **ptest(dat,plt,notitle=F)** |
| 80 | reset() |
| *81 | al15 |
| *82 | al49 |
| 83 | **lims(ctyp,dat,conf,P=numeric(0),Q=numeric(0))** |
| 84 | cpq(P,Q,mu,sig,gt) |
| 85 | fm.lims(dat,conf,P=numeric(0),Q=numeric(0)) |
| 86 | glm.lims(dat,conf,P=numeric(0),Q=numeric(0)) |
| 87 | lrq.lims(dat,conf1,P=numeric(0),Q=numeric(0)) |
| 88 | qrda(dat,conf=.9,J=2,ln=F,labx="",laby="Probability of Response",zee=0) |
| 89 | prtrans(i) |
| 90 | fofL(L) |
| 91 | iofL(L) |
| 92 | bintodec(y) |
| 93 | udli(i) |
| 94 | pfun(pee,n) |
| 95 | zpfun(i) |
| 96 | xlead0(num,dig) |
| 97 | xyllik(rx,ry,m,s) |
| 98 | stopQuietly(...) |
| 99 | calcblim(bl,ul) |
| 100 | mkb0(confv) |
| 101 | unbd(rx,ry,levs,mh1,mh2,es,mlim) |
| 102 | uliknext(rx,ry,levs0,em1,es,em2) |
| 103 | ulik(rx,ry,levs0,em,es) |
| 104 | otherpoint(rx,ry,muhat,levs0,con) |
| 105 | jlik(rx,ry,levs0,ms,op,one23) |
| 106 | jlrcb(dat,notitle=F) |
| 107 | picdat(dat) |
| 108 | simp(dat) |
| 109 | grafl(limx) |
| 110 | clim0(rx,ry,m,s,levb) |
| 111 | clim(rx,ry,m,s,uu,levb) |
| 112 | abllik(data,mu,sig) |
| 113 | lrcb(dat,notitle=F) |
| 114 | mdose.p(obj,p) |
| 115 | mixed(dat) |
| 116 | graf1(limx,t1,k,big,legnd) |
| 117 | lrmax(w,plt=F) |
| 118 | cbs(w,plt,notitle=F) |
| 119 | **wxdat(ic,plt=T)** |
| 120 | **figtab=function(i,cro=F)** |

## Appendix B. Two Phase I Flow Charts

## 3pod

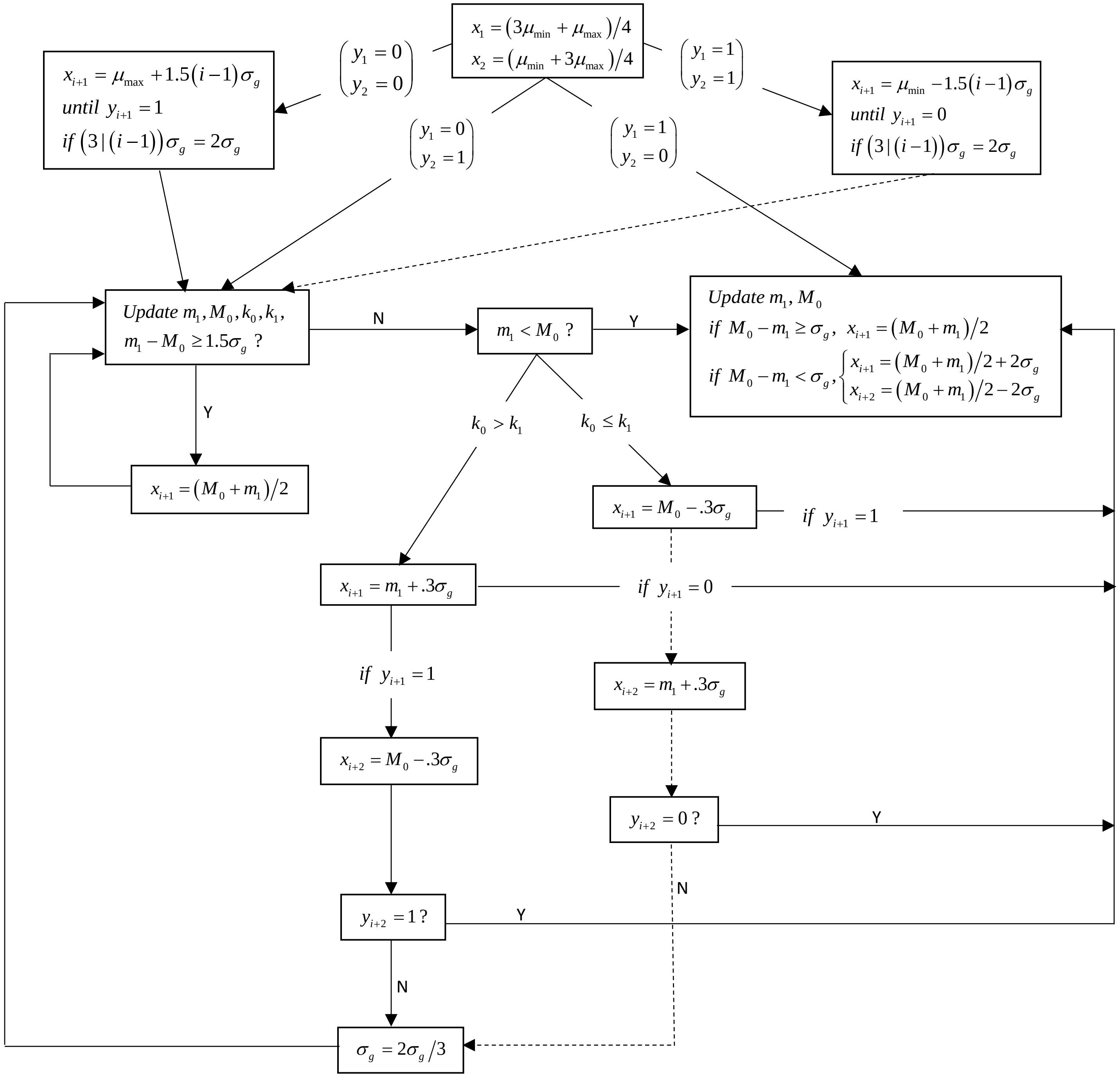

**3pod Phase I Flow Diagram**

**Neyer**

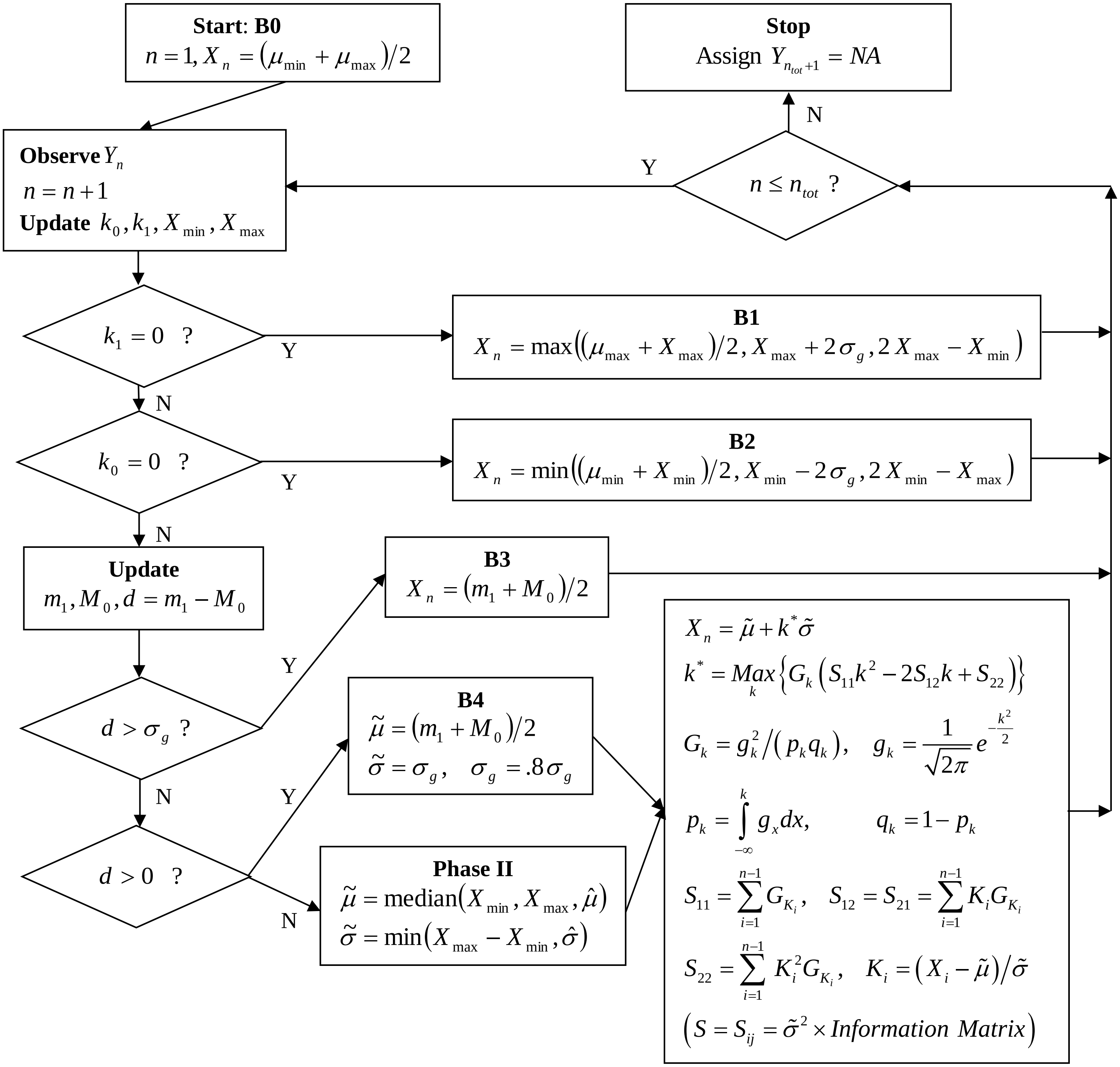

**Neyer Phase I Flow Diagram (with gonogo Block identifiers)**